\documentclass[twocolumn,superscriptaddress,nolongbibliography,amsmath,amssymb,aps,prb]{revtex4-2}
\usepackage[colorlinks=true,urlcolor=blue,citecolor=blue,linkcolor=blue]{hyperref}
\usepackage{amsmath,amssymb,amsthm}
\usepackage{caption}
\usepackage{float}
\usepackage{graphicx,subfigure}
\usepackage{hyperref}
\usepackage{tikz}
\usepackage{url}
\usepackage{adjustbox}
\usepackage[capitalise]{cleveref}
\usepackage{newtxmath}
\usepackage{bm}

\begin{document}
\title{Differentiable programming tensor networks for Kitaev magnets}
\author{Xing-Yu Zhang}
\affiliation{Institute of Physics, Chinese Academy of Sciences, Beijing 100190, China}
\affiliation{Department of Physics, University of Science and Technical of China, Beijing 100080, China}

\author{Shuang Liang}
\affiliation{Institute of Physics, Chinese Academy of Sciences, Beijing 100190, China}

\author{Hai-Jun Liao}
\email{navyphysics@iphy.ac.cn}
\affiliation{Institute of Physics, Chinese Academy of Sciences, Beijing 100190, China}
\affiliation{Songshan Lake Materials Laboratory, Dongguan, Guangdong 523808, China}

\author{Wei Li}
\email{w.li@itp.ac.cn}
\affiliation{Institute of Theoretical Physics, Chinese Academy of Sciences, Beijing 100190, China}
\affiliation{CAS Center for Excellence in Topological Quantum Computation, Chinese Academy of Sciences, Beijing 100190, China}

\author{Lei Wang}
\email{wanglei@iphy.ac.cn}
\affiliation{Institute of Physics, Chinese Academy of Sciences, Beijing 100190, China}
\affiliation{Songshan Lake Materials Laboratory, Dongguan, Guangdong 523808, China}

\begin{abstract}
We present a general computational framework to investigate ground state properties of quantum spin models on infinite two-dimensional lattices using automatic differentiation-based gradient optimization of infinite projected entangled-pair states. The approach exploits the variational uniform matrix product states to contract infinite tensor networks with unit-cell structure and incorporates automatic differentiation to optimize the local tensors. We applied this framework to the Kitaev-type model, which involves complex interactions and competing ground states. To evaluate the accuracy of this method, we compared the results with exact solutions for the Kitaev model and found that it has a better agreement for various observables compared to previous tensor network calculations based on imaginary-time projection. Additionally, by finding out the ground state with lower variational energy compared to previous studies, we provided convincing evidence for the existence of nematic paramagnetic phases and 18-site configuration in the phase diagram of the $K$-$\Gamma$ model.  Furthermore, in the case of the realistic $K$-$J$-$\Gamma$-$\Gamma^\prime$ model for the Kitaev material $\alpha$-RuCl$_3$, we discovered a non-colinear zigzag ground state. Lastly, we also find that the strength of the critical out-of-plane magnetic field that suppresses such a zigzag state has a lower transition field value than the previous finite-cylinder calculations. The framework is versatile and will be useful for a quick scan of phase diagrams for a broad class of quantum spin models. 

%We present a novel computational framework for investigating quantum spin models on infinite two-dimensional lattices using automatic differentiation-based gradient optimization of infinite projected entangled-pair states. The proposed approach exploits the variational uniform matrix product states to effectively contract the tensor network and incorporates automatic differentiation optimization of infinite projected entangled-pair states. We apply this framework to the Kitaev-type model, which is characterized by complex interactions and ordered states. To assess the precision of the proposed method, we compare the results with exact solutions for the Kitaev model and demonstrate that it exhibits a better agreement for various observables than the best imaginary-time evolution full update optimization result available to date. Furthermore, we provide additional evidence for nematic paramagnetic phases in the $K$-$\Gamma$ model, which were calculated directly or by extrapolation. Additionally, we observe that the 18-site configuration exhibited lower energy than the simple update. Moreover, in the case of the realistic $K$-$J$-$\Gamma$-$\Gamma^\prime$ model for the Kitaev material $\alpha$-RuCl$_3$, we identify a non-colinear zigzag ground state. Furthermore, we discover a critical out-of-plane magnetic field strength that suppresses the zigzag state, with a lower transition field value than the previous finite-cylinder calculations.
\end{abstract}
\maketitle

\tableofcontents

\section{Introduction}
\label{sec:introduction}
%\red{need to reorganize 1) Kitaev model -> Kitaev materials -> complex hamiltonians and order with large unit cell 2) calls for reliable methods. However, QMC has sign problem. TNS has difficulty obtaining large unit cell ground state.}
The Kitaev model is a prominent example of frustrated magnets, which consists of anisotropic nearest neighbor Ising interactions with three different types of bonds on a honeycomb lattice. The model has an exact solution for its ground state, which hosts a novel phase called quantum spin liquid (QSL)~\cite{kitaev2006anyons}. Moreover, with the proposal of the existence of Kitaev interaction in real materials~\cite{jackeli2009mott}, Kitaev materials like $\alpha$-RuCl$_3$ have been intensively studied~\cite{PhysRevB.93.214431, 
Winter_2017, doi:10.1146/annurev-conmatphys-033117-053934,takagi2019concept,PhysRevB.95.174429}. A bunch of spin models like the $K$-$\Gamma$ model and the $K$-$J$-$\Gamma$-$\Gamma'$ model with various couplings were proposed and there was no consensus about which one could fully describe the low energy properties of real materials~\cite{PhysRevB.93.214431,Winter_2017, Wei2017theoretical,PhysRevLett.118.107203, PhysRevB.99.249902, PhysRevB.98.094425, PhysRevB.98.060412, PhysRevB.93.075144, li2021identification}. There are two major difficulties in the numerical studies of these spin models in the thermodynamic limit. One is that these models are highly frustrated, which causes sign problem in quantum Monte Carlo methods. Though one could avoid this problem by using the tensor network algorithms, another issue may arise. In these models, the competition between various interactions can lead to the formation of magnetic orders with large unit cells. However, dealing with such situations is not easy because of the realistically limited representation capability of one-dimensional (1D) tensor networks, such as the density matrix renormalization group (DMRG) algorithm. While the DMRG algorithm is capable of mapping a two-dimensional (2D) system to a 1D system, the width of the cylinder that can be studied is limited due to computational complexity~\cite{PhysRevB.83.245104, vodola2015long, PhysRevB.93.075129, gohlke2017dynamics, PhysRevB.98.014418, PhysRevB.97.241110, gohlke2018quantum, lee2020magnetic}. Consequently, an alternative method is urgently required to obtain the ground state of systems with frustration and spin configurations characterized by large unit cells.

To solve 2D lattice models directly in the thermodynamic limit, one can use tensor network wave function ansatz, such as the infinite projected entangled-pair states (iPEPS)~\cite{verstraete2004renormalization}. Furthermore, multisite iPEPS with the large unit cell can represent complex large configuration wave functions directly~\cite{iregui2014probing, PhysRevB.95.024426, PhysRevB.98.184409, PhysRevB.100.165147, lee2020magnetic}. However, there are still two obstructions in the way of conducting groundstate iPEPS calculations with reduced computational cost and improved accuracy.

One obstruction to reducing computational cost is the contraction of the infinite 2D tensor network with a large unit cell. The problem at hand can be addressed through the utilization of the corner transfer matrix renormalization group~\cite{nishino1997corner, PhysRevB.80.094403} (CTMRG) and the variational uniform matrix product states~\cite{zauner2018variational,fishman2018faster, 10.21468/SciPostPhysLectNotes.7} (VUMPS) algorithm. These methods effectively contract the infinite 2D tensor network by incorporating the boundary environments. They differ in two distinct ways. Firstly, there is a discrepancy in the contraction order between CTMRG and VUMPS. Secondly, VUMPS directly addresses the eigenproblem, while CTMRG employs a power method-like iteration. The seamless integration of these two techniques presents an intriguing and promising avenue for capitalizing on their respective advantages.

Their computational cost for large unit cell extensions depends linearly on the size of the unit cell. However, it is barely satisfactory that the large unit cell versions of CTMRG and VUMPS both have some tricky problems. The iterative nature of the CTMRG extension, as described by Ref~\cite{corboz2014competing}, which resembles that of the power method, may affect its rate of convergence to some extent. Moreover, a general principle for optimal approximate contraction has yet to be established. The extension of VUMPS~\cite{nietner2020efficient} in the large cell does not suffer a such problem. But the non-hermitian transform matrix emerging in the large unit cell is not very well recognized and handled properly until Ref~\cite{vanderstraeten2022variational}. What's more, previous work benchmarks their large unit cell in the 2D classical Ising model with $2 \times 2$ unit cell. Whether it works well in quantum systems and larger unit cell is unclear.

The other obstruction is the inadequate optimization of iPEPS towards the ground state. In prior works, the imaginary time evolution (ITE) method, including the simple update~\cite{PhysRevLett.101.090603} or full update method~\cite{PhysRevB.81.165104, PhysRevB.92.035142}, is often employed to update iPEPS. Nonetheless, the ground state iPEPS obtained via these methods is often inaccurate due to the separate update of energy bonds, especially in the case of frustrated systems. 

Moreover, the effectiveness of the ITE method is heavily reliant on the initialization of iPEPS, particularly for large unit cells. A general approach to finding an unconventional ordered phase using ITE is laborious and heavily relies on artificial expertise. Specifically, the classical model is simulated first~\cite{PhysRevB.104.094431}, followed by an attempt to identify a similar configuration in the quantum case with the same parameters via numerous meticulously selected initializations of iPEPS for the large unit cell~\cite{lee2020magnetic, li2022tangle}.

One can leverage differentiable programming techniques for the gradient-based optimization of iPEPS. This approach has been proven to be powerful in spin systems~\cite{liao2019differentiable, 10.21468/SciPostPhys.10.1.012, https://doi.org/10.48550/arxiv.2111.07368}, even in the frustrated Kitaev model~\cite{Liao2019talk, PhysRevLett.129.177201}. Compared to traditional gradient-based tensor methods~\cite{corboz2016variational,vanderstraeten2016gradient}, automatic differentiation (AD) avoids cumbersome graph summations. We observe that optimization, rather than representation ability, is crucial for obtaining the ground state of Kitaev-type frustrated systems because we can obtain the exact ground state energy and correct observables using much smaller iPEPS with virtual bonds $D$ compared to ITE. In principle, once given a forward algorithm to get observables, such as large unit cell VUMPS in this paper, one can get the backward algorithm immediately in the same order of computation complexity. But in practice, we need to carefully design the backward algorithm to guarantee numerical stability and computational efficiency. 
% We present the backward algorithm of large unit cell VUMPS in~\cref{subsec: Optimize iPEPS by automatic dofferentiation}.

In this paper, we will study the iPEPS tensor network ansatz wave functions with large unit cells and their differentiable programming optimization method for the Kitaev-type model. The primary focus of this study is to enhance the optimization capability of iPEPS within comparable dimensions when compared to alternative methodologies, specifically for large unit cell configurations that remain efficient and without artificial effect.

The organization of this paper is as follows. First,~\cref{sec: Methods} reviews the method of ground state iPEPS simulation in the thermodynamic limit and contraction algorithm of large unit cell VUMPS with its carefully designed automatic differentiation optimization. Then, in~\cref{sec: Application} we present the application in the Kitaev-type systems with various competing phases.~\cref{sec: Discussion} discusses possible issues for future development. 

Our paper is accompanied by a ready-to-use Julia package implementing differentiable large unit cell VUMPS calculations~\cite{TeneT.jl, AD-Kitaev}. 

\section{Methods}
\label{sec: Methods}

We will present a method for obtaining the ground state wave functions in the thermodynamic limit using iPEPS when given a Hamiltonian. Our approach involves first calculating the energy through the contraction of an infinite-sized tensor network with randomly initialized iPEPS. This contraction is performed using the VUMPS algorithm and its larger unit cell version. Then, we will optimize the energy to obtain the iPEPS ground state through the gradient calculated by AD. In this process, a well-designed VUMPS backward will be introduced.

\subsection{iPEPS ansatz}
\label{subsec:iPEPS Ansatz and energy contraction}
In general, accurately representing the quantum state of a system with $N$ sites and $d$ states on each site requires $d^N$ parameters, posing significant challenges for numerical simulations, particularly when $N$ is large. Fortunately, the area law of entanglement~\cite{eisert2010colloquium} suggests that a typical quantum ground state is less entangled than an arbitrary state and can be effectively parameterized by tensor networks~\cite{schollwock2005density, perez2006matrix}.
%Generally, to precisely represent a quantum state of a system with $N$ sites and $d$ states on each site, $d^N$ parameters are needed. It results in great difficulties for numerical simulations especially when $N$ is large. Luckily, as the area law~\cite{eisert2010colloquium} suggests, a typical quantum ground state is less entangled than an arbitrary state and could be effectively parameterized by tensor networks~\cite{schollwock2005density, perez2006matrix}.

\begin{figure}[H]
    \centering 
    \includegraphics[width=0.25\textwidth]{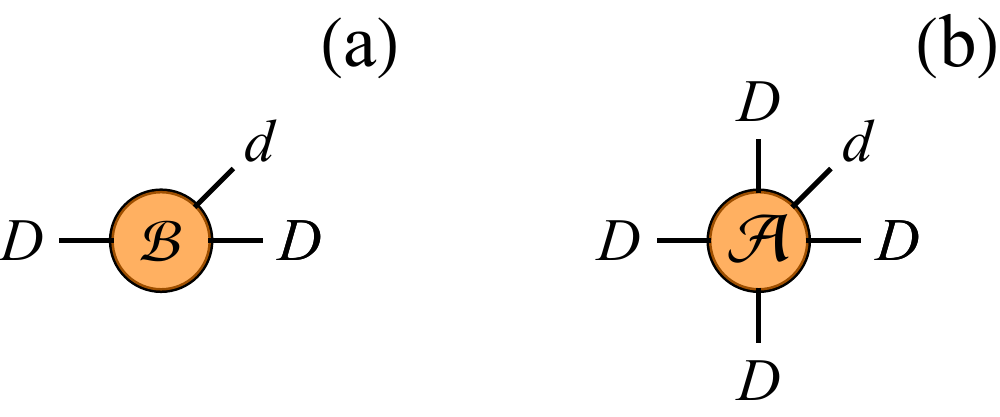}
    \caption{The graphical notation of local tensors. In this notation, each solid shape, such as the above circle, represents a tensor, and each leg (or bond) of it represents an index of the tensor.}
    \label{fig: MPS_and_PEPS}
\end{figure}

The infinite matrix product state (iMPS) and the iPEPS are effective ansatzes for ground states of 1D and 2D systems in the thermodynamic limit respectively. They consist of local tensors that repeat periodically on infinite lattices. The graphical notation of local tensors is shown in~\cref{fig: MPS_and_PEPS}, where $D$ is the virtual bond dimension and the corresponding bonds are contracted, $d$ is the bond dimension of physical legs. 

Consider a 2D translational invariant system with one site in the unit cell on a square lattice, its ground state can be parameterized by the following iPEPS:
\begin{equation*}
    \begin{aligned}
        |\Psi(\mathcal{A})\rangle =&\sum_{\left\{S_{r}\right\}} \operatorname{Tr} \prod_{\mathbf{r}} \mathcal{A}^{S_{r}}[\mathbf{r}]\left| \left\{S_{r}\right\}\right\rangle\\
        =& \adjincludegraphics[valign=c, width=0.2\textwidth]{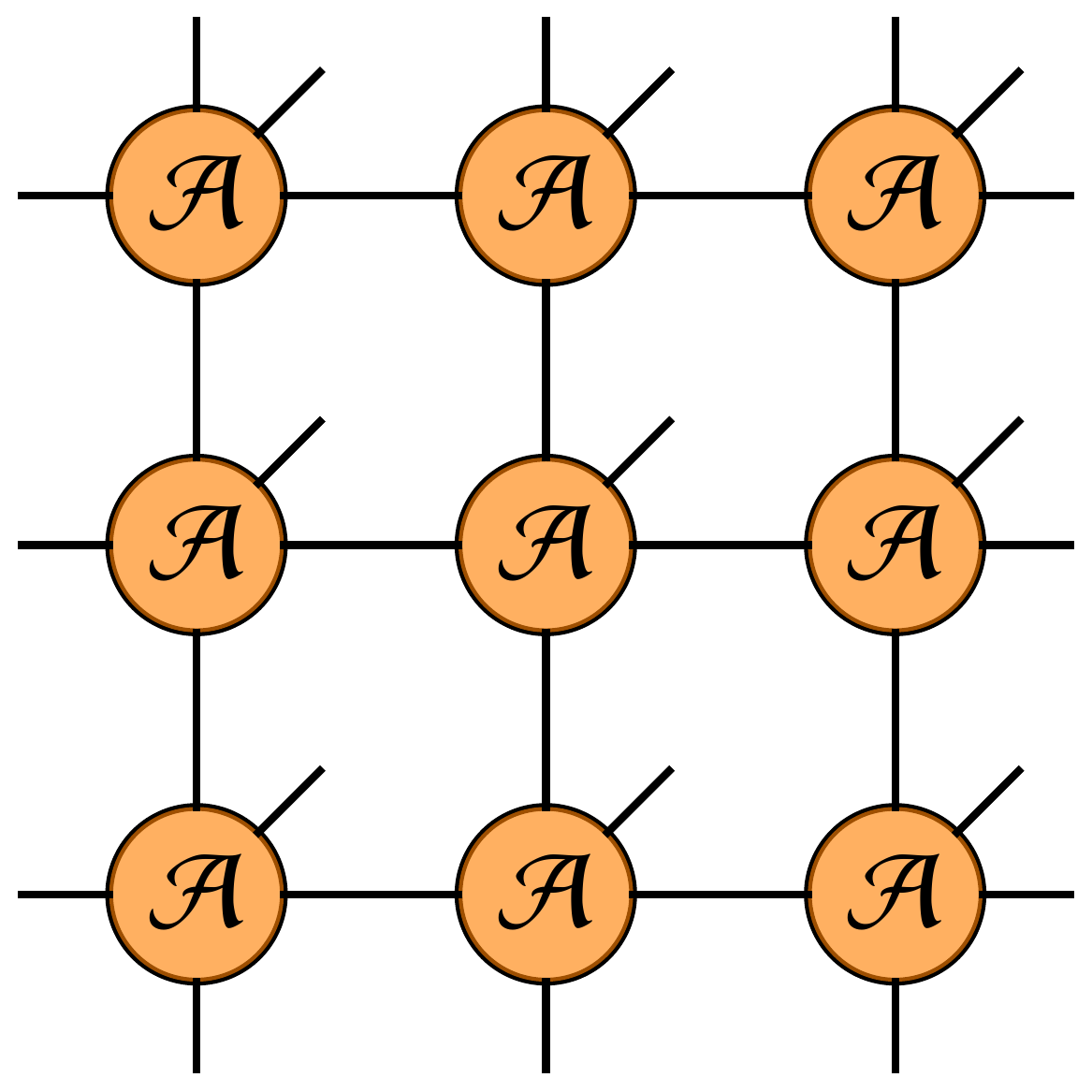}
    \end{aligned}
\end{equation*}
where $\mathcal{A}^{S_{r}}[\mathbf{r}]$ are the local tensors of rank five, with four virtual bonds and one physical bond. $\left|\left\{S_{r}\right\}\right\rangle=\left|S_{\mathbf{r}_{1}}, S_{\mathbf{r}_{2}}, \ldots, S_{\mathbf{r}_{\mathbf{N}}}\right\rangle$ is a many-basis state and $\mathbf{r}_i$ label different lattice sites. $\mathrm{Tr}[\dots]$ symbolizes the contraction of all virtual bonds.

The energy expectation value is calculated by the following equation:
\begin{equation}
    \begin{aligned}
        E(\mathcal{A})=& \langle\Psi(\bar{\mathcal{A}})|H| \Psi(\mathcal{A})\rangle / \langle\Psi(\bar{\mathcal{A}}) \mid \Psi(\mathcal{A})\rangle\\
        =& \adjincludegraphics[valign=c, width=0.4\textwidth]{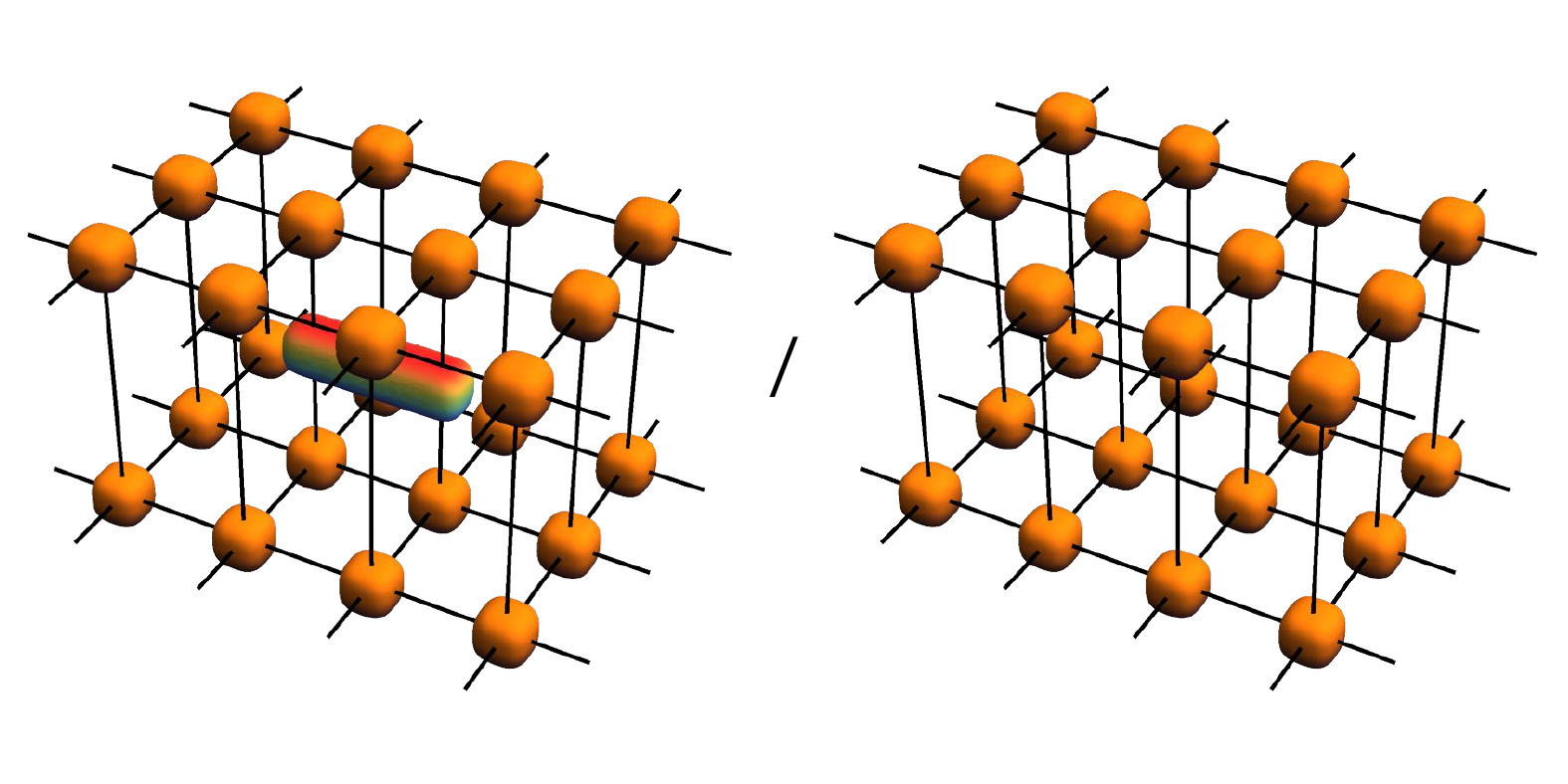}
    \end{aligned}
    \label{equ: iPEPS_energy}
\end{equation}
where $\bar{\mathcal{A}}$ denotes the conjugate of tensor $\mathcal{A}$ and $\adjincludegraphics[valign=c,width=0.05\textwidth]{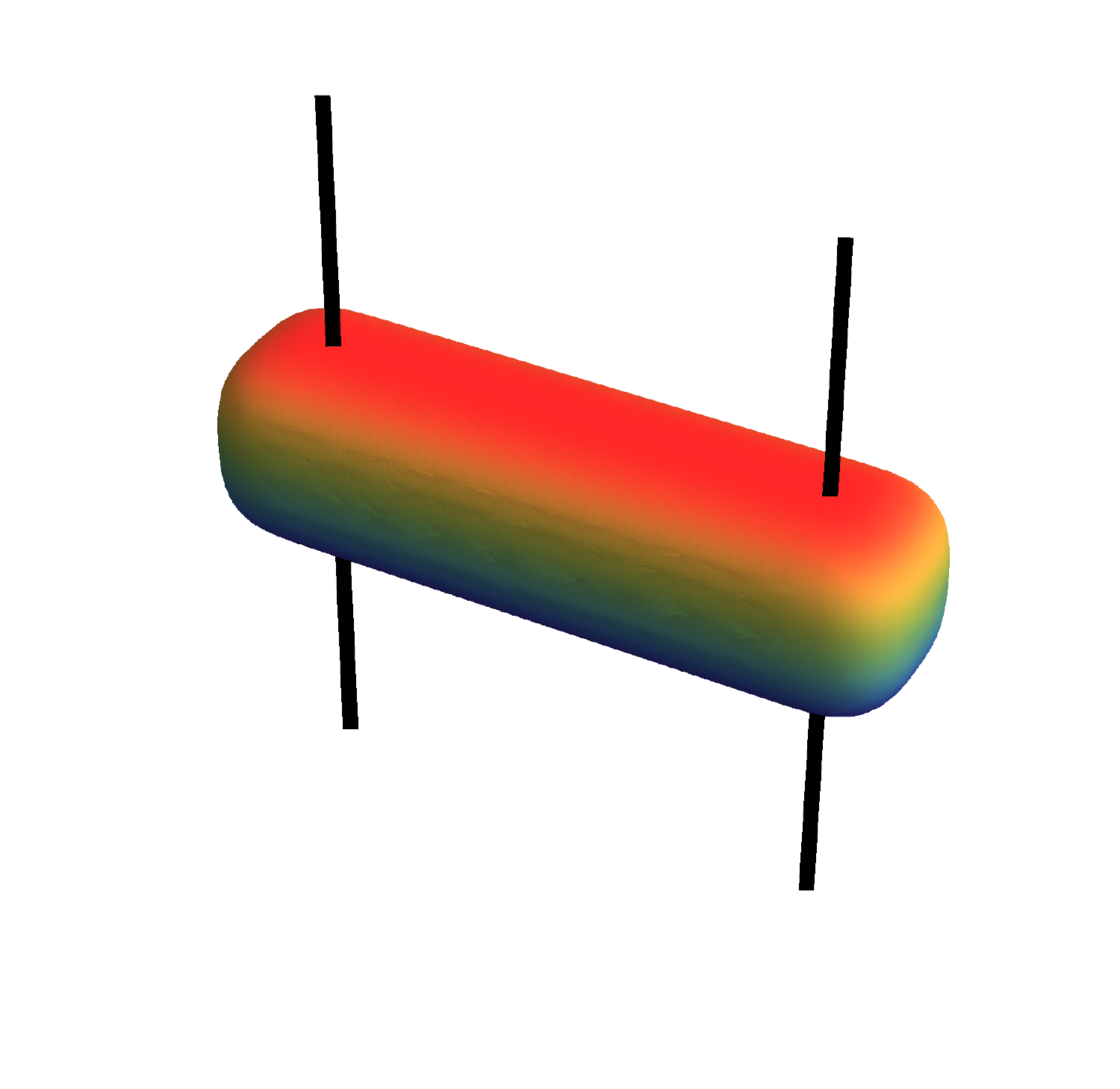}$ represents the Hamiltonian tensor that consists of nearest neighbor interaction terms. 

To contract the infinite tensor networks as shown in~\cref{equ: iPEPS_energy}, we firstly contract vertical physical legs and transform the double-layer tensor network with virtual bond dimension $D$ into a single-layer tensor network whose virtual bond dimension is $D^2$. For example, the denominator in~\cref{equ: iPEPS_energy} becomes
\begin{equation}
    \langle\Psi(\bar{\mathcal{A}}) \mid \Psi(\mathcal{A})\rangle = 
    \adjincludegraphics[valign=c,width=0.2\textwidth]{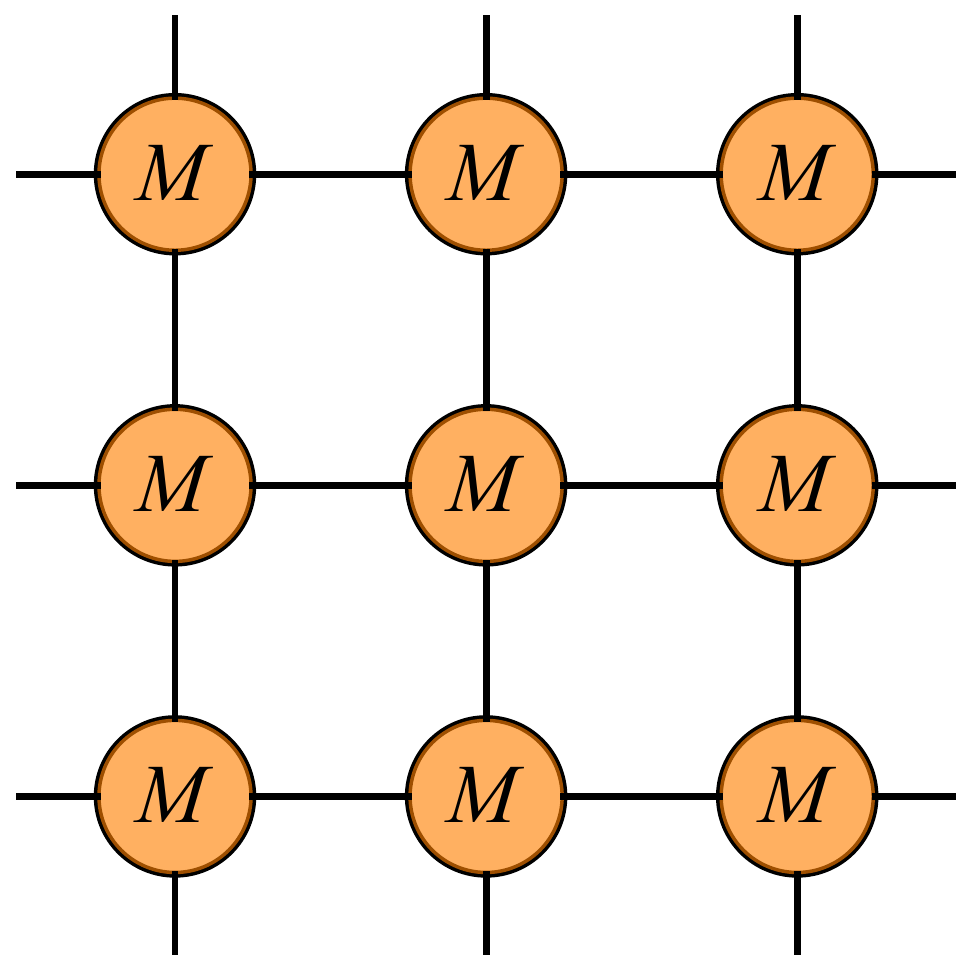}
    \label{equ: M_tensor_network}
\end{equation}
with $M$ being the contraction of $\mathcal{A}$ and $\bar{\mathcal{A}}$:
\begin{equation}
    \includegraphics[width=0.35\textwidth]{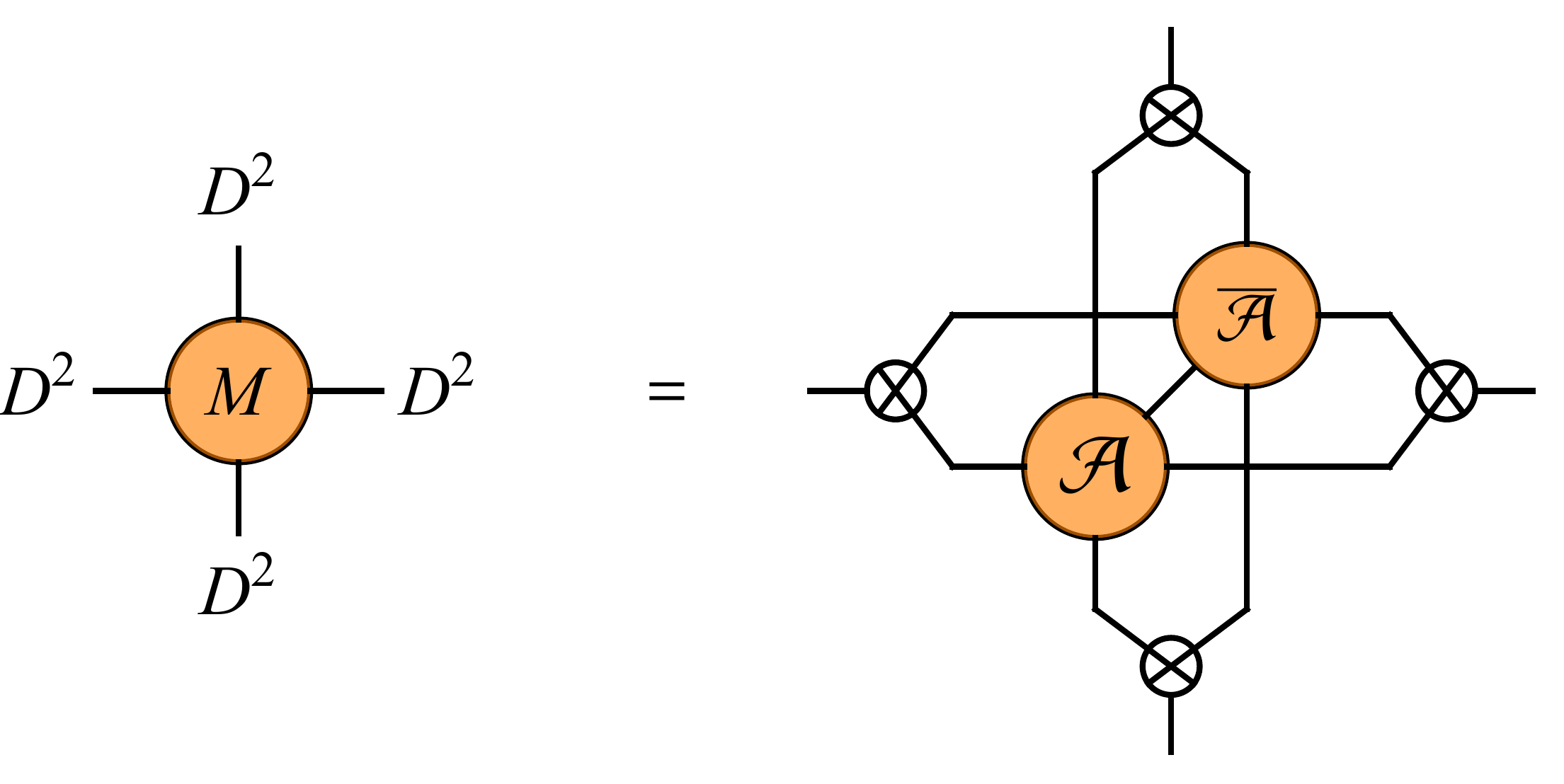}
    \label{equ: AAjoint}
\end{equation}
where $\bigotimes$ denotes the outer product operation that merges two virtual bonds into one.

Then we introduce the \emph{effective boundary environment} $\adjincludegraphics[valign=d,width=0.05\textwidth]{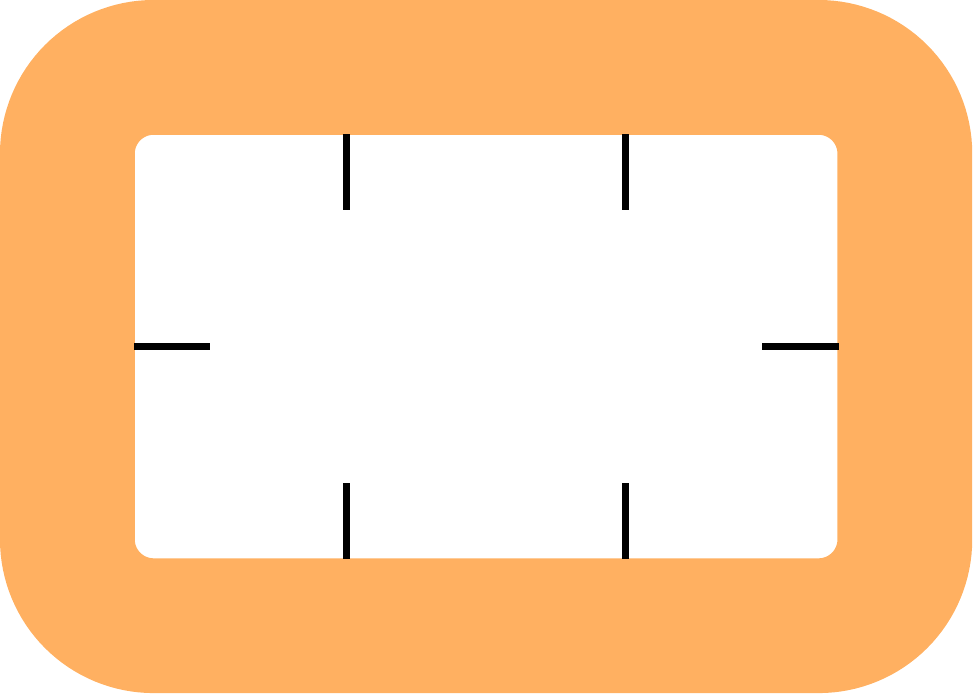}$, which is an approximation of the 2D infinite tensor networks surrounding the local tensor in the center. And ~\cref{equ: iPEPS_energy} can be approximated as:
\begin{equation}
    E = \adjincludegraphics[valign=c,width=0.2\textwidth]{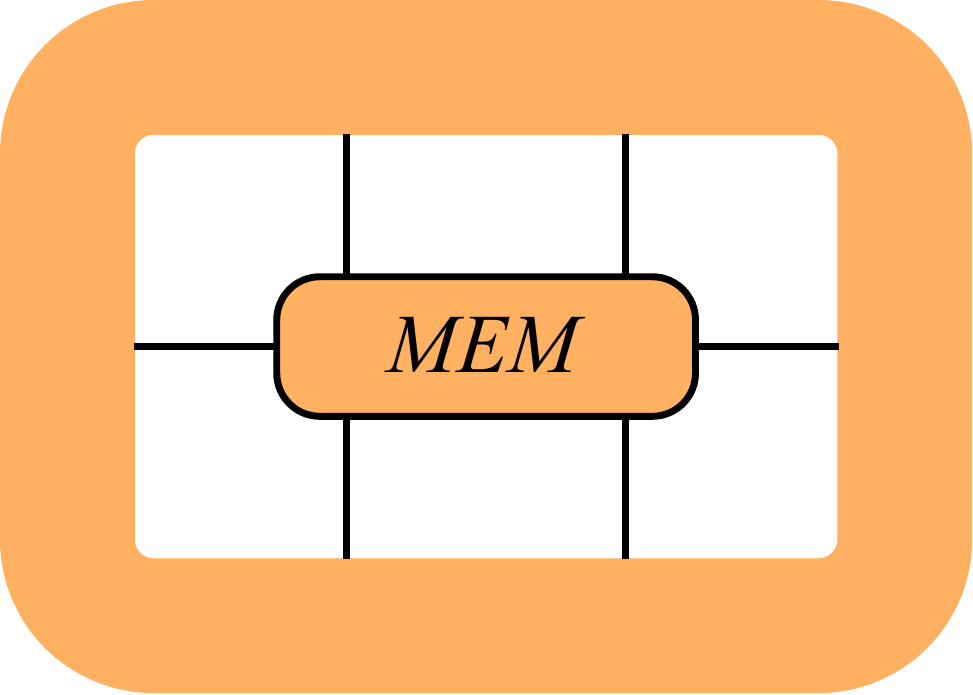} /
    \adjincludegraphics[valign=c,width=0.2\textwidth]{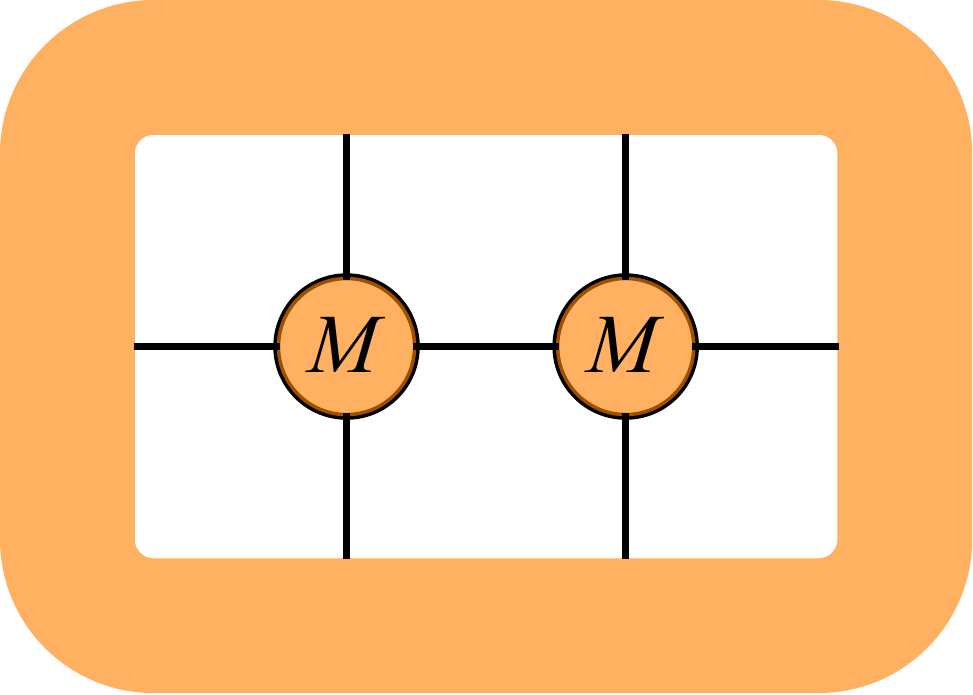}
    \label{equ: E_tensor_network}
\end{equation}
where the contraction of $\mathcal{A}$ tensors and the Hamiltonian tensor $\adjincludegraphics[valign=c,width=0.05\textwidth]{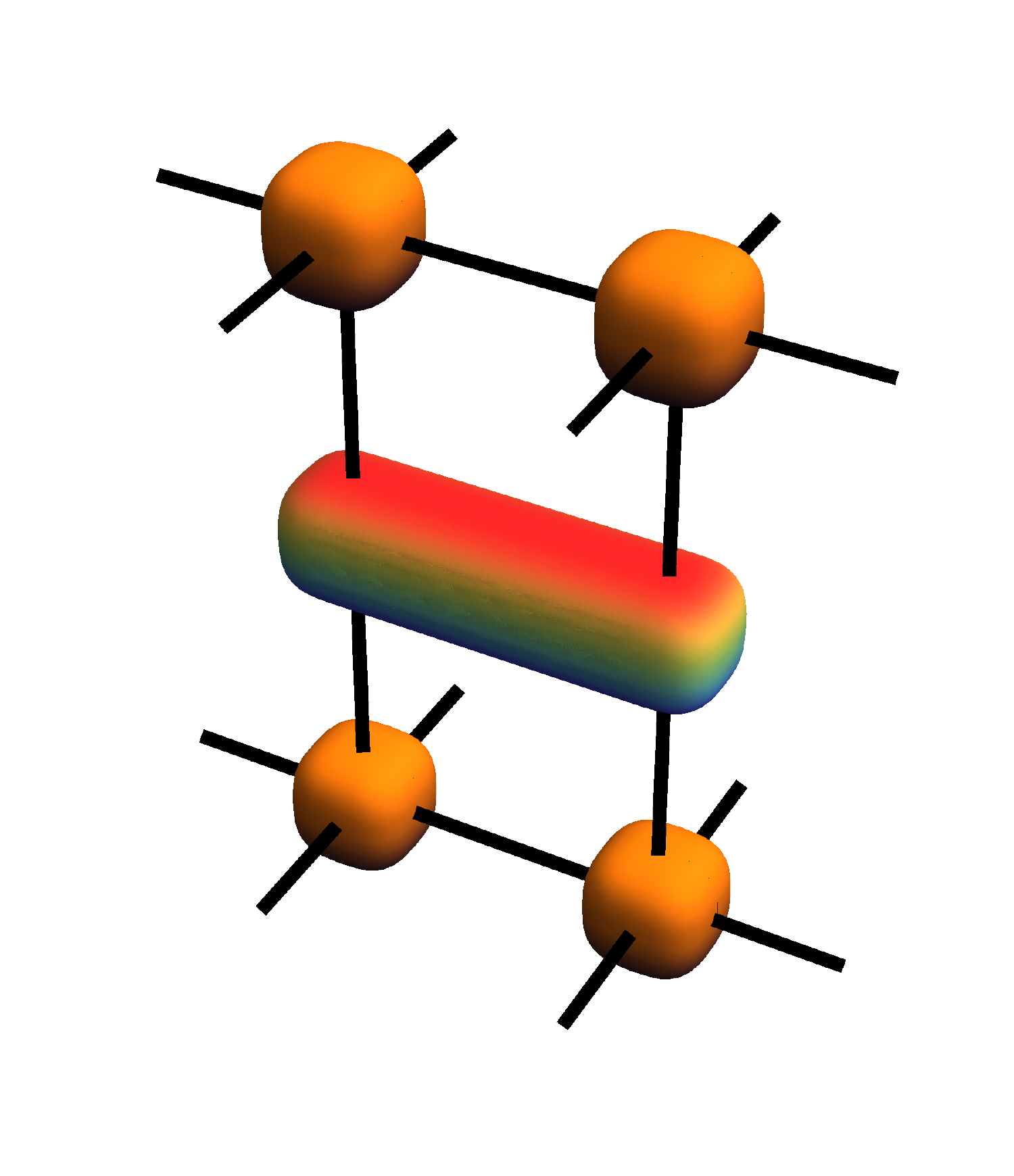}$ is denoted by $MEM$.

Finding the optimal effective boundary environment in~\cref{equ: E_tensor_network} in the key issue in current studies\cite{vanderstraeten2022variational}. On the basic intuition of the power method, this would be equivalent to finding the dominant eigenvectors of transfer matrices. To be specific, if we define an
infinite one-dimensional row transfer matrix $T$:
\begin{equation*}
    \includegraphics[width=0.35\textwidth]{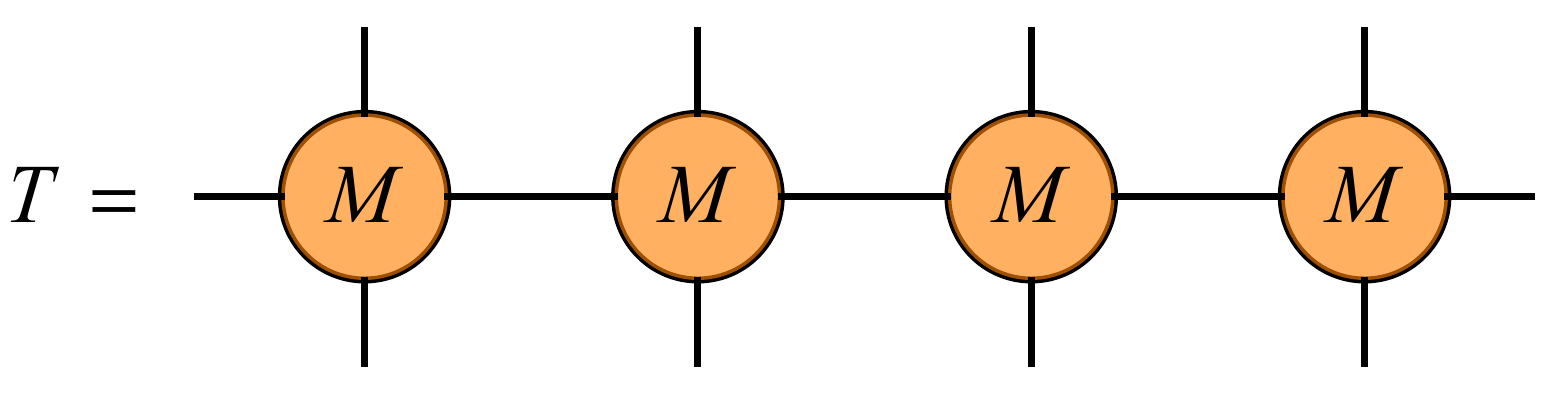}
\end{equation*}
and repeatedly apply it on the upper boundary iMPS until convergence, we would get the following fixed-point equation:
\begin{equation}
    \includegraphics[width=1.0\linewidth]{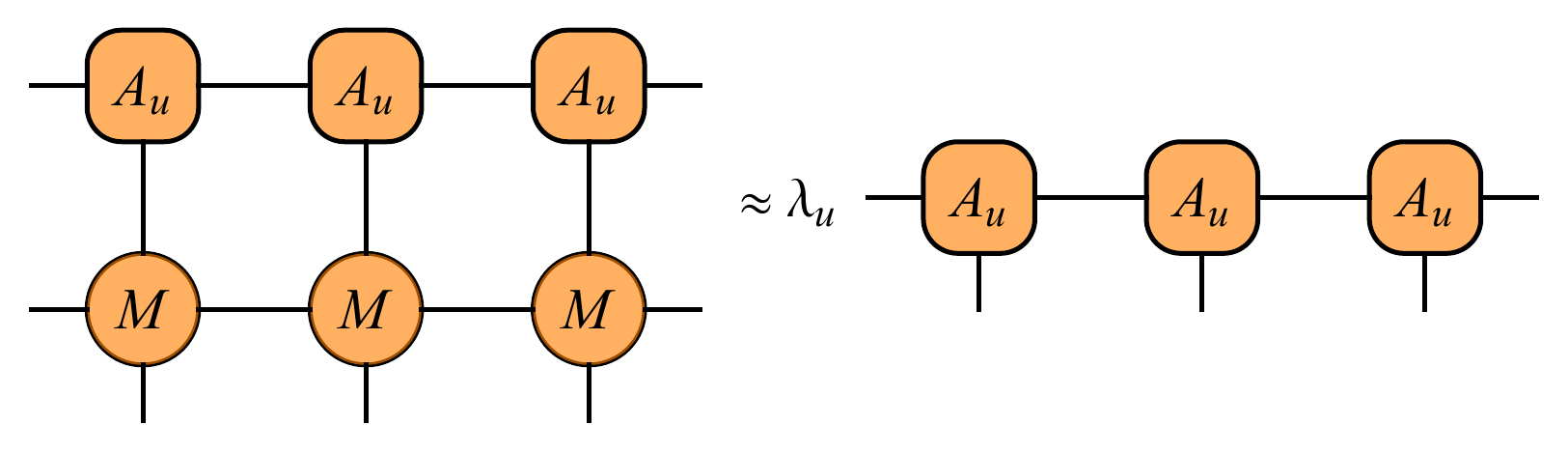}
    \label{equ: 1D_fixed_point_up}
\end{equation}
where $\lambda_{u}$ is the largest eigenvalue of $T$. The unpper boundary iMPS is composed of cell tensor $A_{u}$ with virtual bond dimension $\chi$:
\begin{equation}
    \includegraphics[width=0.1\textwidth]{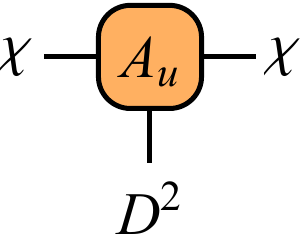}
    \label{equ:iMPS_A}
\end{equation}

Similarly, one can find the lower-boundary environment iMPS:
\begin{equation}
    \includegraphics[width=1.0\linewidth]{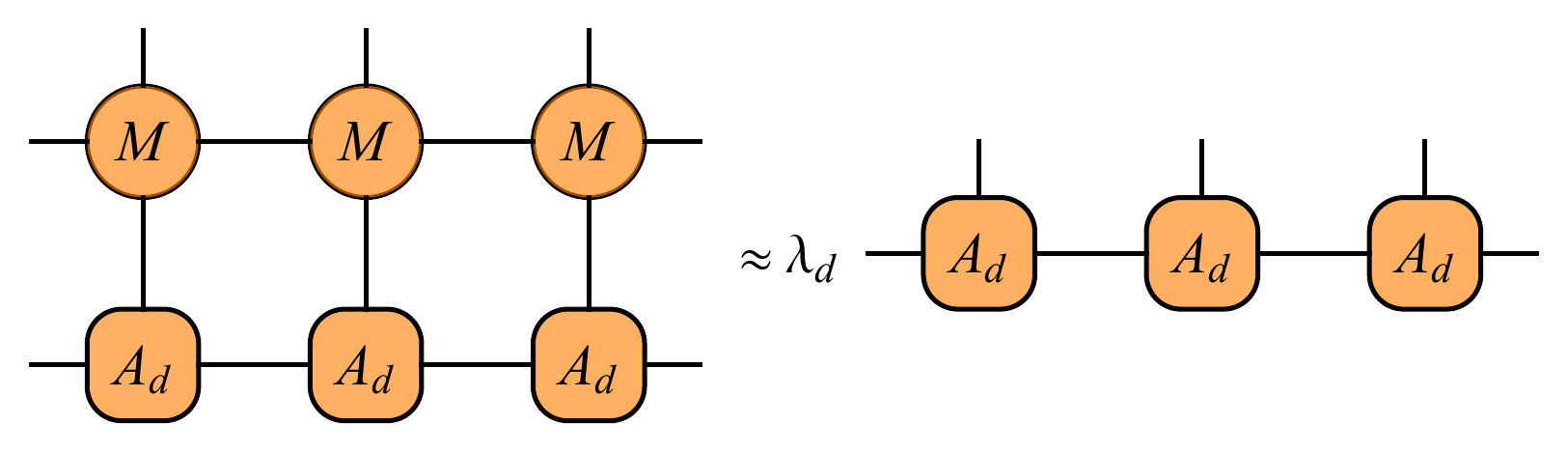}
    \label{equ: 1D_fixed_point_down}
\end{equation}
with $A_d$ being the cell tensor and $\lambda_{d}$ is the corresponding dominant eigenvalue.

Accordingly, one would be able to contract the 2D infinite tensor network, \cref{equ: M_tensor_network} for example, row by row and transform it to a one-dimensional infinite tensor network with three rows:
\begin{equation*}
    \includegraphics[width=1.0\linewidth]{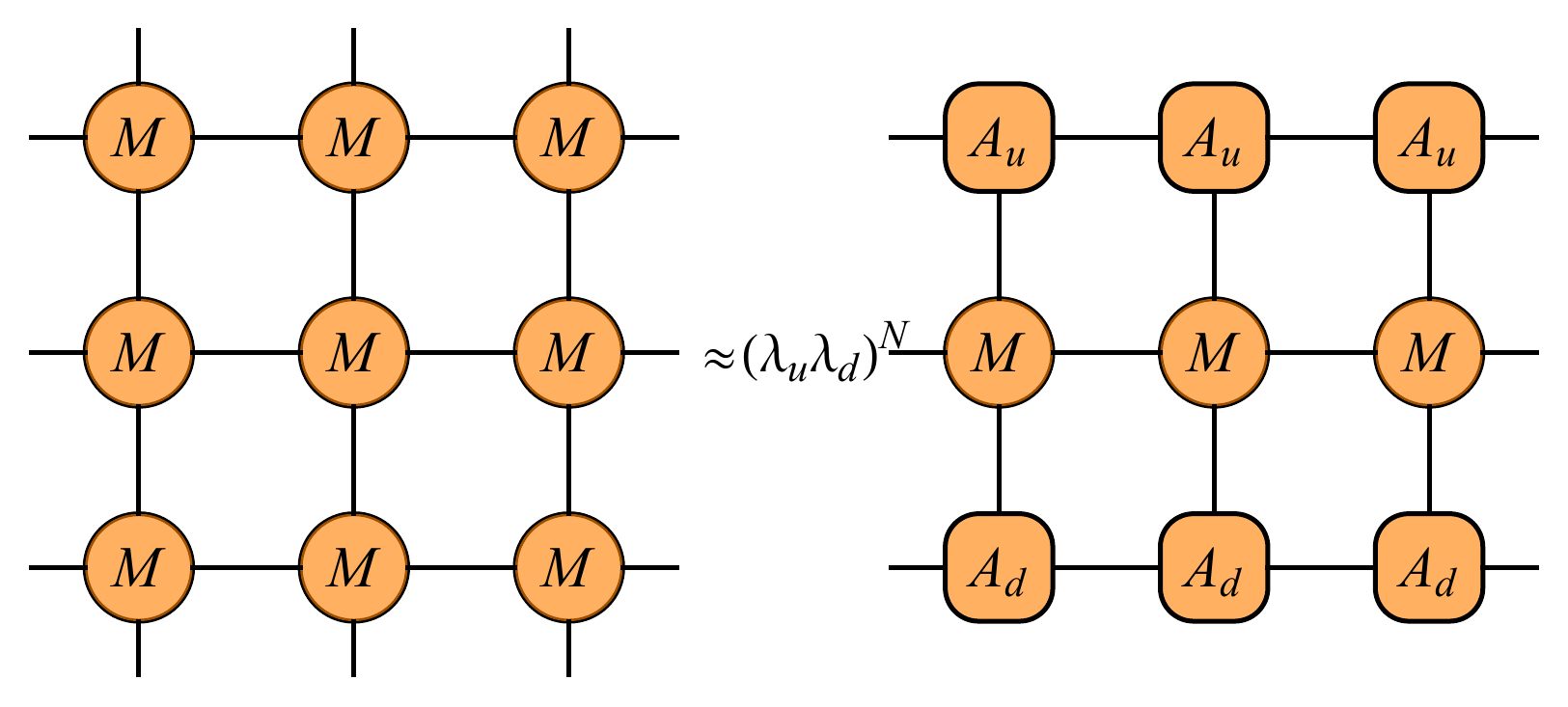}
\end{equation*}
where $N$ denotes the number of rows applied.

This infinite tensor network could also be contracted easily in the same way as discussed before by defining a column transfer matrix $\adjincludegraphics[valign=c,width=0.02\textwidth]{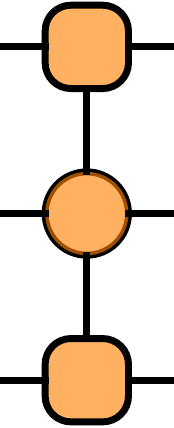}$, and finding its left and right dominant eigenvectors. The whole 2D infinite tensor network is therefore contracted as:
\begin{equation*}
    \includegraphics[width=1.0\linewidth]{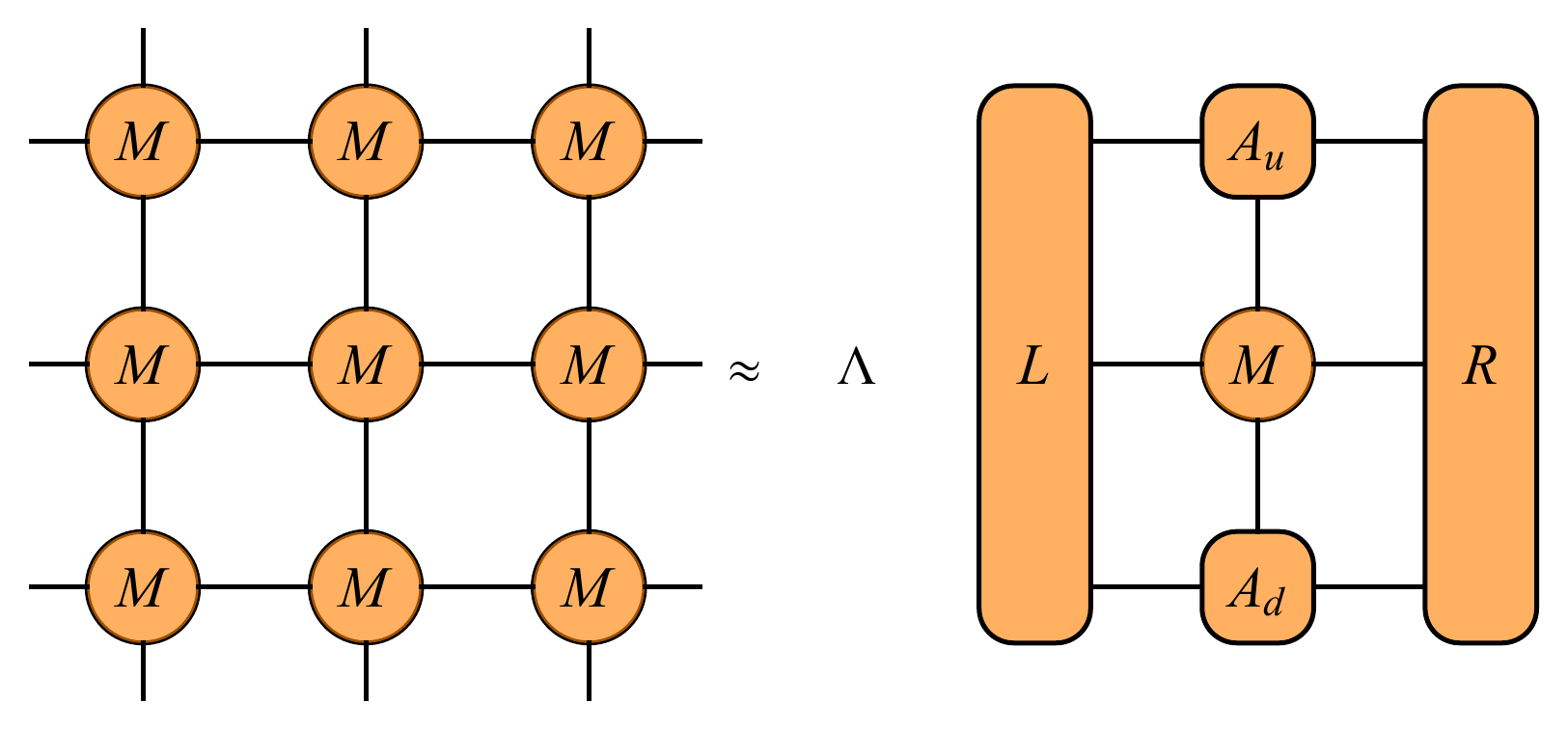}
\end{equation*}
As a result, the \emph{effective boundary environment} in~\cref{equ: E_tensor_network} can be written as:
\begin{equation}
    \includegraphics[width=1.0\linewidth]{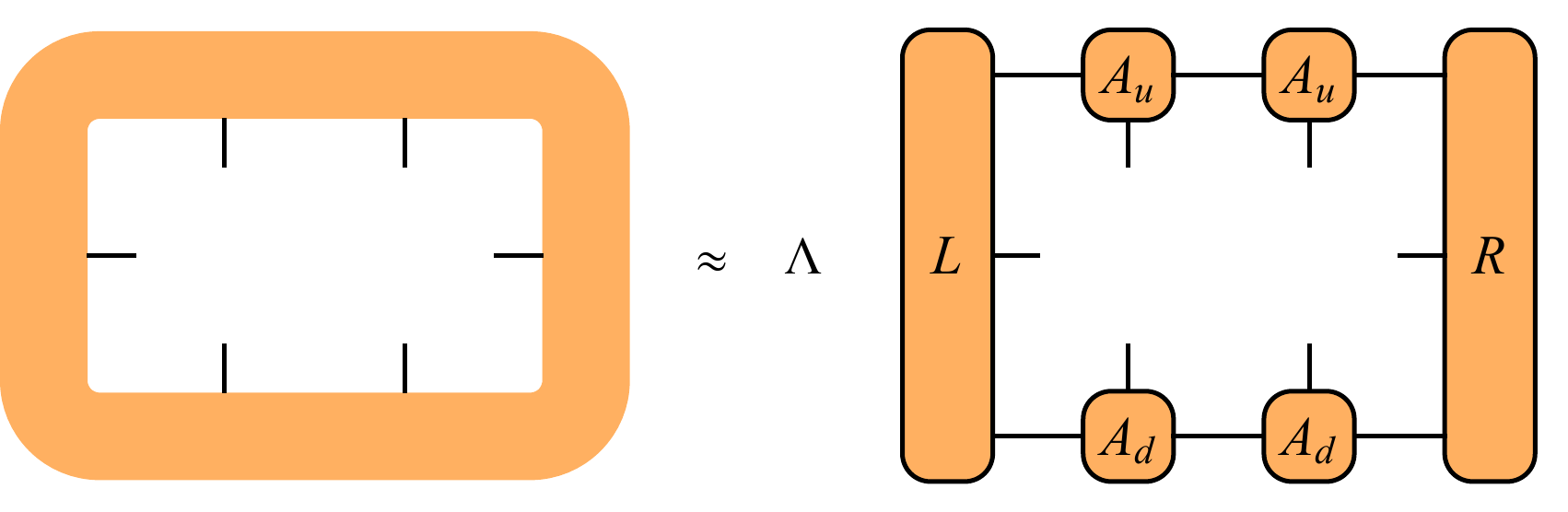}
    \label{equ: 1x1_iPEPS_environment}
\end{equation}
where $\Lambda$ is an unimportant coefficient consisting of the multiplication of dominant eigenvalues and it would be canceled in the numerator and the denominator of~\cref{equ: E_tensor_network}.

%We obtain the energy of arbitrary iPEPS by contracting the infinite 3D tensor network, as shown in~\cref{equ: iPEPS_energy}, which is equivalent to contracting the infinite 2D tensor network, as shown in~\cref{equ: M_tensor_network}. To transform the infinite tensor network contraction into a finite form, we use the effective boundary environment, as shown in~\cref{equ: E_tensor_network}, which are the dominant eigenvectors of transfer matrixes, as shown in~\cref{equ: 1D_fixed_point_up} and~\cref{equ: 1D_fixed_point_down}.

\subsection{Tensor network contraction: VUMPS}
\label{subsec:tensor-network-contraction-vumps}
%Apart from power iterations, more advanced methods have also been developed 
%to find the effective boundary environment, such as~\cite{corboz2011stripes, corboz2014competing,fishman2018faster}. 
Apart from power iterations, more sophisticated techniques have been developed for determining the effective boundary environment, as demonstrated in Refs.~\cite{corboz2011stripes, corboz2014competing,fishman2018faster}. Among these methods, VUMPS algorithm which iterates fixed-point functions based on the variational principle has attracted much attention due to its high efficiency and extensibility~\cite{zauner2018variational, fishman2018faster, 10.21468/SciPostPhysLectNotes.7}. In this section, we will first introduce VUMPS method that is applicable in the case where the transfer matrix is hermitian. Then discuss its extension to the non-hermitian case. Last, talking about its implementation in more general cases where there is more than one lattice site in a unit cell. 

%To contract the infinite tensor network in~\cref{equ: M_tensor_network} is equivalent to finding the effective boundary environment, as shown in~\cref{equ: 1x1_iPEPS_environment}. The key point to find the environment is how to get the approximation of~\cref{equ: 1D_fixed_point_up} and~\cref{equ: 1D_fixed_point_down}. One can always find it by power method, as~\cref{equ: 1x1_iPEPS_environment}. But it is inefficient and inaccurate. The VUMPS algorithm gets a better approximation by iterating the fixed point of the variational method, and we will present it and its large unit cell version in this section.

% The reason why we use the VUMPS algorithm instead of CTMRG~\cite{corboz2011stripes} is that the VUMPS algorithm is more efficient and more accurate than CTMRG~\cite{fishman2018faster}. In detail, firstly, the main computation cost is matrix multiplication in VUMPS and SVD in CTMRG. And matrix multiplication benefits more than SVD when calculated in GPU~\cite{fatahalian2004understanding, lahabar2009singular}. Then compared with CTMRG, which backward includes adjoint of SVD, the backward of VUMPS including adjoint of QR and dominant eigensolver is more stable. What's more, due to artificial large cell division in CTMRG, we think the large cell extension of VUMPS~\cite{nietner2020efficient,vanderstraeten2022variational} is more natural and convincing than the extension of CTMRG~\cite{corboz2014competing}.

\subsubsection{Hermitian VUMPS}
\label{subsubsec: Hermitian VUMPS}
Inspired by the DMRG method, the VUMPS algorithm~\cite{zauner2018variational} was initially developed to find the ground states of 1D systems. Later, taking advantage of the \emph{canonical form} and fixed-point functions, this approach was adapted to contract the infinite two-dimensional square lattice tensor networks with hermitian transfer matrices~\cite{vanhecke2021tangent}. We will present this algorithm as a prelude, a reader who is familiar with it could move to the next subsection.

Before anything else, let's first discuss the hermiticity of the transfer matrix. A direct result of a hermitian transfer matrix is that the lower environment is the complex conjugate of the upper environment, which means $A_d = \bar{A}_u$. This means only one side environment is required and we would see that this also be the key to reformulating the dominant eigenvalue problem to a variational optimization problem. This hermiticity is closely related to the symmetries of physical systems, especially the special reflection that makes:
\begin{equation}
    \includegraphics[width=0.2\textwidth]{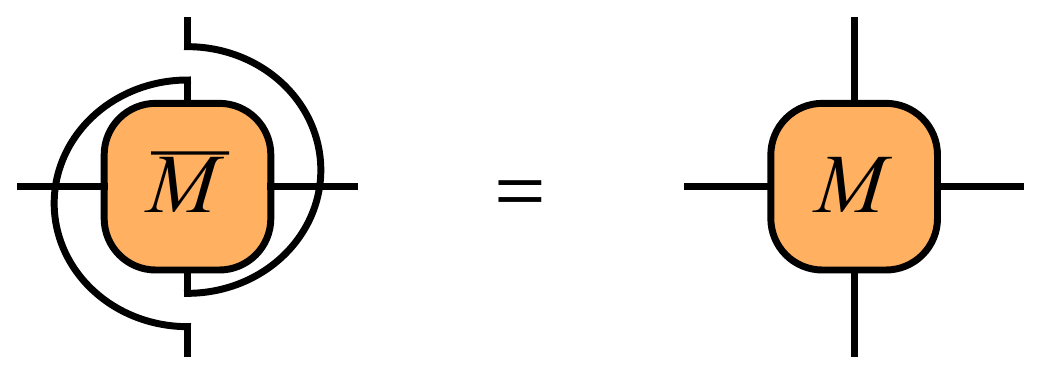}
    \label{squ: M_reflection}
\end{equation}
This defines the Hermitian adjoint of the matrix product operator in the physical bond.

Another crucial concept is the \emph{canonical form} of iMPS. For a given tensor $A$, an iMPS is uniquely defined, but the converse is not true. This can be seen from the fact that a local \emph{gauge transformation}:
\begin{equation*}
    \includegraphics[width=0.3\textwidth]{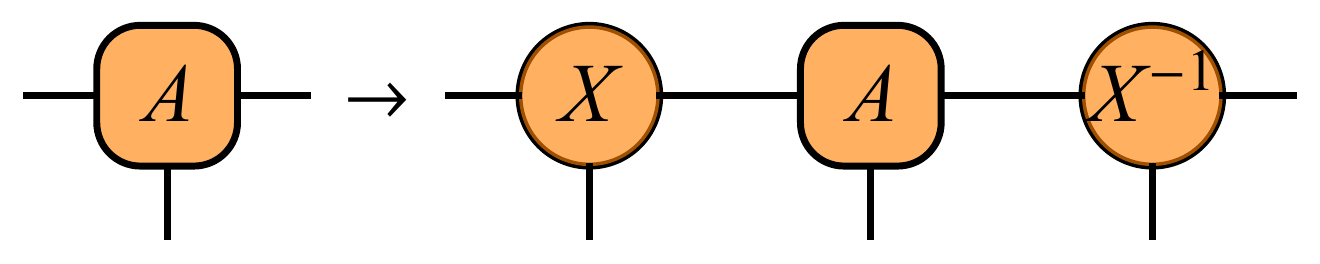}
\end{equation*}
leaves the iMPS invariant. Therefore, one can always choose a preferable gauge and the canonical representation is the most commonly used. In this representation, the left-orthonormal tensor $A_L$ and the right-orthonormal tensor $A_R$ obey the following conditions:
\begin{equation}
    \includegraphics[width=0.4\textwidth]{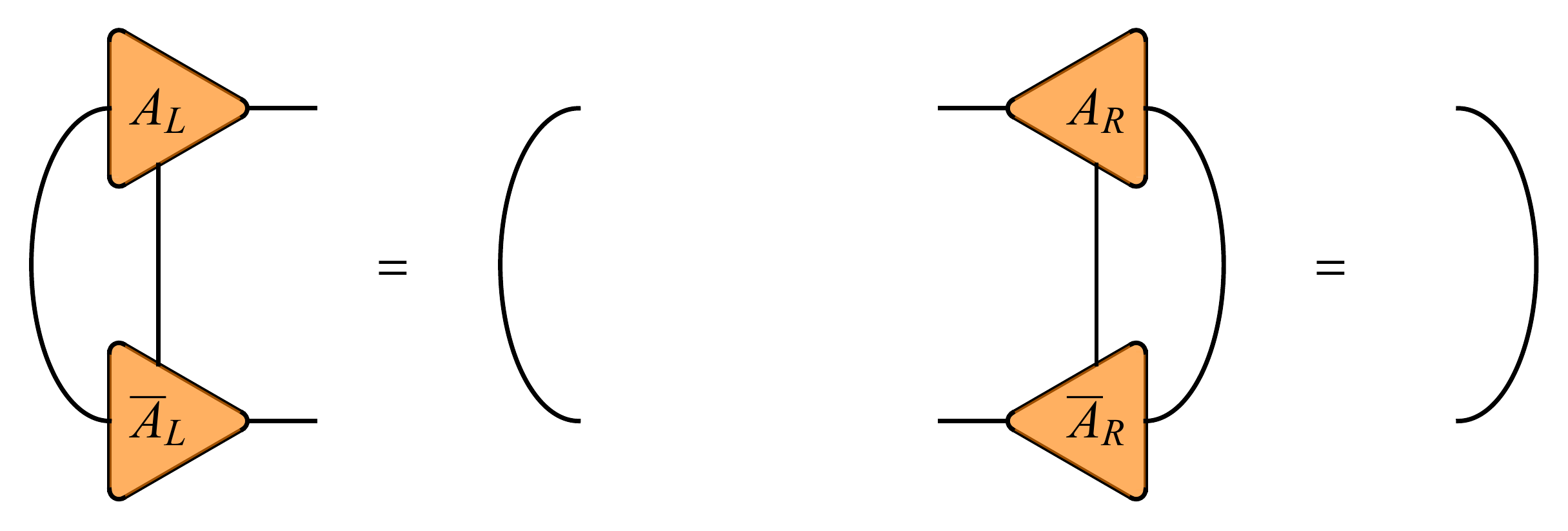}
    \label{equ:1x1_canonical_form}
\end{equation}
Suppose the gauge transform between $A_L$ and $A_R$ is implemented by tensor $C$,  center-site tensor $A_C$ can thus be defiend:
\begin{equation}
    \includegraphics[width=0.45\textwidth]{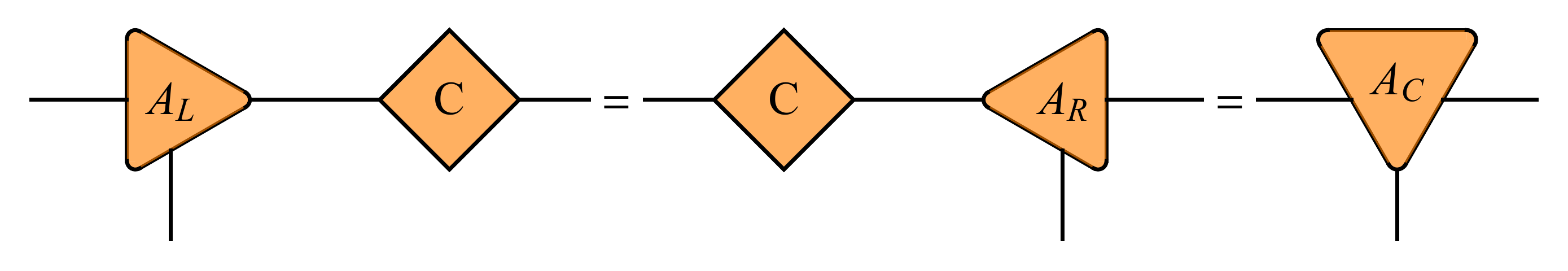}
    \label{equ:1x1_ALCARAC}
\end{equation}
Consequently, the canonical form of an iMPS has the following form:
\begin{equation*}
    \includegraphics[width=1.0\linewidth]{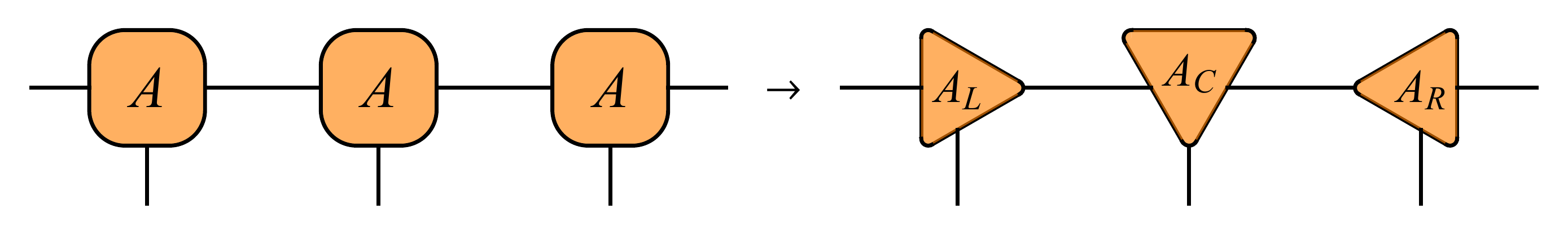}
\end{equation*}
with one $A_C$ tensor in the middle and an infinite number of $A_L$ and $A_R$ tensors on the left and right side of it respectively. The tensor network contraction could be simplified when using canonical representation, and we are going to use it from now on. For more information, please check Ref~\cite{vanhecke2021tangent} for details about getting the canonical form and ~\cite{TeneT.jl} for code implementation.

It's time to present the VUMPS algorithm. Just to reiterate, our goal is to find the maximum eigenvalue of the following eigenvalue equation:
\begin{equation}
    T|\psi(A)\rangle=\lambda_{\mathrm{row}}|\psi(A)\rangle
    \label{equ: transfer matrix eigenvalue equation}
\end{equation}
or graphacally~\cref{equ: 1D_fixed_point_up} and~\cref{equ: 1D_fixed_point_down}, where $|\psi(A)\rangle$ is the boundary iMPS composed of $A$ tensor. In the case that the transfer matrix ($T$) is hermitian, the variational principle applies and finding the maximum eigenvalue is equivalent to solving:
\begin{equation}
    A=\underset{A}{\arg \min }\left(-\log \frac{\langle\psi(\bar{A})|T| \psi(A)\rangle}{\langle\psi(\bar{A}) | \psi(A)\rangle}\right)
    \label{equ: transfer matrix variational problem}
\end{equation}

Instead of the gradient-based optimization method, an iteration method of updating the fixed-point function at zero gradients of~\cref{equ: transfer matrix variational problem} until convergence is used in the VUMPS algorithm, which is proved to be more efficient. When using the canonical representation, the fixed point function can be simplified as:
\begin{equation}
    \includegraphics[width=0.32\textwidth]{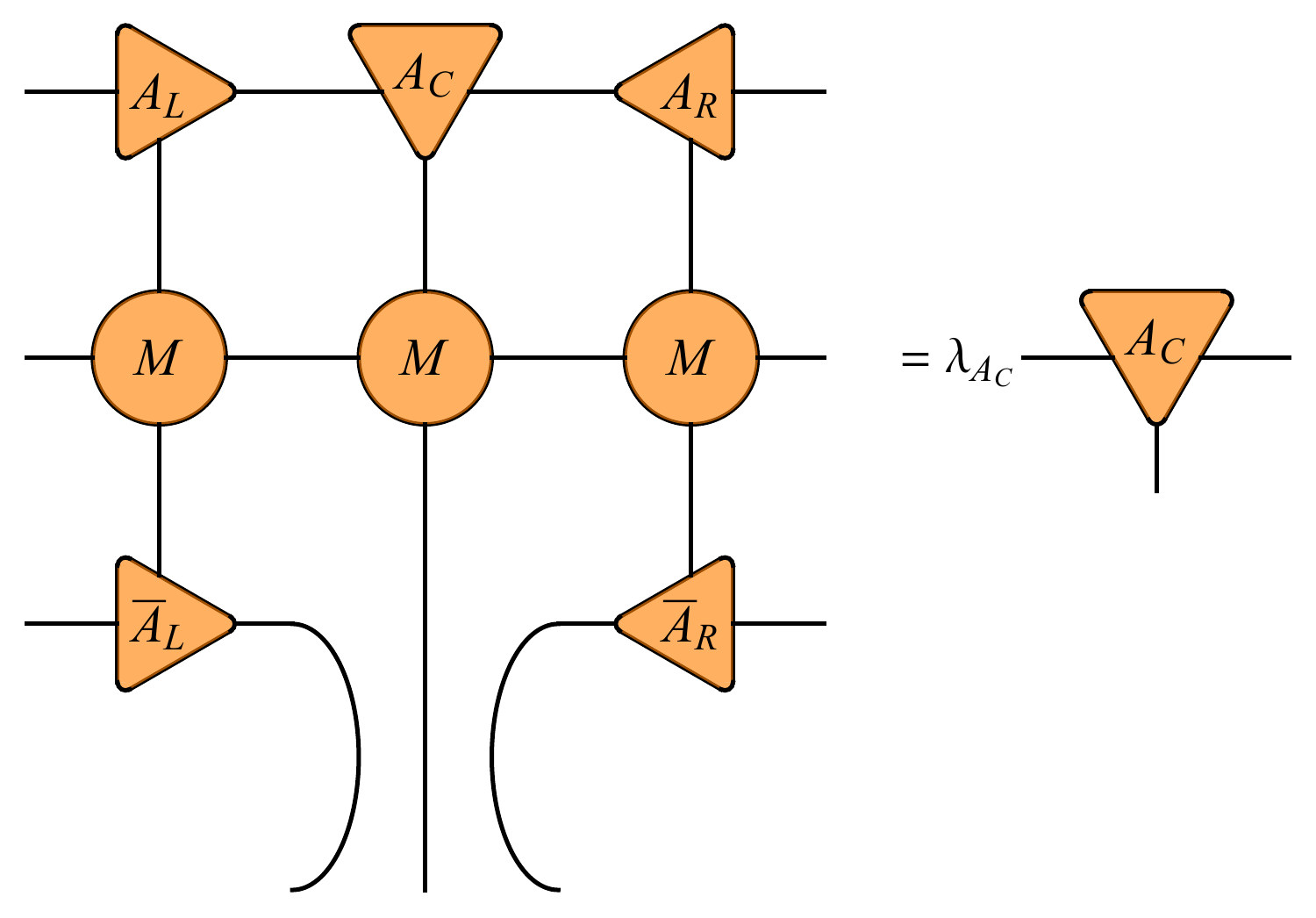}
    \label{fig:1x1_fixed_point_AC}
\end{equation}

where $\lambda_{A_c}$ is the corresponding dominant eigenvalue. Apply $\adjincludegraphics[valign=c,width=0.07\textwidth]{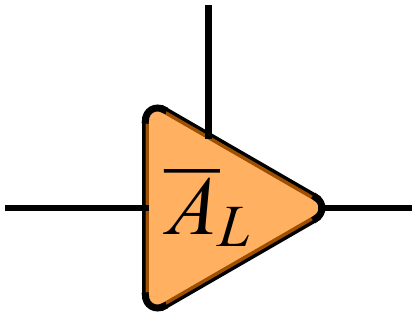}$ to the above equation, and let $\lambda_L$ be the largest eigenvalue of transfer matrix $\adjincludegraphics[valign=c,width=0.02\textwidth]{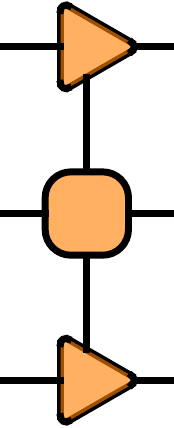}$ when applying it to the left environment, we can get another fixed point function:
\begin{equation}
    \includegraphics[width=0.32\textwidth]{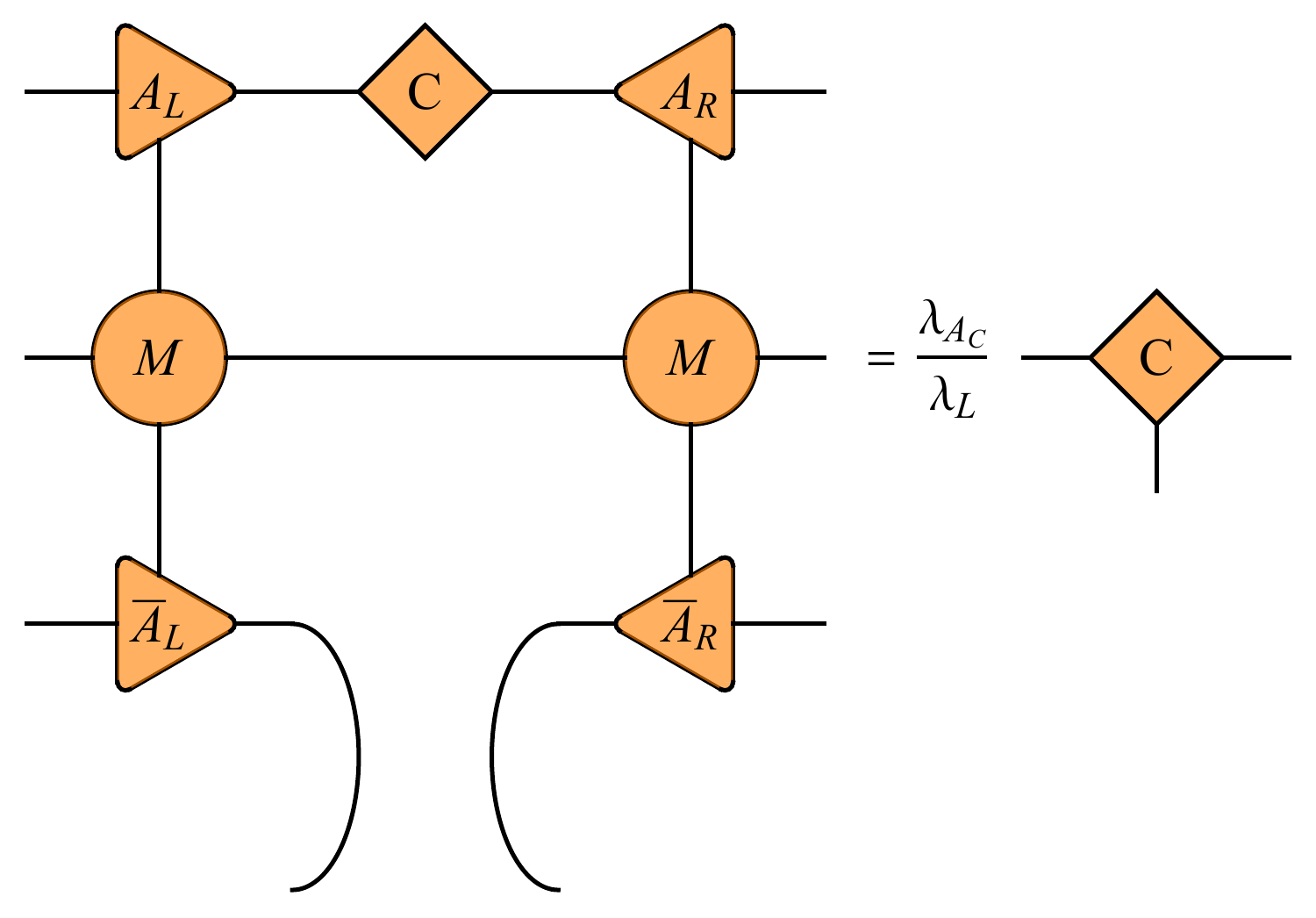}
    \label{fig:1x1_fixed_point_C}
\end{equation}

\begin{figure}[H]
    \centering 
    \includegraphics[width=1.0\linewidth]{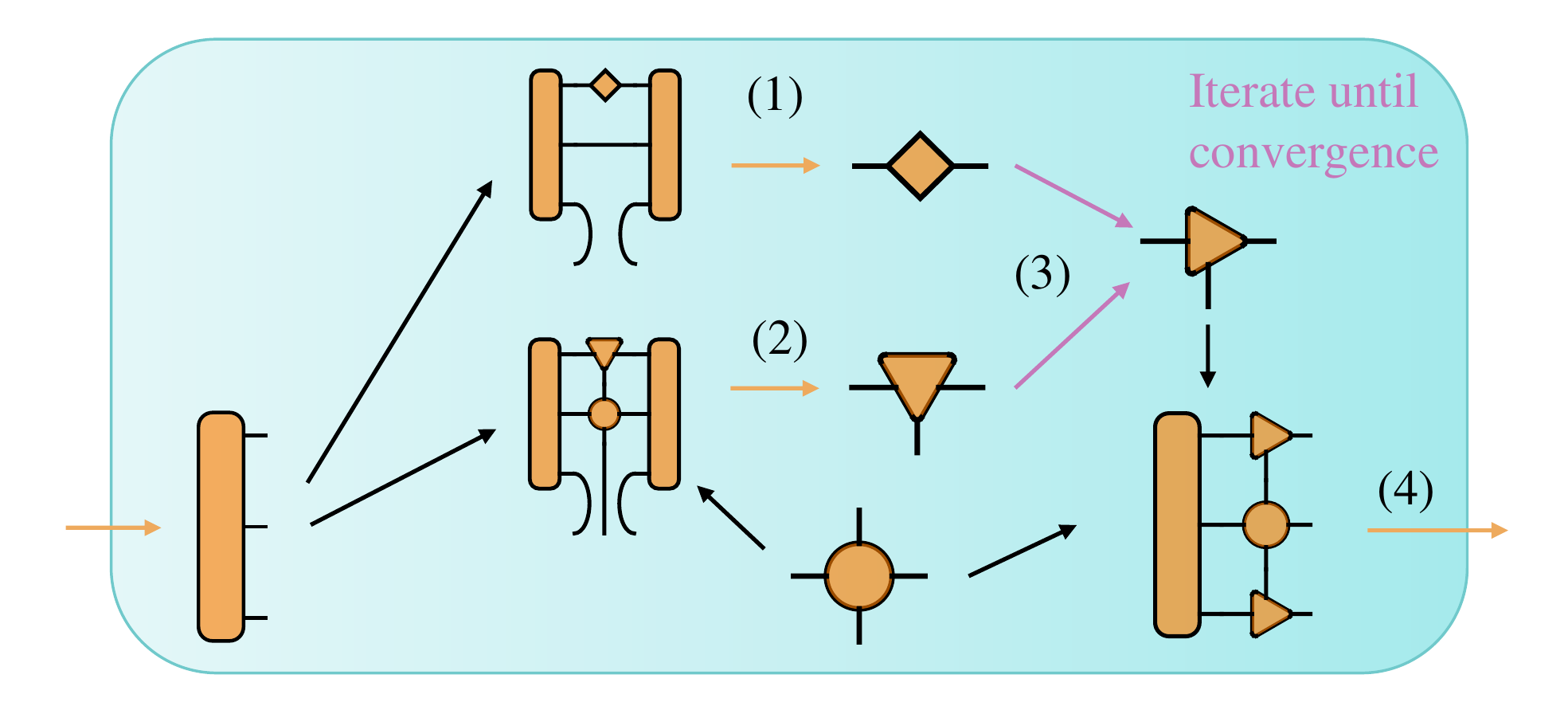}
    \caption{Hermitian VUMPS steps: black arrows are tensor contraction, orange arrows are dominant eigensolvers, and purple arrows are QR decomposition. (1)(2)(4) get dominant eigenvector $A_C$,$C$ and $L$,$R$ by their tensor contraction mapping. (3) QR decomposition of $A_C$ and $C$ to get $A_L$ as~\cref{equ: QR_AC_C}}
    \label{fig: VUMPS_computation_graph_original}
\end{figure}
One can use an iterative dominant eigensolver, such as Arnoldi\cite{arnoldi1951principle}, to solve these eigenvalue problems.

The iteration process is shown in~\cref{fig: VUMPS_computation_graph_original}. We start by randomly initiating a iMPS local tensor and find its left/right environment $L/R$ in the canonical representation and then go through steps $L, R \rightarrow  A_C, C \rightarrow A_L, A_R $ until convergence. It should be stressed here that ~\cref{equ:1x1_ALCARAC} is not applied in the step $A_C, C \rightarrow A_L, A_R$, because $C$ may become singular when its dimension is high and this may cause instability in practical calculations of $A_L = A_C\cdot C^{-1}$. Instead, we apply the QR decomposition of $A_C$ and $C$, and get the inversion from the transpose:
\begin{equation}
\begin{aligned}
    A_L &= A_C \cdot C^{-1} \\
       &= Q_{A_C} \cdot R_{A_C} \cdot R_{C}^{-1} \cdot Q_{C}^{-1} \\
       &= Q_{A_C} \cdot R_{A_C} \cdot R_{C}^{-1} \cdot Q_{C}^{\dagger}  \\
       &\approx Q_{A_C} \cdot Q_{C}^{\dagger}
    \label{equ: QR_AC_C}
\end{aligned}
\end{equation}
The last approximation has a low error when VUMPS converges. So we can define $\| R_{A_C}-R_{C} \|$ as the VUMPS algorithm error. Furthermore, $A_L$ automatically meet~\cref{equ:1x1_canonical_form} canonical gauge condition because $Q_{A_C}$ and $Q_{C}$ are hermitian. $A_R$ can be solved in the same way.

Finally, the \emph{effective boundary environment} in~\cref{equ: E_tensor_network} corresponding to a hermitian transfer matrix can be written as:
\begin{equation}
    \includegraphics[width=1.0\linewidth]{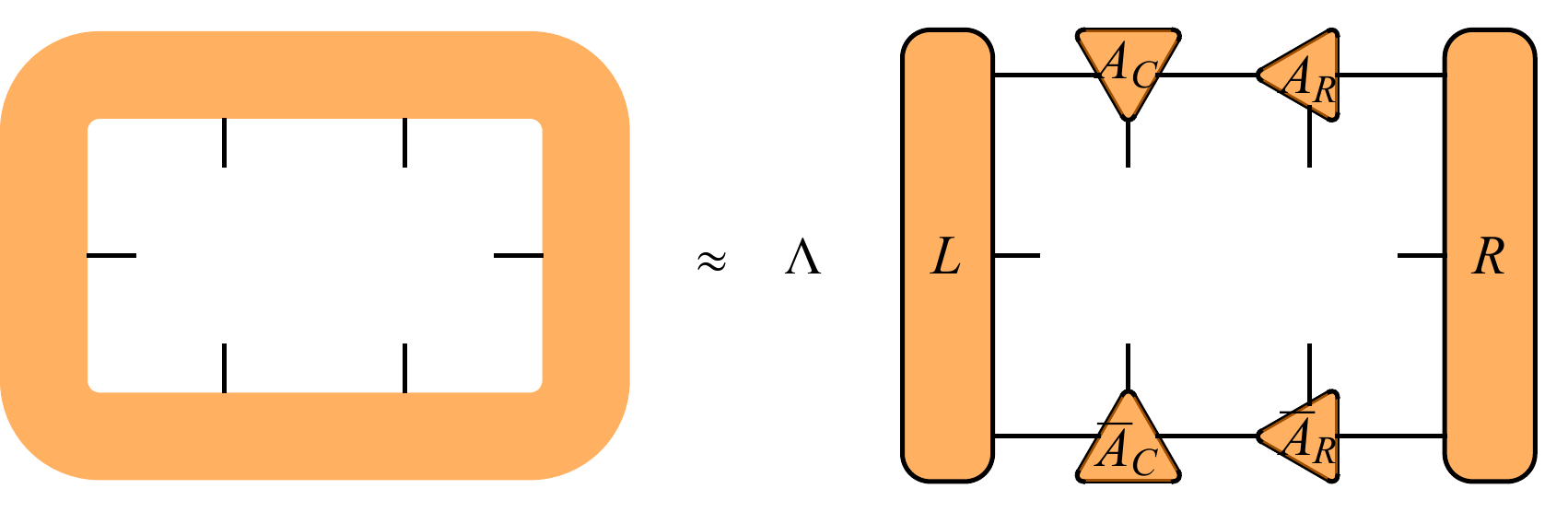}
    \label{equ: 1x1_M_canonical_environment}
\end{equation}
%We obtain the effective boundary environment by Hermitian VUMPS in the canonical form, as shown in~\cref{equ: 1x1_M_canonical_environment}.

\subsubsection{Non-hermitian VUMPS}
\label{subsubsec: Non-hermitian VUMPS}
In actual physical models, the hermitian condition of the transfer matrix can not always be fulfilled, especially in the case where there is more than one site in a unit cell~\cite{PhysRevResearch.3.013041, vanderstraeten2022variational}. So the VUMPS algorithm must be extended to handle more general situations. Though the variational principle is no longer applied here, one can always use the power method to find dominant eigenvectors~\cite{vanhecke2021tangent}. As preparation for more complicated cases where there are multi-site unit cells, we limit ourselves to the single-site unit cell case and assume $T$ is non-hermitian in this section, so the reader could get a better understanding of this method.

%In practice calculation, such as systems without up and down reflection symmetry~\cite{PhysRevResearch.3.013041} or with larger than the $3 \times 3$ unit cell \footnote{We will mention it in~\cref{subsubsec: Large Unit Cell version of VUMPS}. We can limit the transfer matrix to be Hermitian in a $2 \times 2$ unit cell~\cite{vanderstraeten2022variational}, but there is no generalization for unit cells larger than $3 \times 3$.}, the non-hermitian transfer matrix is ineluctable. So the question is how we use VUMPS in the non-hermitian case. This section is a preparation for the large unit cell version of VUMPS.

%The VUMPS algorithm presented in~\cref{fig: VUMPS_computation_graph_original} is only useable when the transfer matrix is up and down hermitian, as mentioned in~\cref{squ: M_reflection}. This is because the reformulation of~\cref{equ: transfer matrix eigenvalue equation} to~\cref{equ: transfer matrix variational problem} is only right when $T$ is hermitian conjugate.

When the transfer matrix is non-hermitian, we could not get all environment tensors at once as in the hermitian case. We should follow the procedure in \cref{subsec:iPEPS Ansatz and energy contraction} and calculate firstly the environment tensors horizontally and then vertically.

We start by discussing a single power step. Given the upper boundary iMPS with bond dimension $\chi$ at $t$-th step, and applying the transfer matrix $T$ to it, the bond dimension would be increased. Our goal is to find another iMPS with bond dimension $\chi$ to approximate it and the new one would serve as the initial iMPS for the $t+1$-th step. The process is represented by the following equation:
\begin{equation*}
    \includegraphics[width=0.4\textwidth]{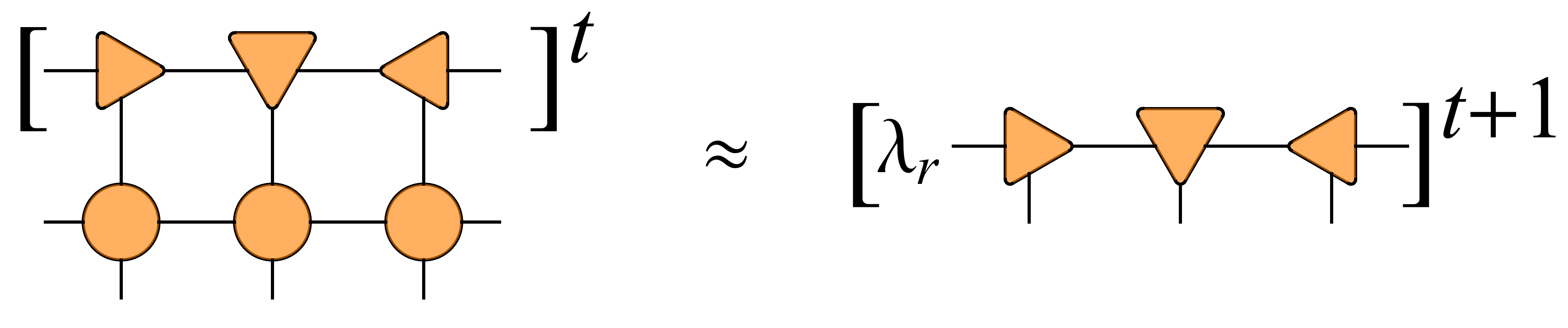}
\end{equation*}
where $\lambda_r$ is the ratio. We simplify our notation by removing all letters in the tensor notation without ambiguity for the reason that different tensors are denoted by different shapes as in~\cref{equ:1x1_ALCARAC}. And the superscript outside of the parentheses labels the power step.

As a consequence, one should maximize the fidelity between two boundary iMPS $|\psi(A^{t})\rangle$ and $|\psi(A^{t+1})\rangle$ within a power step. That leads to a variational optimization for $A^{t+1}$~\cite{vanhecke2021tangent}:
\begin{equation}
    \underset{A^{t+1}}{\arg \max }\left(\log \frac{\langle\psi(\bar{A}^{t+1})|T| \psi(A^t)\rangle \langle\psi(\bar{A}^t)|T^{\dagger}| \psi(A^{t+1})\rangle}{\langle\psi(\bar{A}^{t+1}) | \psi(A^{t+1})\rangle}\right)
\end{equation}
It has similar form with ~\cref{equ: transfer matrix variational problem} but to replace $T \rightarrow T| \psi(A^t)\rangle \langle\psi(\bar{A}^t)|T^{\dagger}$. Likewise, the corresponding fixed point function could also be written as:
\begin{equation}
    \includegraphics[width=0.32\textwidth]{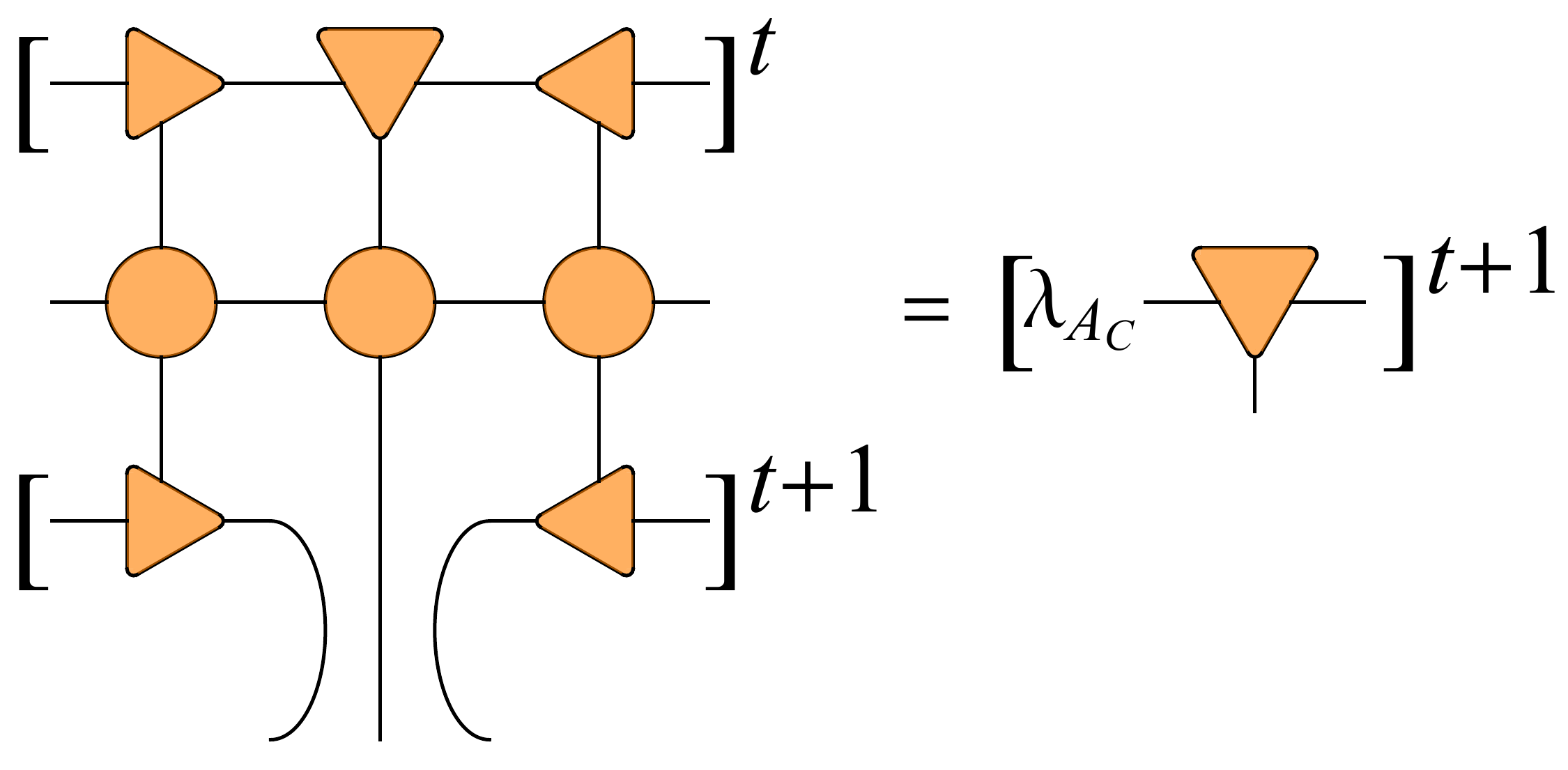}
\end{equation}
\begin{equation}
    \includegraphics[width=0.32\textwidth]{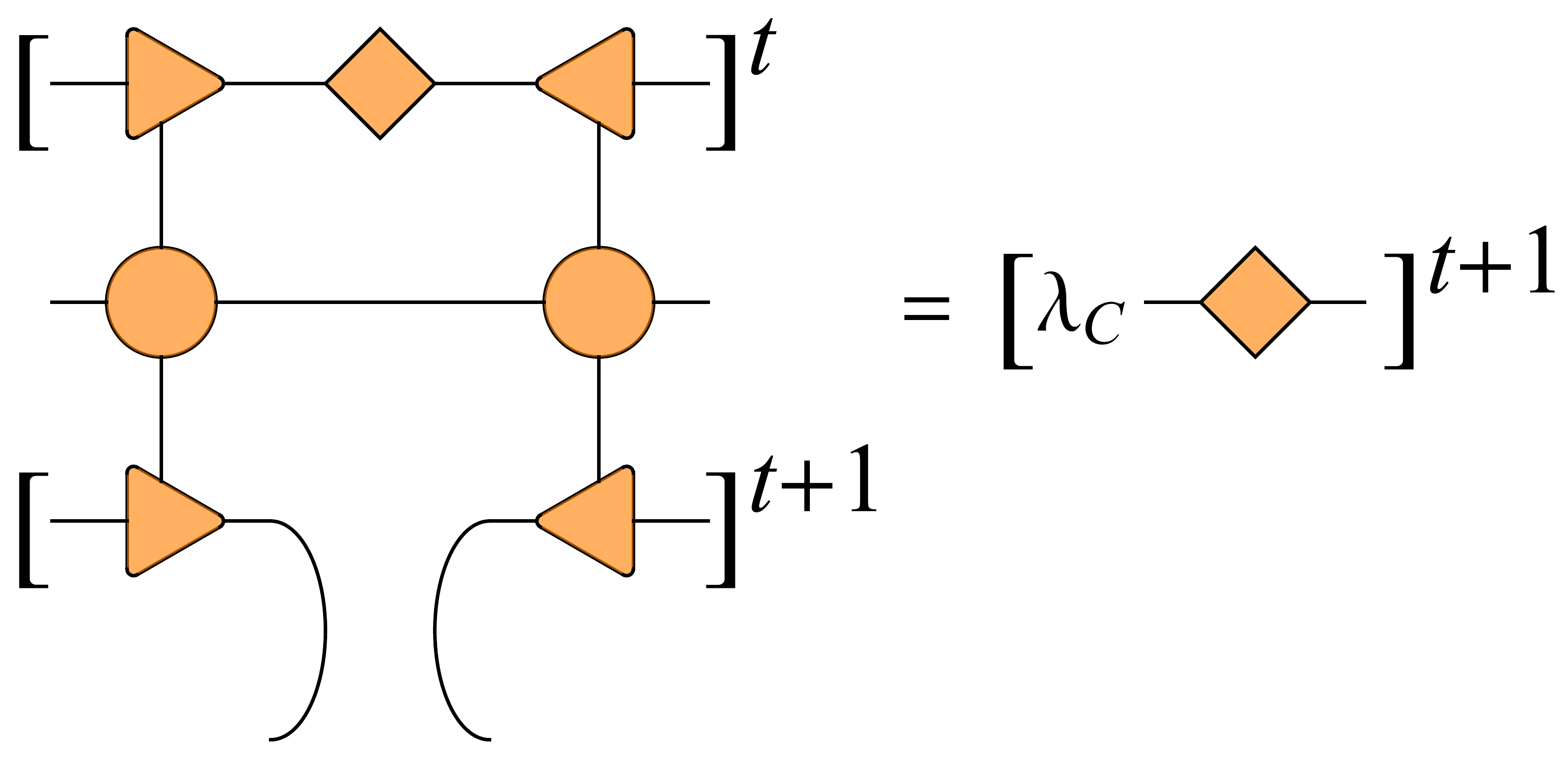}
\end{equation}

\begin{figure}[H]
    \centering 
    \includegraphics[width=1.0\linewidth]{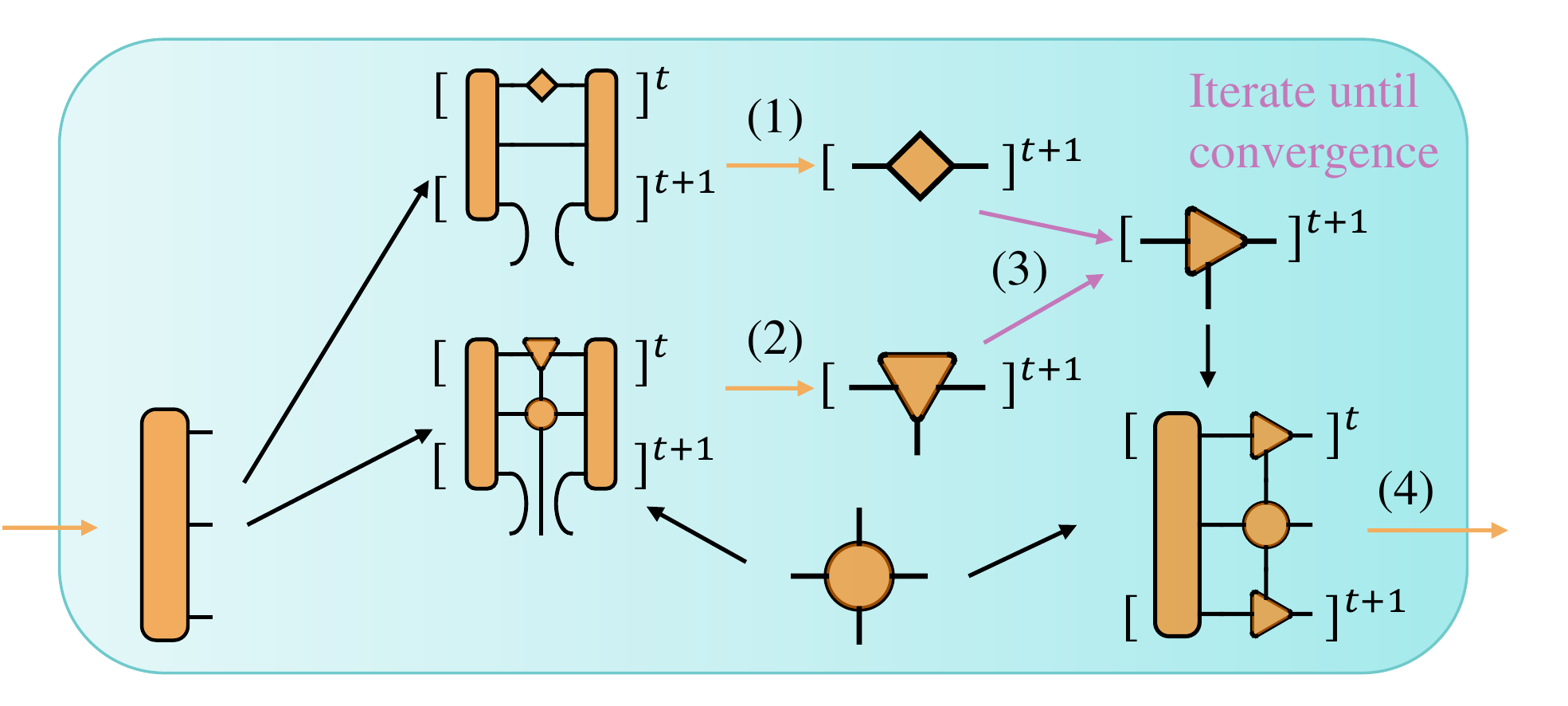}
    \caption{non-hermitian VUMPS steps: different colors and steps (1)(2)(3)(4) meanings are the same as~\cref{fig: VUMPS_computation_graph_original}. The difference is that we use superscript for $t$-th iteration in the power method.}
    \label{fig: VUMPS_computation_graph_non-hermitian}
\end{figure}
It is nothing but to replace $\bar{A}_L$ and $\bar{A}_R$ in ~\cref{fig:1x1_fixed_point_AC} and~\cref{fig:1x1_fixed_point_C} by their value in the next power step. Following the process in~\cref{fig: VUMPS_computation_graph_non-hermitian}, one can get the optimal upper boundary iMPS.

It is worth noting that, in the case of a non-hermitian transfer matrix, the upper and the lower environments may not exhibit identical characteristics. As a result, it is necessary to repeat the aforementioned calculation to obtain the lower iMPS. This would ultimately result in a computational cost twice as large as that in the hermitian case.

With upper and lower environment tensors in hand, it is easy to contract the infinite three-row tensor and to find the left and right environment tensors. The \emph{effective boundary environment} in~\cref{equ: E_tensor_network} then be written as:
\begin{equation}
    \includegraphics[width=1.0\linewidth]{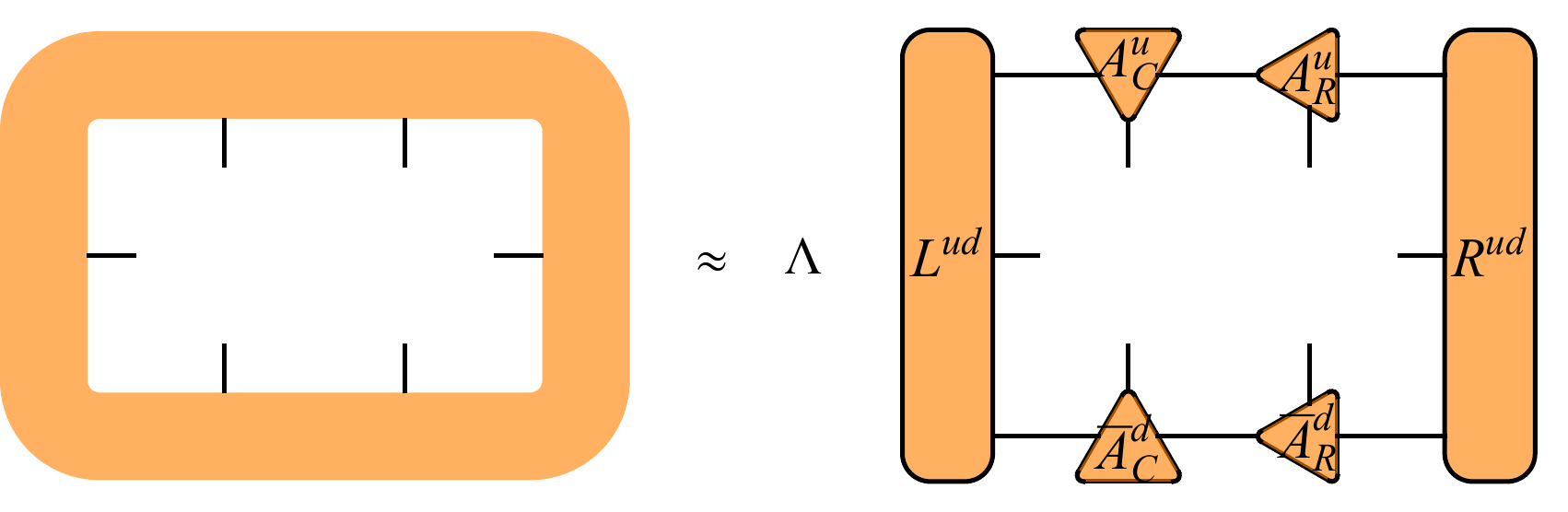}
    \label{equ: 1x1_power_M_environment}
\end{equation}

%We obtain the non-Hermitian effective boundary environment by applying up and down VUMPS in a sense of the power method, as shown in~\cref{equ: 1x1_power_M_environment}.

\subsubsection{Large Unit Cell version of VUMPS}
\label{subsubsec: Large Unit Cell version of VUMPS}
Now we are ready to present the algorithm to deal with general cases with multi-site unit cells. And We developed an easy-used package TeneT.jl specifically for this purpose~\cite{TeneT.jl}. 

Suppose the size of a unit cell of an infinite 2D tensor network on a square lattice is  $N_i \times N_j$, where $N_i$ and $N_j$ are numbers of lattice sites on each column and row respectively. Take $N_i = N_j=3$ as an example, the unit cell could be graphically represented as: 
\begin{equation}
    \includegraphics[width=0.2\textwidth]{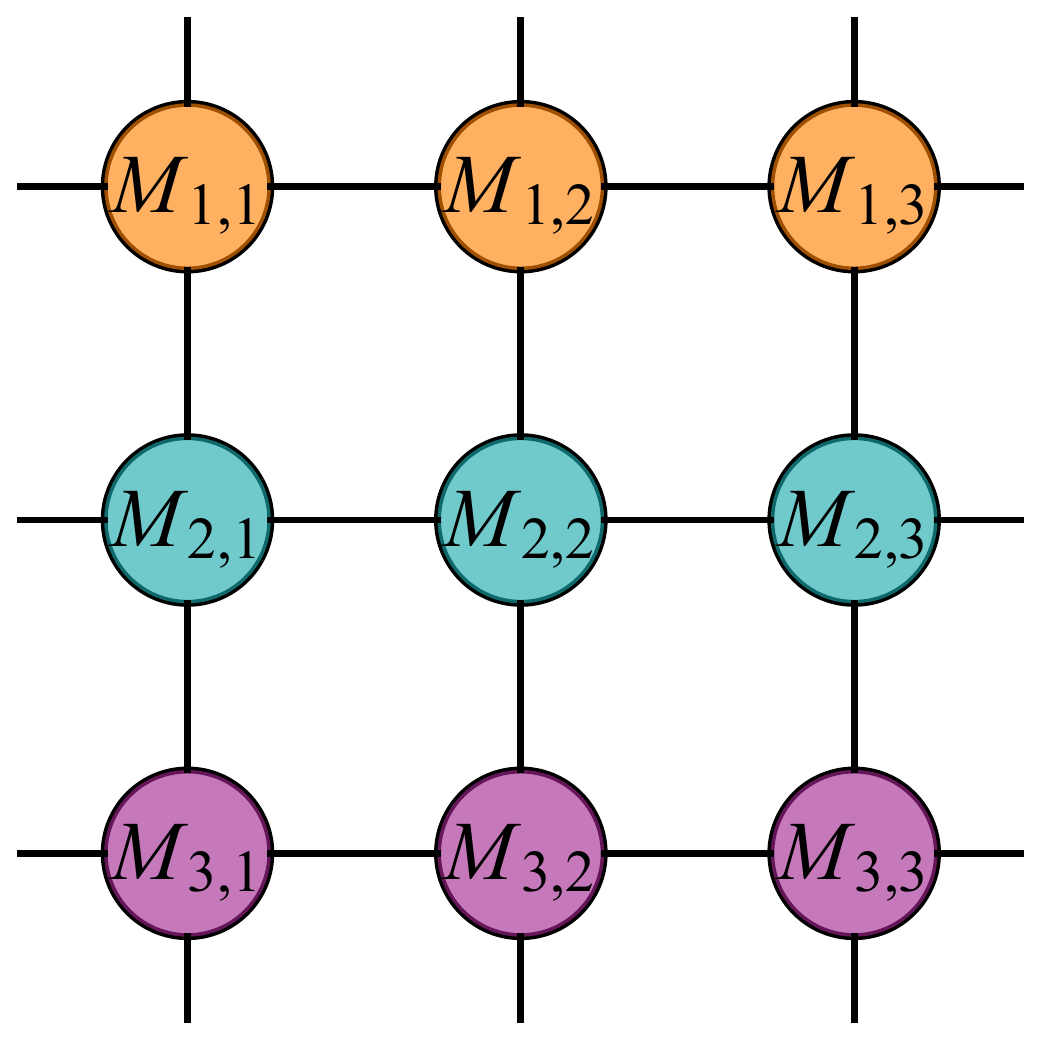}
\end{equation}
where lattice sites are labeled by their row and column indices. Each row is colored differently for easy distinction. Three transfer matrices $T_1$, $T_2$ and $T_3$ can be defined accordingly. We can see even if the transfer matrix $T_1 T_2 T_3$ is hermitian, the alternation transfer matrix $T_2 T_3 T_1$ and $T_3 T_1 T_2$ are non-hermitian. So we need to find the upper and the lower environment.

Apparently, there are six fixed point equations. Three of them are for upper environments:
\begin{equation}
    \includegraphics[width=1.0\linewidth]{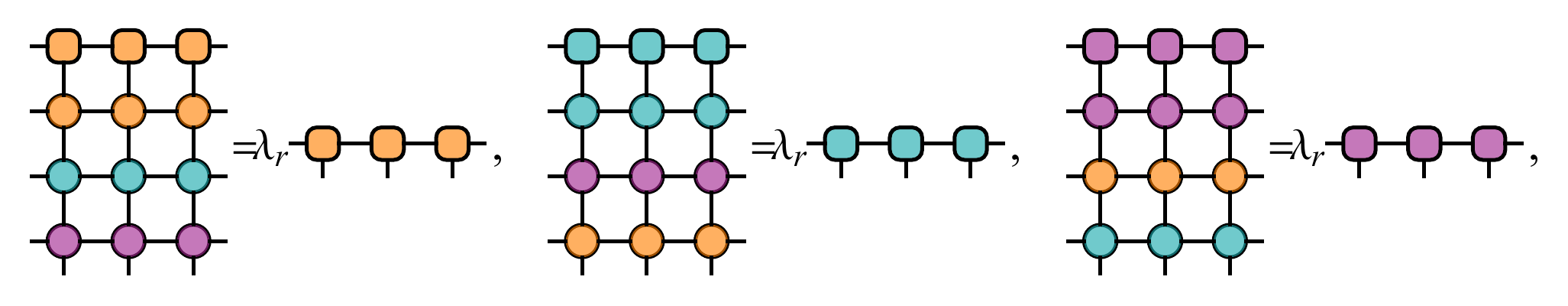}
\end{equation}
in which the boundary iMPS are defined as:
\begin{equation}
    \includegraphics[width=0.2\textwidth]{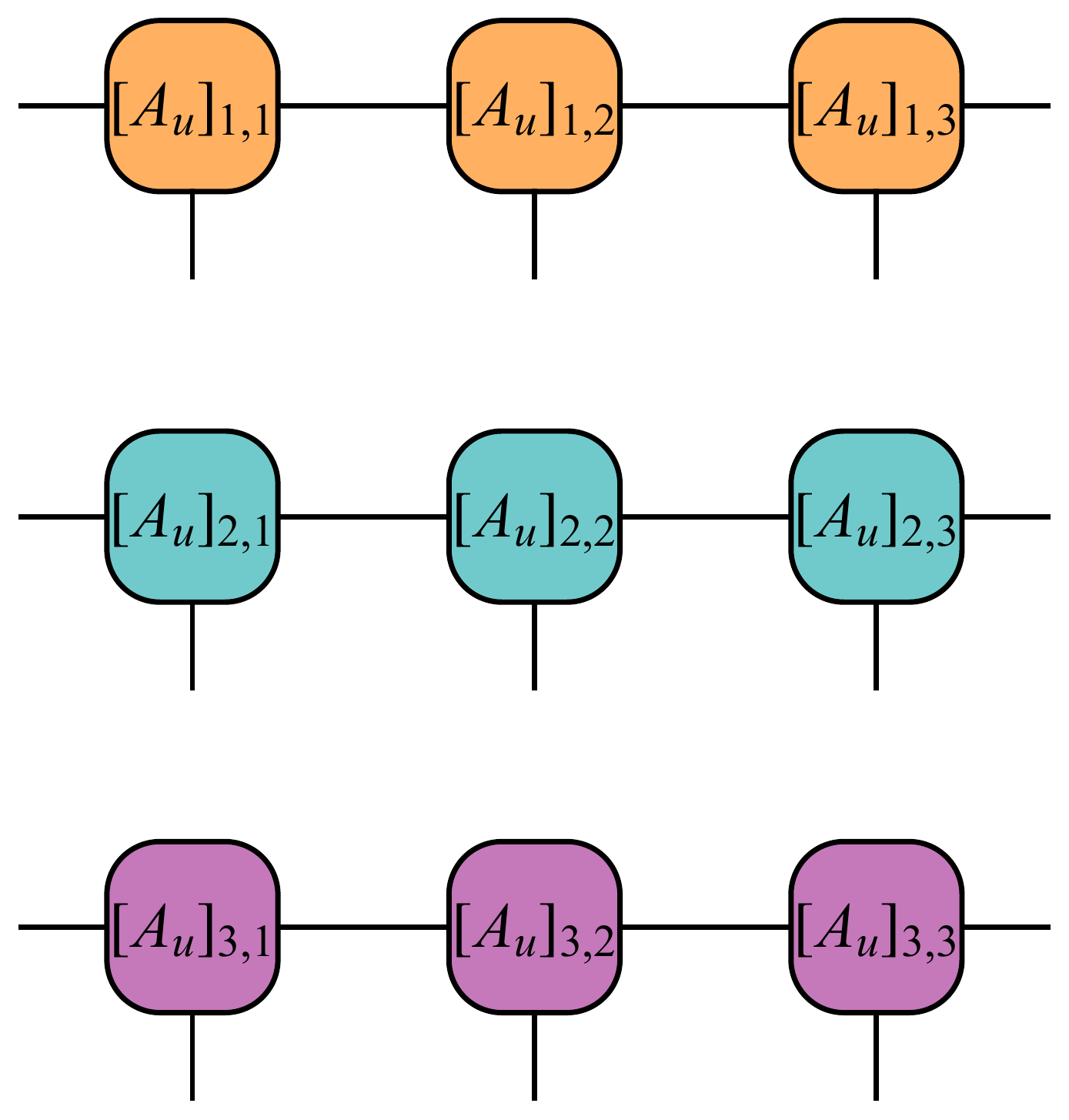}
\end{equation}
While the other three are for lower environments: 
\begin{equation}
    \includegraphics[width=1.0\linewidth]{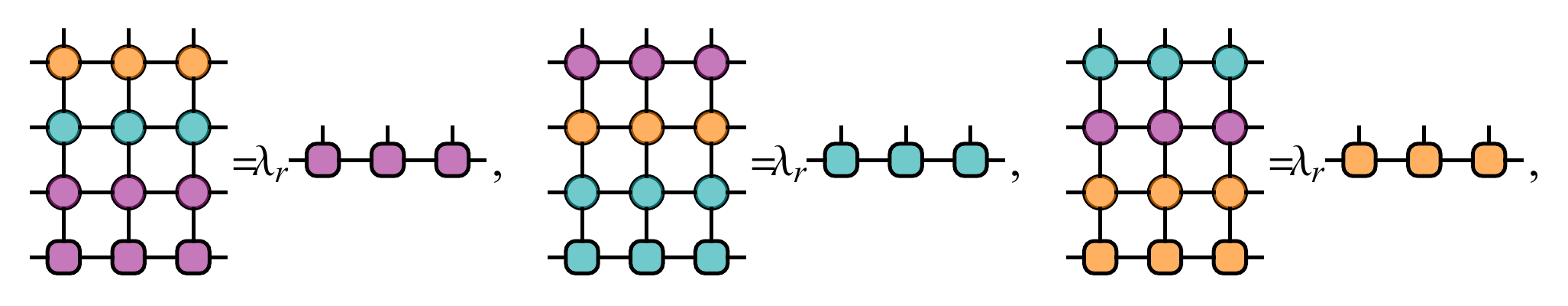}
\end{equation}
in which the boundary iMPS are:
\begin{equation}
    \includegraphics[width=0.2\textwidth]{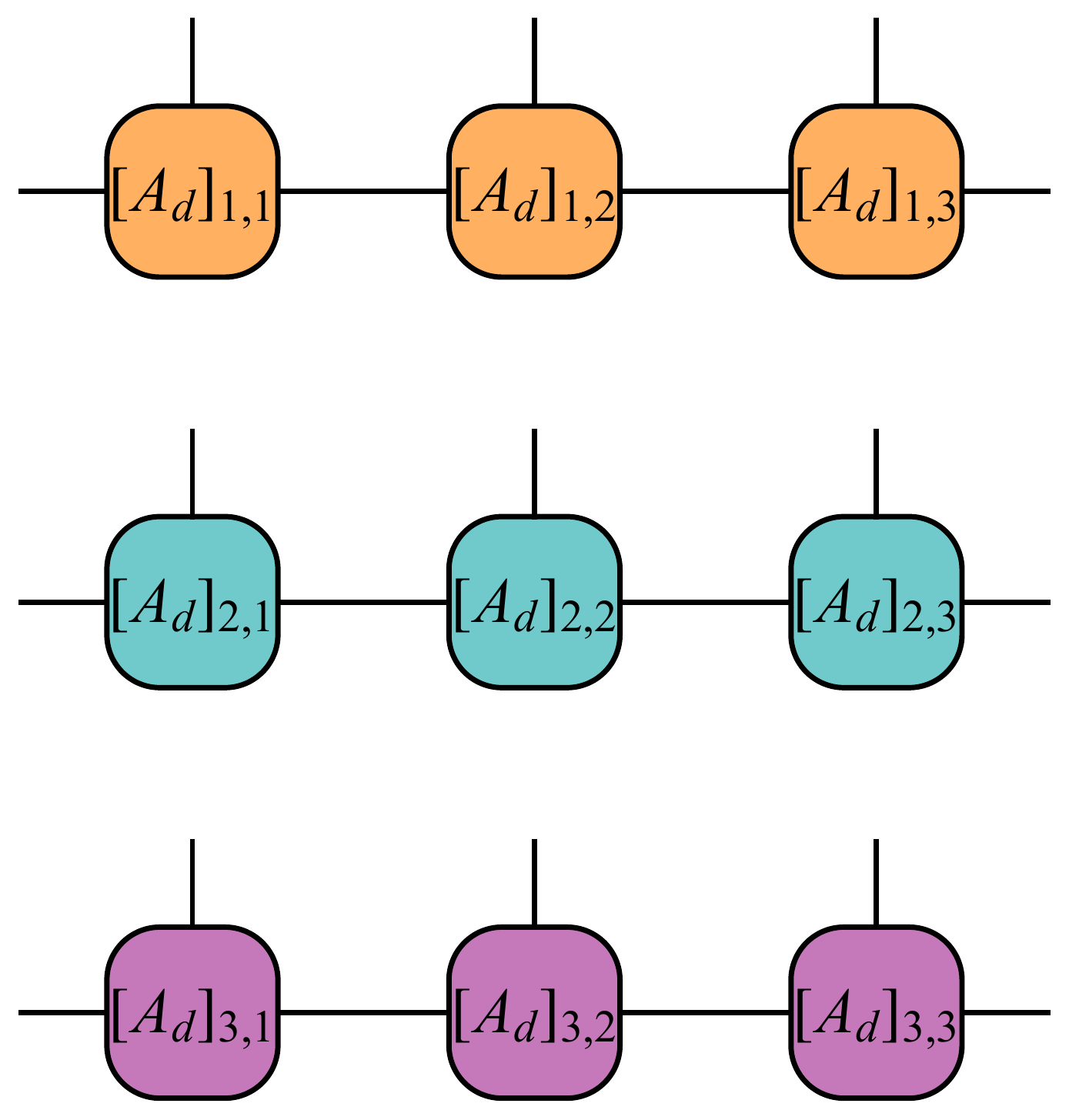}
\end{equation}
The maximum eigenvalue $\lambda_r$ is the same in all six equations because of translation invariance while corresponding eigenvectors for different transfer matrices are not equivalent.

If we straightforwardly apply the method discussed in the previous section on these six equations, the computation complexity would be $3 \times 3 \times 3$ times of that in the single unit cell case, where $3 \times 3$ complexity increase comes from the increasing number of parameters within a unit cell and the last $3$ times is a result of three different pair of environments. However, we can use a trick to solve the all upper or lower environments simultaneously and reduce the computation complexity to only $3 \times 3$ of that in the single unit cell case. Details are shown below. 

It is easy to transform the six equations above into six smaller equations:
\begin{equation}
    \includegraphics[width=1.0\linewidth]{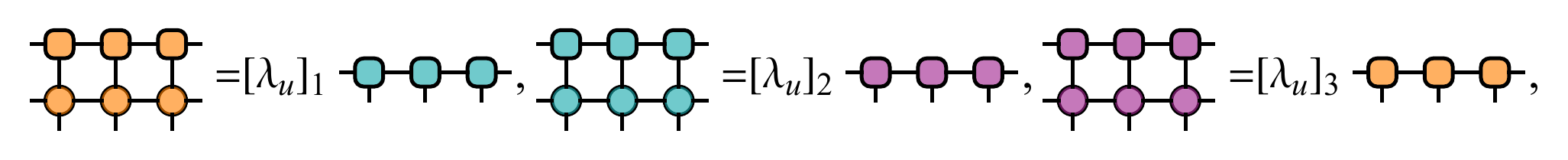}
\end{equation}
And:
\begin{equation}
    \includegraphics[width=1.0\linewidth]{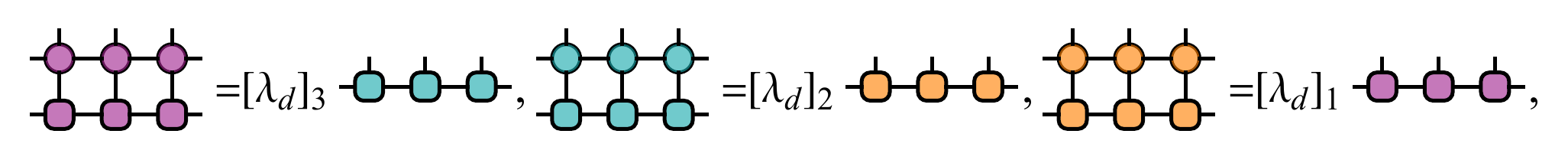}
\end{equation}
with $\lambda_r = [\lambda_u]_1 [\lambda_u]_2 [\lambda_u]_3 = [\lambda_d]_1 [\lambda_d]_2 [\lambda_d]_3$, and all $\lambda $s can be different. The price we pay is that these equations are no longer eigenequations and a dominant eigensolver can not be applied anymore. Luckily, an eigenproblem could be reconstructed if we write the group of equations in the following form:
\begin{equation}\label{equ: up-environment-eigen-function}
    \left(\begin{array}{lll}
            &        & T_3 \\
    T_1 &        &         \\
            &T_2 &
    \end{array}\right)\left(\begin{array}{l}
    \left[A_u\right]_{1,:} \\
    \left[A_u\right]_{2,:} \\
    \left[A_u\right]_{3,:}
    \end{array}\right)=\lambda_u\left(\begin{array}{l}
    \left[A_u\right]_{1,:} \\
    \left[A_u\right]_{2,:} \\
    \left[A_u\right]_{3,:}
    \end{array}\right)
\end{equation}
\begin{equation}\label{equ: down-environment-eigen-function}
    \left(\begin{array}{lll}
            & \tilde{T}_2 &          \\
            &         & \tilde{T}_3  \\
    \tilde{T}_1 &         &
    \end{array}\right)\left(\begin{array}{l}
    \left[A_d\right]_{1,:} \\
    \left[A_d\right]_{2,:} \\
    \left[A_d\right]_{3,:}
    \end{array}\right)=\lambda_d\left(\begin{array}{l}
    \left[A_d\right]_{1,:} \\
    \left[A_d\right]_{2,:} \\
    \left[A_d\right]_{3,:}
    \end{array}\right)
\end{equation}
where we fix $\lambda_u = [\lambda_u]_1 = [\lambda_u]_2 = [\lambda_u]_3$ and $\lambda_d = [\lambda_d]_1 = [\lambda_d]_2 = [\lambda_d]_3$ for convince. And the colon means the whole row of tensors and the tilde means exchange upper and lower tensor indices.

\iffalse
At last, we can fix $\lambda = [\lambda_u] = [\lambda_d]$, get the whole eigen equations:
\begin{small}
\begin{equation}
    \left(\begin{array}{llllll}
            &         & T_3   & & &\\
            T_1   &        &  & & &\\
                      &T_2 &  & & &\\
    & & &         & \tilde{T}_2 &          \\
    & & &         &         & \tilde{T}_3  \\
    & & & \tilde{T}_1 &         &
    \end{array}\right)\left(\begin{array}{l}
    \left[A_u\right]_{1,:} \\
    \left[A_u\right]_{2,:} \\
    \left[A_u\right]_{3,:} \\
    \left[A_d\right]_{1,:} \\
    \left[A_d\right]_{2,:} \\
    \left[A_d\right]_{3,:}
    \end{array}\right)=\lambda \left(\begin{array}{l}
    \left[A_u\right]_{1,:} \\
    \left[A_u\right]_{2,:} \\
    \left[A_u\right]_{3,:} \\
    \left[A_d\right]_{1,:} \\
    \left[A_d\right]_{2,:} \\
    \left[A_d\right]_{3,:}
    \end{array}\right)
\end{equation}
\end{small}
\fi

It is worth mentioning that there is no memory problem even though a large matrix consisting of the transfer matrices is present here since the matrix is not constructed explicitly in practice. We regard \cref{equ: up-environment-eigen-function} and \cref{equ: down-environment-eigen-function} as linear maps which are combinations of three fixed point equations as shown above. They can be solved simultaneously by a generic eigensolver such as Arnoldi\cite{arnoldi1951principle}. 
%In practice, we define the linear map rather than construct the large matrix, then use an iterative dominant eigensolver, such as Arnoldi, to solve the eigenproblem, so there is no memory problem. Another advantage is we get the up and down environment simultaneously avoiding different max eigenvalues.

Before coming into the power process, let's explicitly write down the canonical form of the boundary iMPS in large unit cell\cite{nietner2020efficient}:
\begin{equation}
    \includegraphics[width=1.0\linewidth]{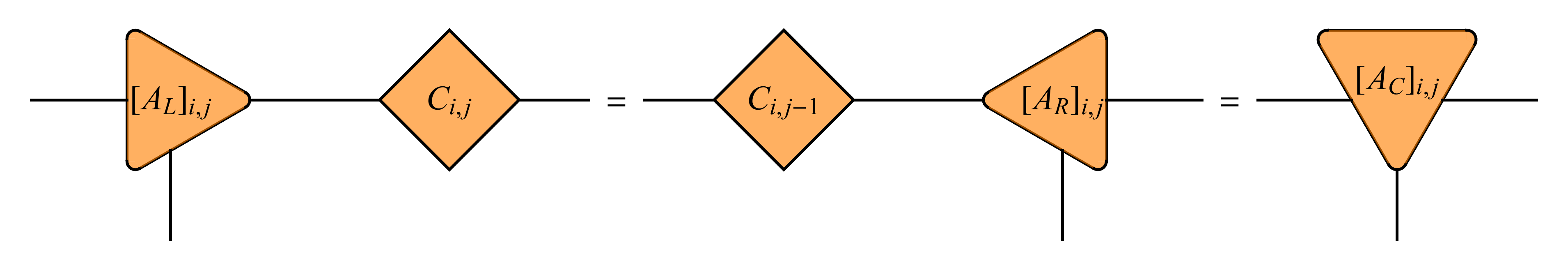}
    \label{equ:ALCARAC}
\end{equation}
with:
\begin{equation}
    \includegraphics[width=0.4\textwidth]{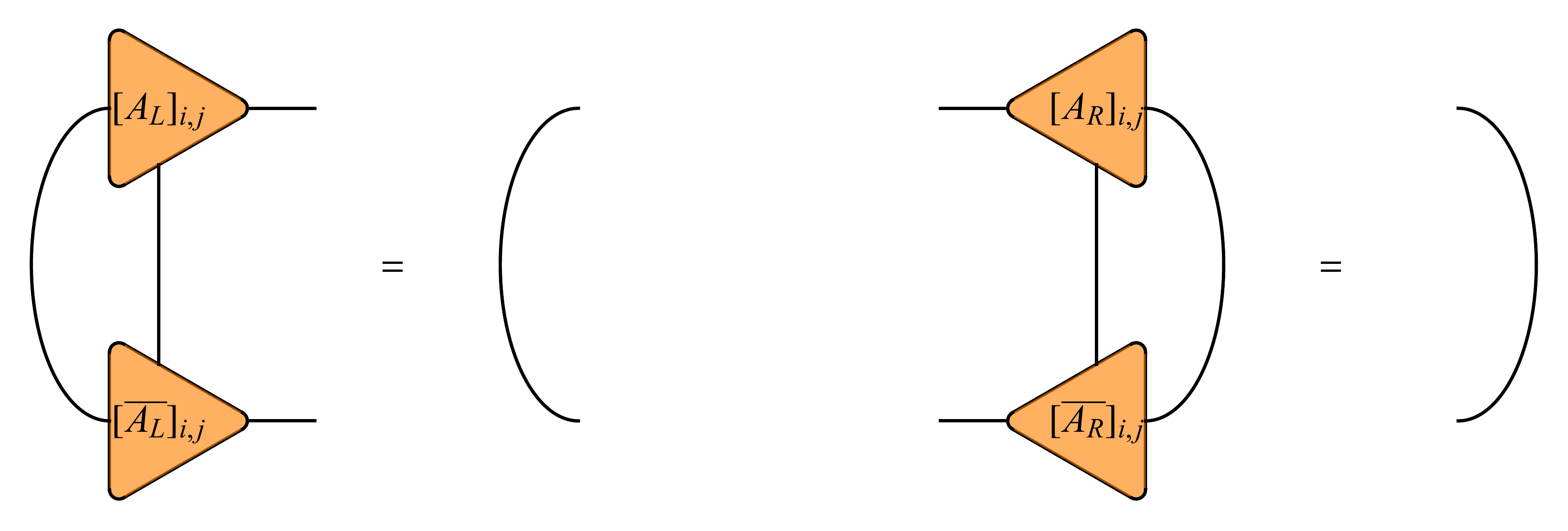}
    \label{equ:canonicalALAR}
\end{equation}
where the index of each tensor is carefully labeled. 

In this case, applying one-step power means to find:
\begin{equation}
    \includegraphics[width=0.4\textwidth]{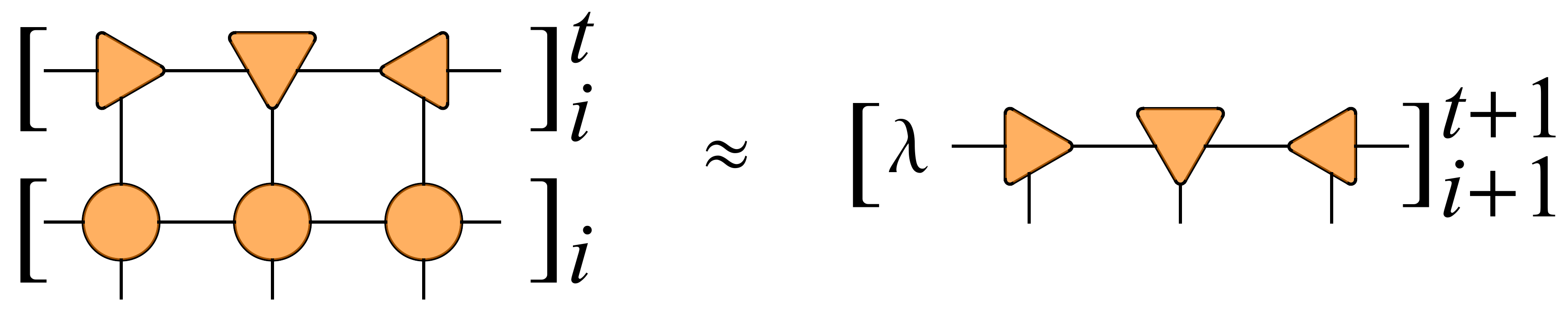}
    \label{equ: NixNj_one_step_power}
\end{equation}
where the superscript again labels the $t$-th iteration in the power method and the subscript indicates the equation is for the $i$-th row transfer matrix, $i$ and $i+1$ cyclically permutate between $1$ and $N_i = 3$. For each function in \cref{equ: NixNj_one_step_power}, the corresponding fixed point equation is:
\begin{equation}
    \includegraphics[width=0.32\textwidth]{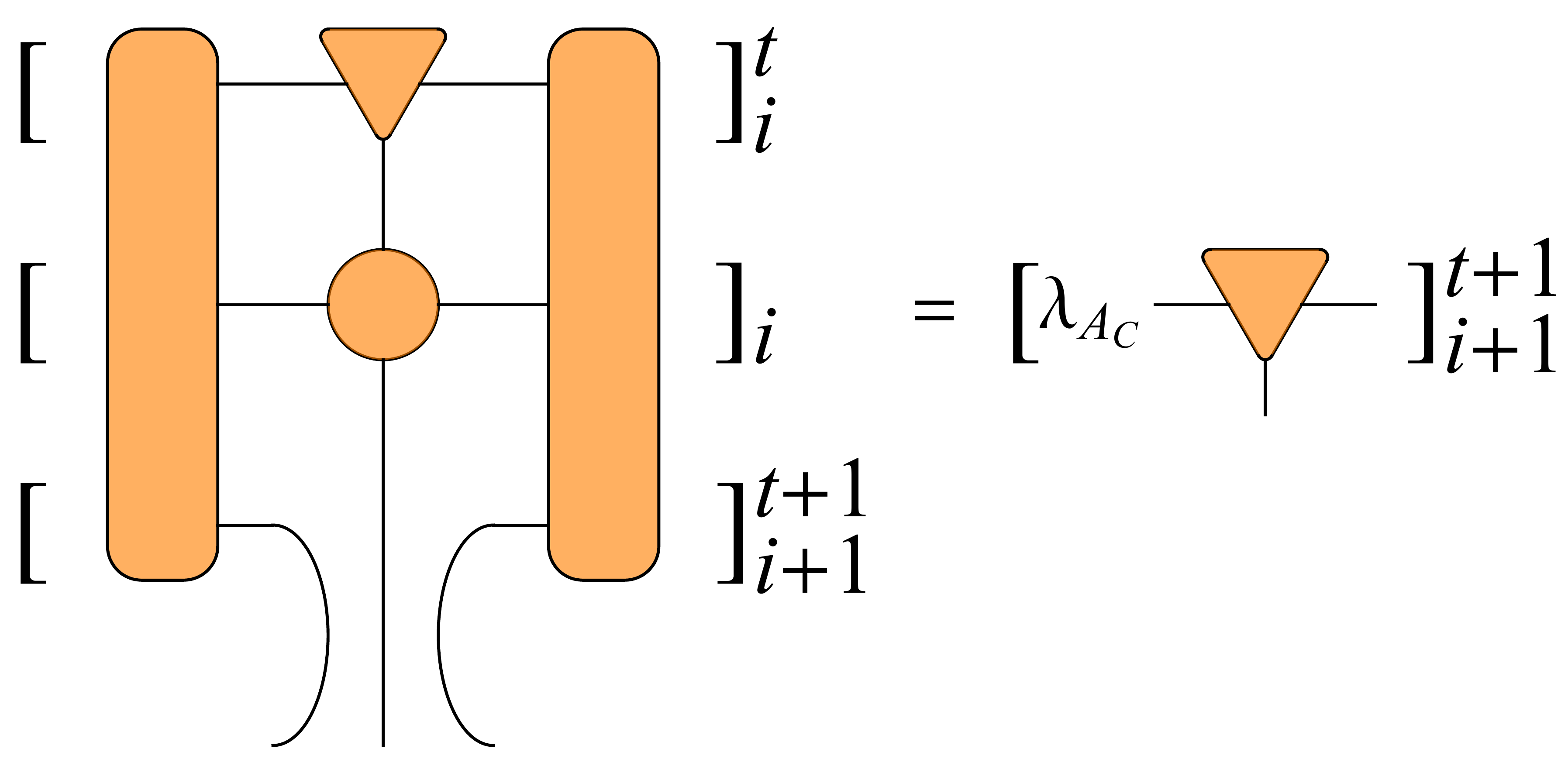}
    \label{equ: NixNj_fixed_point_AC}
\end{equation}
\begin{equation}
    \includegraphics[width=0.32\textwidth]{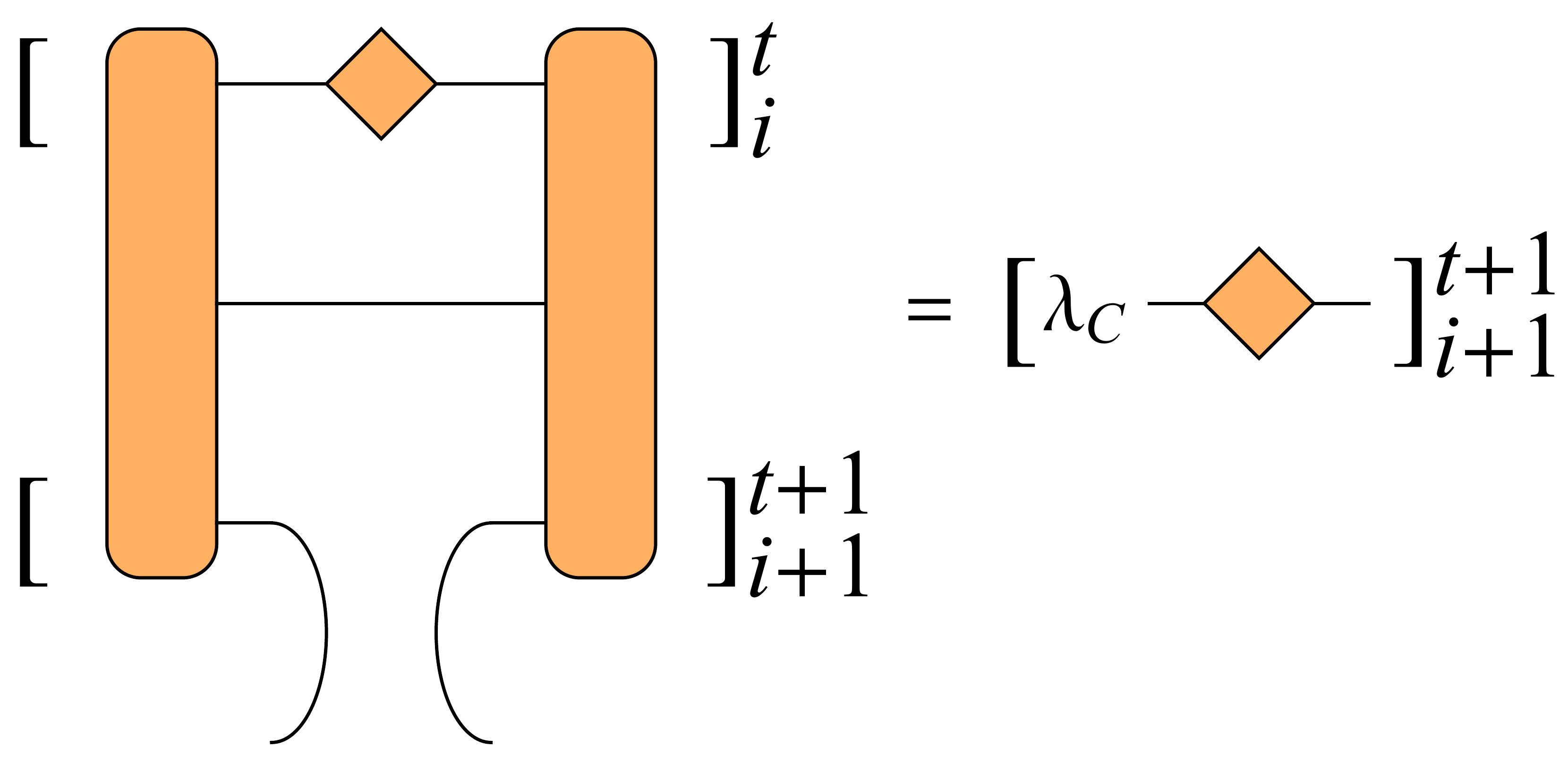}
    \label{equ: NixNj_fixed_point_C}
\end{equation}

\begin{figure}[H]
    \centering 
    \includegraphics[width=1.0\linewidth]{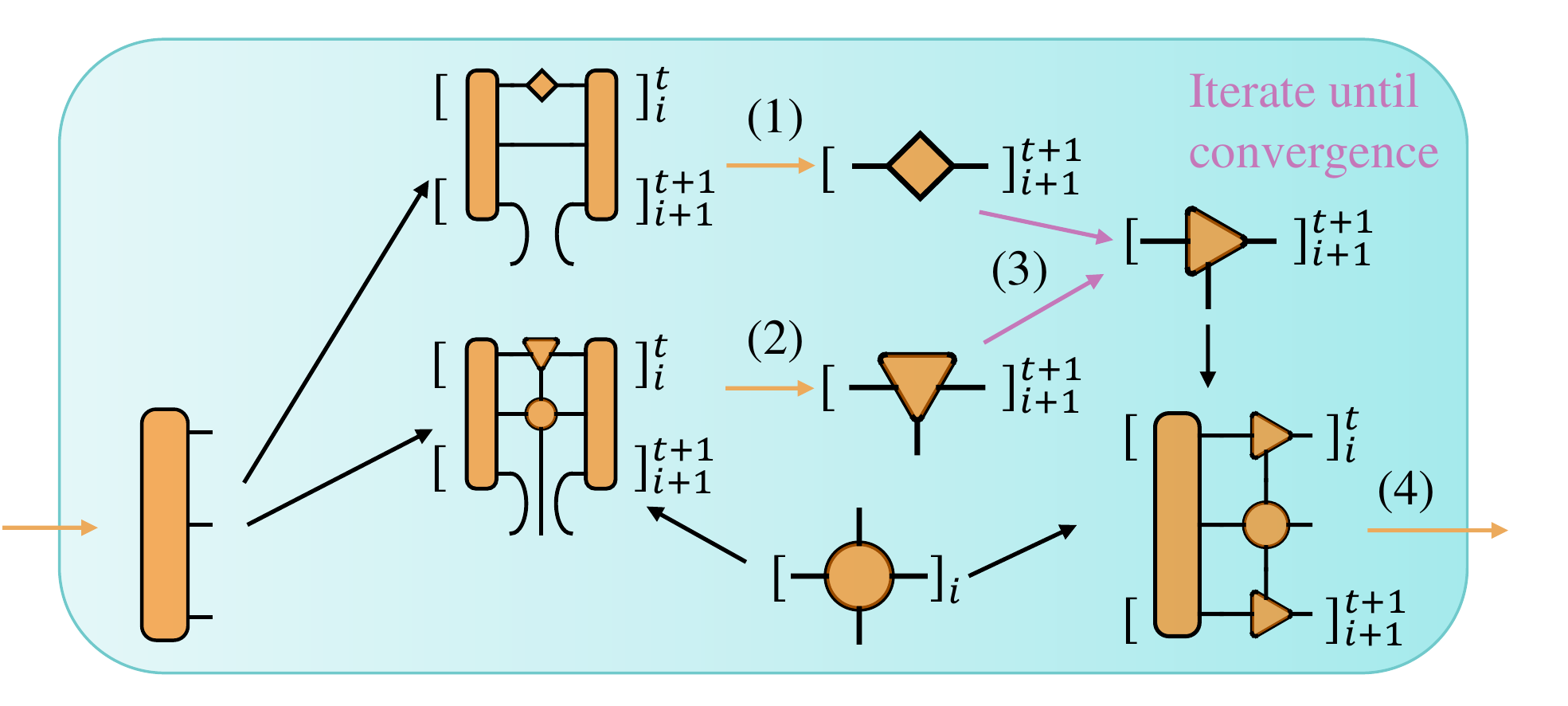}
    \caption{large unit cell VUMPS steps: different colors and steps (1)(2)(3)(4) meanings are the same as~\cref{fig: VUMPS_computation_graph_original}. The difference is that we use superscript for $t$-th iteration in the power method and subscript for $i$-th row transfer matrix.} 
    \label{fig: VUMPS_computation_graph_large_unit_cell}
\end{figure}

The iteration process for large unit cell cases is presented in~\cref{fig: VUMPS_computation_graph_large_unit_cell}

Compared to previous work~\cite{nietner2020efficient}, we enhance the algorithm by iterating the fixed points function~\cref{equ: NixNj_fixed_point_AC} and~\cref{equ: NixNj_fixed_point_C} with not only row index $i$, but also power index $t$. This is based on the non-Hermitian VUMPS method~\cite{vanhecke2021tangent}. In practical applications, our improved iterative approach has demonstrated superior performance compared to the original method.

\subsection{Automatic differentiation optimization}
\label{subsec: Automatic differentiation optimization}

The ground state is obtained by optimizing the ground state energy directly using gradient-based optimization on~\cref{equ: iPEPS_energy}. We calculate its gradient by automatic differentiation, as described in~\cite{liao2019differentiable}, then we optimize the iPEPS using
a quasi-Newton L-BFGS algorithm~\cite{nocedal2006numerical}. Compared to numerical differentiation, automatic differentiation can compute derivatives to machine precision and has the same complexity as energy evaluation \cite{BAUR1983317}. Moreover, in comparison with traditional gradient-based tensor methods such as those proposed in \cite{corboz2016variational,vanderstraeten2016gradient}, automatic differentiation eliminates the need for cumbersome graph summation. 

\subsubsection{Basics of AD}
Automatic differentiation is fundamentally dependent on utilizing the computation graph framework, where nodes within the graph represent data points, and the arrows signify the data flow through computational steps. For example, the simple linear chain computation graph $\theta \rightarrow T_1 \rightarrow T_2 \rightarrow \mathcal{L}$ illustrates the process of a vector parameter $\theta$ transforms into a scalar output $\mathcal{L}$. This process is commonly referred to as forward evaluation.

In order to obtain the gradient of $\mathcal{L}$ with respect to $\theta$, the chain rule is utilized as:
\begin{equation}
    \frac{\partial \mathcal{L}}{\partial \theta}=\frac{\partial \mathcal{L}}{\partial T_2} \frac{\partial T_2}{\partial T_1} \frac{\partial T_1}{\partial \theta}
    \label{equ: simple chain rule}
\end{equation}

In most cases where the input dimension is higher than the output dimensions, such as in deep neural networks' loss function or the free energy in physics, it is more efficient to compute~\cref{equ: simple chain rule} from left to right. The corresponding computation graph is $\bar{\theta} \leftarrow \bar{T}_1 \leftarrow \bar{T}_2 \leftarrow \bar{\mathcal{L}}$. This process involves defining the adjoint variable as $\bar{T} = \partial \mathcal{L} / \partial T$, which denotes the gradient of the final output $\mathcal{L}$ with respect to the variable $T$. The calculation commences from $\bar{\mathcal{L}}=1$ to $\bar{\theta}$, i.e., the gradient, and is conducted through multiple vector-Jacobian product operations that are represented by left arrows. This technique is referred to as reverse-mode automatic differentiation or backward evaluation.

It's worth noting that computation graphs of iterative tensor network calculations can be quite cumbersome to analyze. Fortunately, modern machine learning frameworks, e.g., PyTorch\cite{paszke2017automatic}, Jax\cite{frostig2018compiling}, and Zygote~\cite{innes2018don}, support automatic differentiation on arbitrary complex computations graphs. In principle, once the forward calculation is established, the backward calculation should become readily available. Nonetheless, an optimized efficient forward program may not be differentiable in practice due to in-place memory usage or limitations of the basic library. To solve this difficulty, one can build differentiable primitives by composing many elementary operations like addition, multiplication, and math functions together as a single entity, followed by defining the adjoint of this primitive. Once all primitives in the initial program become differentiable, the whole program becomes differentiable, ensuring that efficiency remains identical to that of the forward computation by careful design.

\subsubsection{VUMPS primitives}
\label{subsubsec: VUMPS primitives}
\begin{figure}[H]
    \centering 
    \includegraphics[width=0.45\textwidth]{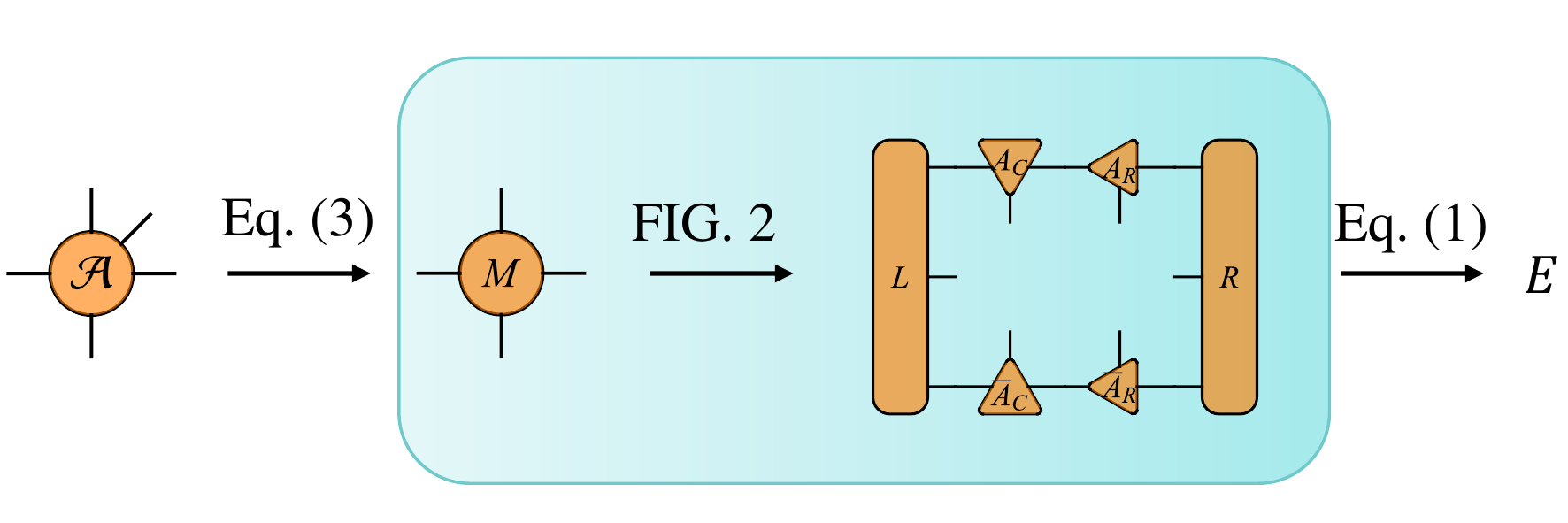}
    \caption{Whole computation graph from iPEPS to energy, where \cref{equ: AAjoint} is constructing $M$ tensor, \cref{fig: VUMPS_computation_graph_original} is the steps of VUMPS to get environment tensors and \cref{equ: iPEPS_energy} is the 3D contraction for energy} 
    \label{fig: whole_computation_graph}
\end{figure}

The computation graph to get iPEPS contraction energy is shown in~\cref{fig: whole_computation_graph}, and it relies on its subgraph of VUMPS, as shown in~\cref{fig: VUMPS_computation_graph_original}. In the VUMPS process, the QR decomposition and dominant eigensolver are non-differentiable directly. Therefore we regard them as primitives and discuss how to obtain their adjoints.

The adjoint of QR decomposition~\cite{seeger2017auto} is
\begin{equation}
    \bar{\mathcal{M}}=[\bar{Q}+Q \operatorname{Hermitian}(R\bar{R}^\dagger-\bar{Q}^\dagger Q)] (R^\dagger)^{-1}
\end{equation}
where $\operatorname{Hermitian}$ is to replace upper triangular matrix elements with the conjugate of the lower triangular matrix.

The adjoint of the dominant eigensolver is given by paper~\cite{xie2020automatic}. The key point is to solve a set of restricted linear equations. The dominant eigenequation is defined as:
\begin{equation}
    \boldsymbol{l}^{T} \mathcal{M}=\lambda \boldsymbol{l}^{T}, \quad \mathcal{M} \boldsymbol{r}=\lambda \boldsymbol{r}, \quad \boldsymbol{l}^{T} \boldsymbol{r}=1
    \label{equ: dominant eigenequation}
\end{equation}
Adjoint of backward is:
\begin{equation}
    \bar{\mathcal{M}}=\bar{\lambda} \boldsymbol{l} \boldsymbol{r}^{T}-\boldsymbol{l} \xi_{\boldsymbol{l}}^{T}-\xi_{\boldsymbol{r}} \boldsymbol{r}^{T}
    \label{equ: AD dominant eigensolver}
\end{equation}
where $\xi_{\boldsymbol{l}}$ and $\xi_{\boldsymbol{r}}$ are solved by:
\begin{equation}
    \begin{aligned}
    (\mathcal{M}-\lambda I) \xi_{l} &=\left(1-\boldsymbol{r} \boldsymbol{l}^{T}\right) \bar{\boldsymbol{l}}, & & \boldsymbol{l}^{T} \boldsymbol{\xi}_{l}=0 \\
    \left(\mathcal{M}^{T}-\lambda I\right) \xi_{r} &=\left(1-\boldsymbol{l} \boldsymbol{r}^{T}\right) \bar{\boldsymbol{r}}, & & \boldsymbol{r}^{T} \xi_{r}=0
    \end{aligned}
    \label{equ: AD dominant eigensolver linear function}
\end{equation}
The above equations should be solved in space that is orthogonal to $\boldsymbol{l}$ or $\boldsymbol{r}$, but considering $\mathcal{M}-\lambda I$ is not full rank, we only need to solve the equation using an iterative algorithm, such as GMRES~\cite{saad1986gmres}, with initial guess in this orthogonal space.

In the backward of VUMPS, we only need one side dominant eigenvector, $\boldsymbol{l}$ or $\boldsymbol{r}$, because we solve the environment separately. So~\cref{equ: AD dominant eigensolver} can be simplified. Firstly, suppose that only $\boldsymbol{l}$ is used, $\bar{\lambda}=0$, $\bar{\boldsymbol{r}}=0$, and the calculation result is gauge invariant, so~\cite{xie2020automatic}
\begin{equation}
    \boldsymbol{l}^{T} \bar{\boldsymbol{l}}=\boldsymbol{r}^{T} \bar{\boldsymbol{r}}=0
\end{equation}
\cref{equ: AD dominant eigensolver linear function} can be simplified as:
\begin{equation}
    \begin{aligned}
    (\mathcal{M}-\lambda I) \boldsymbol{\xi}_{l} &=\bar{\boldsymbol{l}}, & & \boldsymbol{l}^{T} \boldsymbol{\xi}_{l}=0 \\
    \left(\mathcal{M}^{T}-\lambda I\right) \boldsymbol{\xi}_{r} &=0, & & \boldsymbol{r}^{T} \boldsymbol{\xi}_{r}=0
    \end{aligned}
    \label{equ: simplified AD dominant eigensolver linear function}
\end{equation}
Assuming $\xi_{r} \neq 0$, $\xi_{r}$ must be a left eigenvector of $\mathcal{M}$, which conflicts with $\boldsymbol{r}^{T} \xi_{r}=0$, so $\xi_{r} = 0$. Finally, we get adjoint of left dominant eigensolver only with forward of $\boldsymbol{l}$:
\begin{equation}
    \bar{\mathcal{M}}=-\boldsymbol{l} \boldsymbol{\xi}_{\boldsymbol{l}}^{T}
\end{equation}
Similarly, we can get adjoint of the right dominant eigensolver only with forward of $\boldsymbol{r}$:
\begin{equation}
    \bar{\mathcal{M}}=-\boldsymbol{\xi}_{\boldsymbol{r}} \boldsymbol{r}^{T}
\end{equation}

\subsubsection{Technical comments}
\label{subsubsec: Technical suggestions}
To minimize computational expenses, it is advisable to refrain from constructing the adjoint $\bar{\mathcal{M}}$ tensor for the backward of dominant eigensolver, just as we should avoid constructing $\mathcal{M}$ in the forward. Instead, we can directly construct the adjoint of these two steps, which is the contraction of $\mathcal{M}$ and the dominant eigensolver.

Specifically, when to get the VUMPS environment, we do not construct $\mathcal{M}$ directly. As an illustration, to obtain $L$, the six-rank tensor $\mathcal{M}$, represented by $\adjincludegraphics[valign=c,width=0.02\textwidth]{./ALMAL.pdf}$, is subject to contraction by $A_L^u$, $M$, and $A_R^d$, resulting in significant memory consumption. Instead, we can define a linear map of this eigenequation, i.e., contract $L$, $A_L^u$, $M$, $A_L^d$ in sequence, then solve dominant eigenequation by an iterative method like Krylov subspace method~\cite{arnoldi1951principle, lehoucq1998arpack, KrylovKit.jl}. Performing this linear mapping with an optimal contraction order will result in a four-rank tensor and decrease memory usage.

\begin{equation}
    \centering 
    \includegraphics[width=0.25\textwidth]{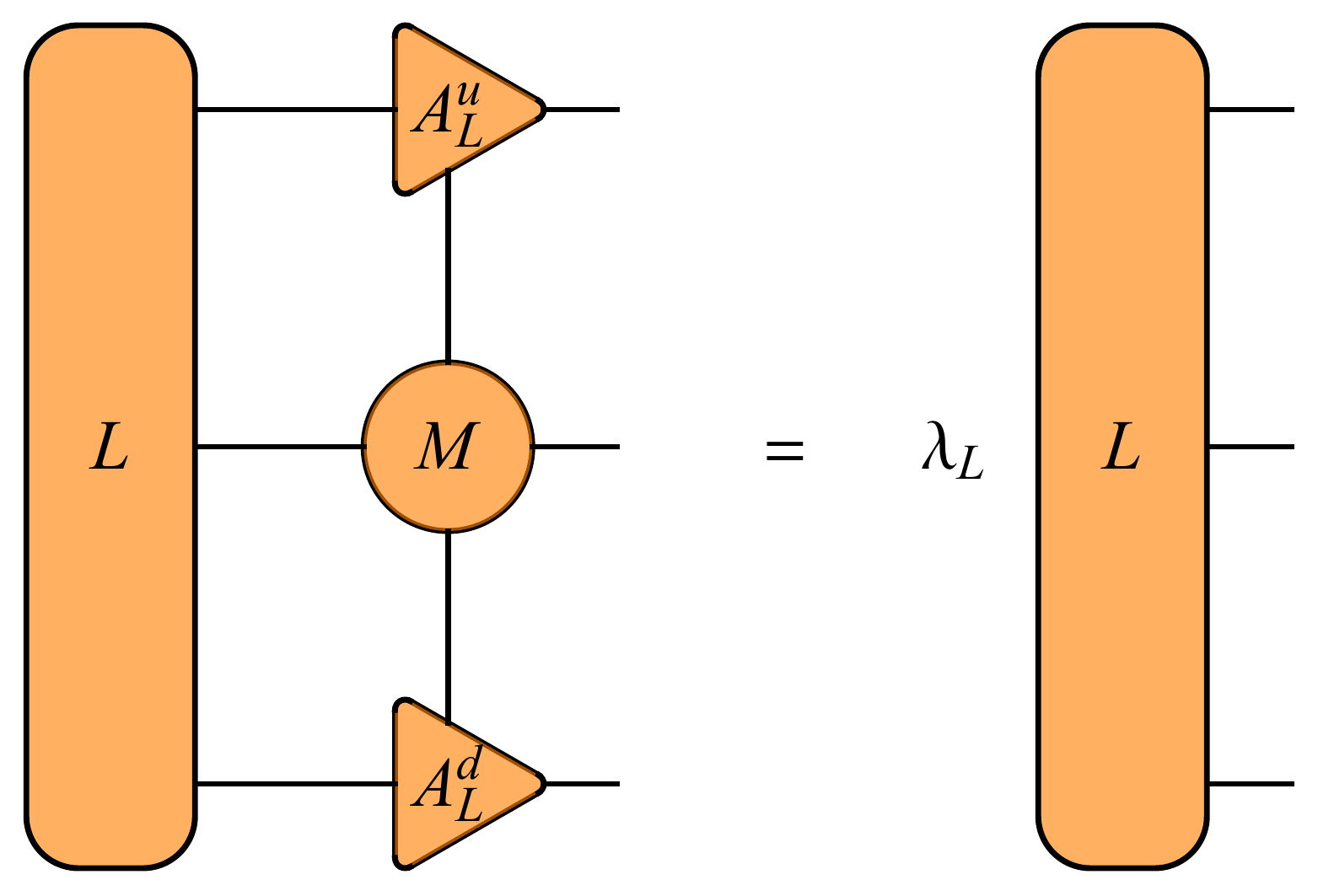}
\end{equation}

Similarly, we do not get $\bar{\mathcal{M}}$ directly in the backward propagation. We get $\bar{M},\bar{A}_L^u,\bar{A}_L^d$ according to calculation graph $M,A_L^u,A_L^d \longrightarrow \mathcal{M} \longrightarrow L$. Specifically, solve $\boldsymbol{\xi}_{l}$ by the Krylov method using first equation of \cref{equ: simplified AD dominant eigensolver linear function}, then use the contract graph below to get $\bar{A}_L^u,\bar{M},\bar{A}_L^d$:
\begin{equation}
    \centering 
    \includegraphics[width=0.28\textwidth]{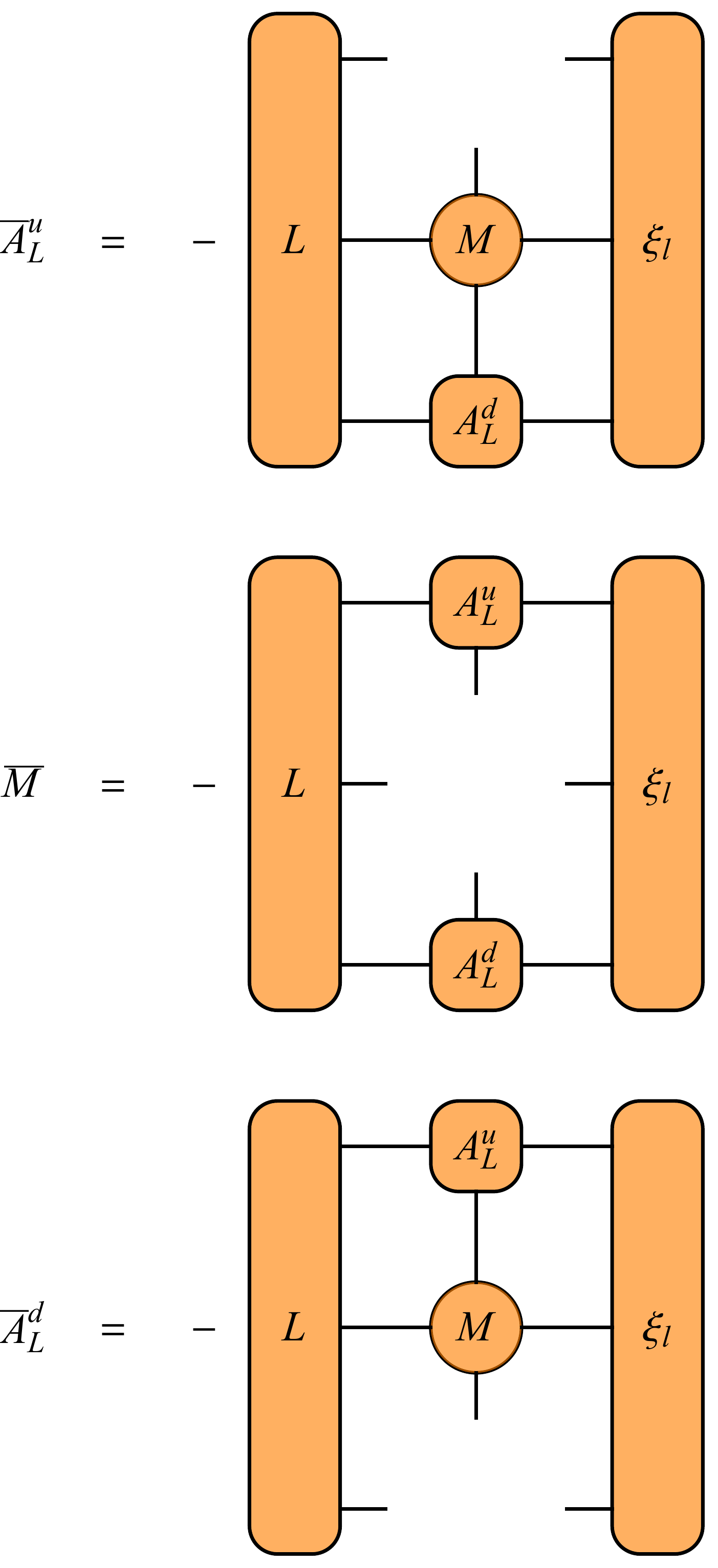}
    \label{fig:adjoint_ALuMALd}
\end{equation}

Finally, to ensure the stability and efficiency of backward, there are some crucial tricks:
\begin{itemize}
    \item The adjoint of QR decomposition include calculation inversion of a matrix, and mix gauge inversion $C^{-1}$ of VUMPS is singular, which means $\det (C)\approx 0$. We solve linear equations by adding a small number($10^{-12}$) to the diagonal of $C$ rather than calculating inversion. This trick can stabilize the calculation at the cost of introducing a small error in the gradient, which has very little effect on the final result experimentally.
    \item Caused by singularity above, linear equation~\cref{equ: simplified AD dominant eigensolver linear function} is not stable because $\bar{\boldsymbol{l}}$ is not orthogonal to $\boldsymbol{l}$. So, We should subtract part of $\bar{\boldsymbol{l}}$ which is not orthogonal to $\boldsymbol{l}$.
    \item We keep saving and loading the VUMPS environment in the process of optimization for two reasons. One is that inspired by annealing, the iPEPS in two successive steps are only slightly different, and so do the environment tensors. Therefore, loading environment tensors from the previous step as initiation will make the algorithm converge faster. The other is that with fewer forward steps, the gradient explosion is avoided and saves memory usage in the backward process.
    \item To address the largest memory usage issue arising from the last energy contraction, we employ checkpointing. 
\end{itemize}

\section{Applications}
\label{sec: Application}

%\subsection{Basic setup}
%\label{subsec: Basic setup}
As a first application of the developed method, we consider the 
$K$-$J$-$\Gamma$-$\Gamma^\prime$~\cite{PhysRevLett.112.077204, 
Winter_2017,PhysRevLett.105.027204} Hamiltonian, 
\begin{equation}
\begin{aligned}
\hat{H}=&\sum_{\left \langle i,j\right \rangle_{\gamma} } K S_{i}^{\gamma} S_{j}^{\gamma}+J\mathbf{S}_i\cdot \mathbf{S}_j +\Gamma\left(S_{i}^{\alpha} S_{j}^{\beta}+S_{i}^{\beta} S_{j}^{\alpha}\right) \\
    &+\Gamma^{\prime}\left(S_{i}^{\alpha} S_{j}^{\gamma}+S_{i}^{\gamma} S_{j}^{\alpha}+S_{i}^{\beta} S_{j}^{\gamma}+S_{i}^{\gamma} S_{j}^{\beta}\right)
    \end{aligned}
    \label{equ: Kitaev-type Hamiltonian}
\end{equation}
where $\mathbf{S}_i=\{S_i^x, S_i^y, S_i^z\}$ are the pseudo spin-1/2 operators 
at site $i$, and $\left \langle i,j\right \rangle_{\gamma}$ denotes the nearest-neighbor 
pair on the $\gamma$ bond, with $\{\alpha, \beta, \gamma\}$ being $\{x,y,z\}$ 
under a cyclic permutation. $K$ is the Kitaev coupling,  $J$ is the Heisenberg 
term, $\Gamma$ and $\Gamma^\prime$ the off-diagonal couplings. 

The Hamiltonian \cref{equ: Kitaev-type Hamiltonian}, containing additional terms to the pure Kitaev interaction, is can be used to describe realistic materials such as iridium oxides and $\mathrm{\alpha}$-$\mathrm{RuCl_3}$. 
%In detail~\cite{doi:10.1146/annurev-conmatphys-031115-011319}, Kitaev interactions $K$ are caused by iridium $5d$ orbitals linked on the edge-shared octahedra~\cite{PhysRevB.96.115103}. 
For example, there is a Heisenberg term $J$ caused by $5d-5d$ superexchange interaction. Moreover, there are $\Gamma$ and $\Gamma^\prime$ terms which are superexchange interactions with reduced symmetry due to spin-orbit couplings and lattice distortions in the materials. 
%Another symmetry allowed interactions for second and third-nearest neighbors have been also discussed~\cite{yamaji2014first, sizyuk2014importance}.

The  $K$-$J$-$\Gamma$-$\Gamma^\prime$ Hamiltonian exhibits a rich phase diagram consisting of ferromagnetic (FM), antiferromagnetic (Néel), zigzag (ZZ), stripy phases~\cite{iregui2014probing}, as well as more complex phases with 6-site, 18-site unit cells~\cite{lee2020magnetic,li2022tangle}. Our method developed above is ideally suitable to investigate such a problem. 
%To carry out numerical simulations for these phases need large unit cells, which are non-trivial for previous iPEPS studies. They always use the full update for simple phases and the simple update for complex phases because of computation complexity, but sometimes the simulations are unconvincing and self-contradictory. 

\begin{figure}[H] 
\centering
\includegraphics[width=0.5\textwidth]{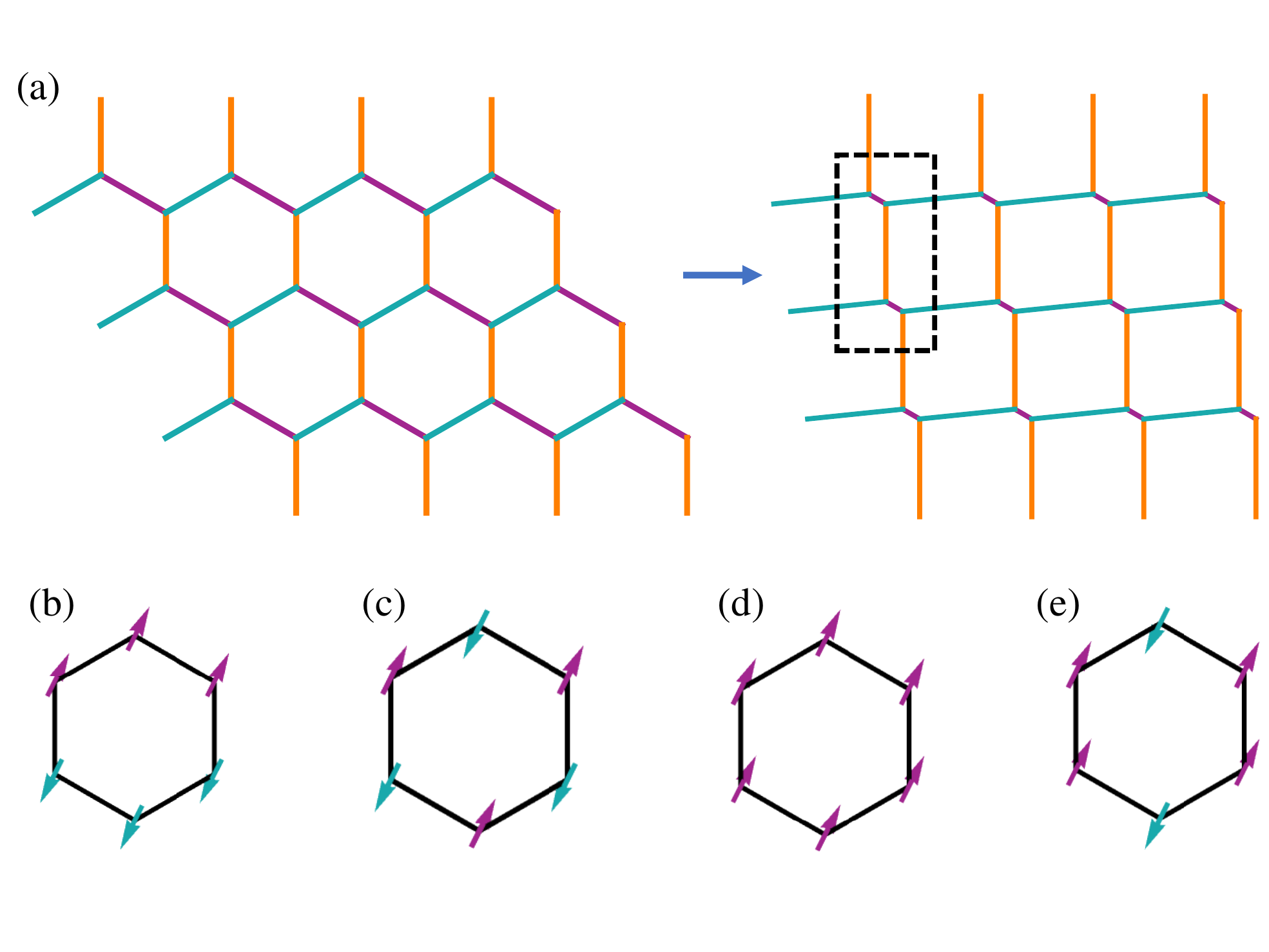} 
\caption{(a) Transformation of the honeycomb lattice to the square lattice 
by merging two sites in a unit cell of the original honeycomb lattice. %$S^y, S^x, S^z$ 
$x$-, $y$-, and $z$-bonds are represented by purple, cyan and orange lines, 
respectively. The dashed black box is $2 \times 1$ unit cell and four phases 
are illustrated: (b) zigzag, (c) Néel, (d) ferromagnetic and (e) stripy phase} 
\label{fig: 2x1_iPEPS_ansatz} 
\end{figure}

First of all, we group two sites in the unit cell of the honeycomb lattice so we will still deal with iPEPS defined on a square lattice with physical bond dimension $d=2^2$, as shown in ~\cref{fig: 2x1_iPEPS_ansatz}(a). Similarly, when considering complex spin-ordered phases, one could similarly group several unit cells to represent a large magnetic unit cell, for example, the black box in~\cref{fig: 2x1_iPEPS_ansatz}(a) represents a magnetic unit cell that consists of four lattice points.~\footnote{In contrast with brick-wall lattice~\cite{iregui2014probing}, this transformation merging two sites contains 
twice the physical bonds but half the number of unit cells. In practice, we find such a scheme is more effective and thus use it in later simulations.}  

\subsection{Pure Kitaev model}
\label{sec: pure kitaev}
We start with benchmark calculations of the pure Kitaev model with only 
$K=-1$ and no other terms.
%, the model in~\cref{equ: Kitaev-type Hamiltonian} becomes the famous Kitaev honeycomb model with equal bond couplings(B phase~\cite{kitaev2006anyons}).
The ground state is exactly solvable~\cite{kitaev2006anyons}, and is shown to 
be a gapless quantum spin liquid without magnetic order. On the other hand, 
%directly solving the original 
numerical simulations of the Kitaev spin model is nevertheless rather challenging due to high degrees of spin frustrations.

\begin{figure}[H]
    \centering     
    \includegraphics[width=0.5\textwidth]{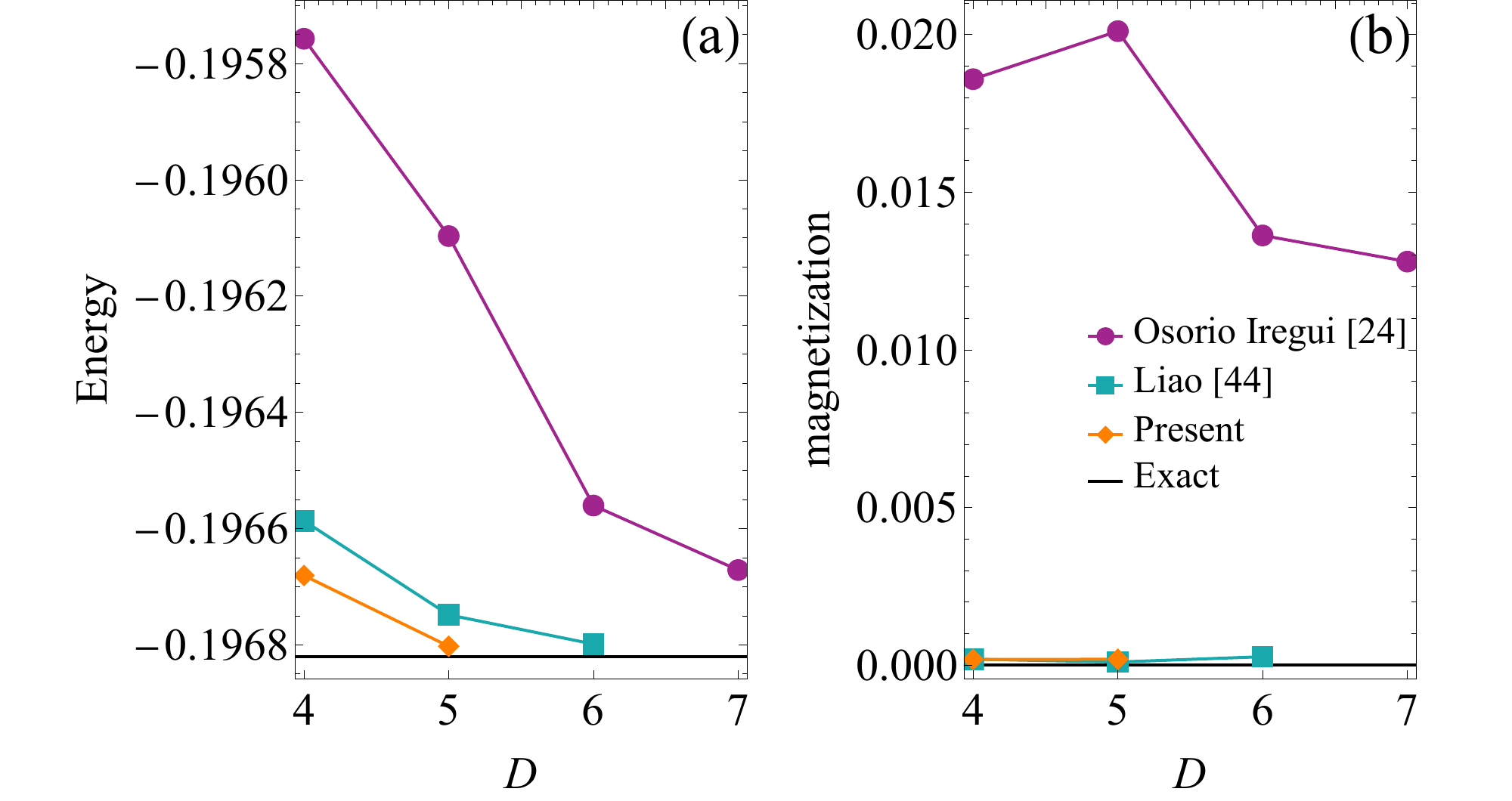}
    \caption{The groundstate energy and magnetization of the pure Kitaev model ($K=-1$) versus $D$. We obtained more accurate results than the ITE and achieved results comparable to CTMRG.}
    \label{fig: Kitaev_energy_mag}
\end{figure}

\cref{fig: Kitaev_energy_mag}(a) shows that with $D=5$ ($\chi=120$) one can obtain a variational energy $E=-0.196805$ which is in fourth decimal precision agreement with the exact result $-0.19682$. The variational energy is significantly lower than the previous imaginary-time projection-based optimization (full update) up to $D=7$ ($\chi=60$)~\cite{iregui2014probing} and comparable to the carefully constructed iPEPS $-0.19681$ with $D=8$~\cite{PhysRevLett.123.087203}.  More importantly, as shown in Fig.~\ref{fig: Kitaev_energy_mag}(b), the variational optimization gives essentially zero magnetization (as small as $10^{-4}$). Also shown in the figures are the unpublished results~\cite{Liao2019talk} by differentiating through a CTMRG contraction, which is similar to the recent result~\cite{PhysRevB.107.054424}. These benchmarks show that even with moderate bond dimensions, one can still obtain highly accurate results on the pure Kitaev model with a more thorough optimization. 

The dominant computational complexity of the VUMPS method is equivalent to CTMRG. The computational complexity can be expressed as $\mathcal{O}(D^6\chi^3)$. We performed simulations using an NVIDIA A100-40G GPU. For a pure Kitaev model, it took approximately 4 hours at $D/\chi=4/80$ and 17 hours at $D/\chi=5/120$. 

What is noteworthy is that the AD method uses smaller $D$ but larger $\chi$ than ITE to get lower energy. This exactly evidences that the optimization ability of AD is better than ITE and the key point to solving the Kitaev model is optimization ability rather than representation ability. It means representation ability, i.e. the amount of entanglement, of small $D$ is enough to carry the complexity of the Kitaev ground state and large boundary environment bond dimension $\chi$ is needed. The reason why we need larger $\chi$ is that our groundstate iPEPS at the same $D$ is more complex (thus supposed to be highly entangled) than ITE, so we need environments with larger $\chi$, thus more entanglement, to conduct an accurate contraction. To demonstrate this, we examine the convergence of both the energy and the correlation length as a function of the bond dimension $\chi$ in~\cref{sec: The convergence versus the boundary bond dimension}. Based on the above results, we find the optimization ability of AD brings iPEPS to an unattainable place for ITE. At the same time, the contraction ability is also crucial and we also overcome this with AD by a sense of annealing as mentioned in Sec.~\ref{subsubsec: Technical suggestions}. 

\subsection{Kitaev-$\Gamma$ model}
\label{subsec: KG model}
We consider the $K$-$\Gamma$ model with ferromagnetic Kitaev interaction which was proposed to capture the essence of the spin couplings in realistic material RuCl$_3$~\cite{Wei2017theoretical,PhysRevLett.118.107203}. It was found that there are nematic paramagnetic (NP) and magnetically ordered phases with 18-site unit cells~\cite{lee2020magnetic,li2022tangle} in addition to the Kitaev spin liquid phase (KSL) in the $K$-$\Gamma$ model. 

First, we found that the KSL phase survives at $\Gamma=0.03$. Similar to the 
comparison on the pure Kitaev model in~\cref{sec: pure kitaev}, we obtain exact 
zero magnetization with and lower energy compared to previous iPEPS studies 
with imaginary-time projection optimization~\cite{lee2020magnetic}. 
\begin{figure}[H] 
    \centering
    \includegraphics[width=0.475\textwidth]{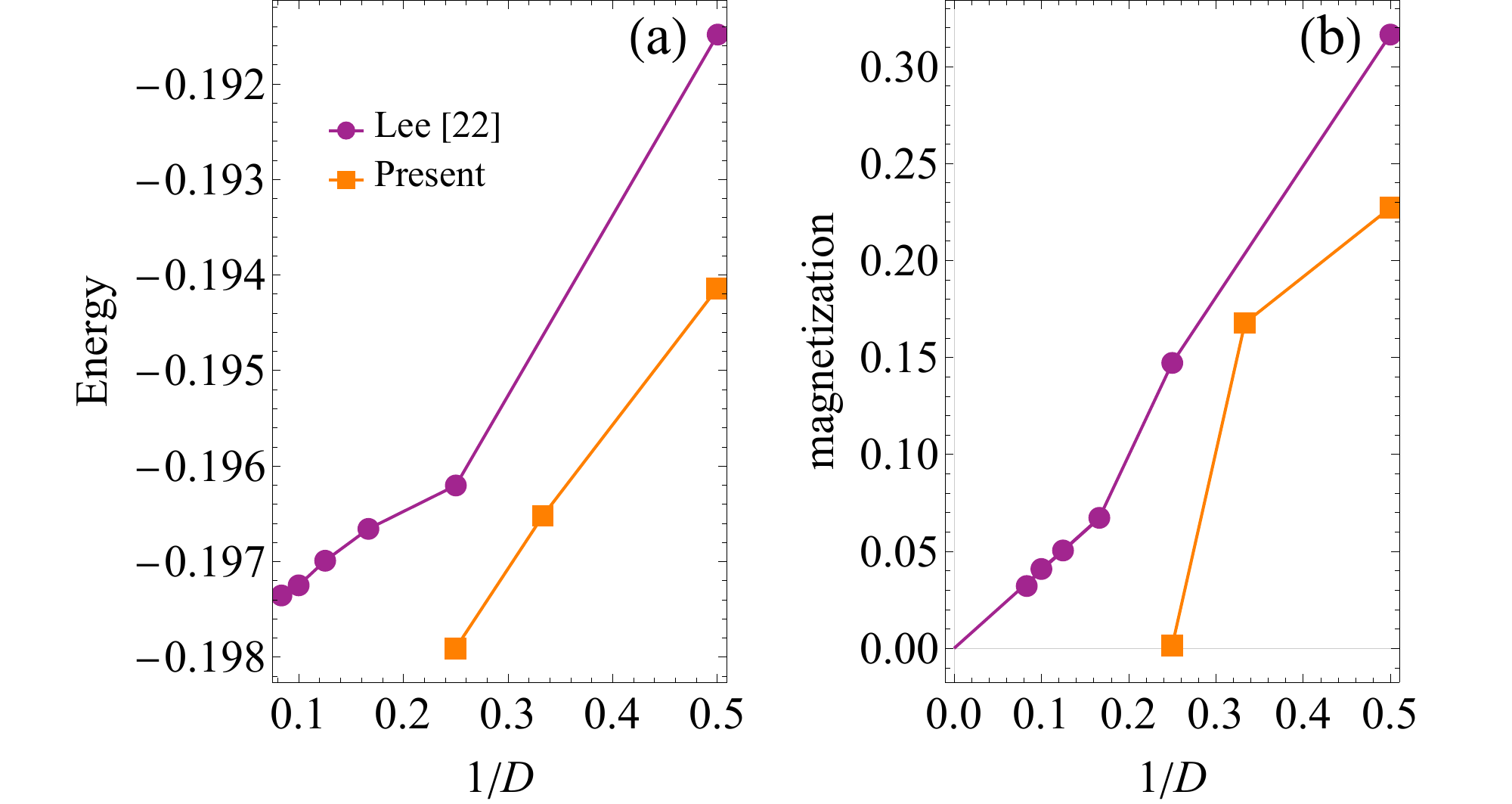}      
    \caption{The groundstate energy and magnetization of the $K$-$\Gamma$ model 
    at $\Gamma=0.03$. We obtain lower energy and zero magnetization directly 
    compared to \cite{lee2020magnetic}.} 
    \label{fig:K-Gamma_0.03}
\end{figure}

\begin{figure}[H] 
    \centering
    \includegraphics[width=0.5\textwidth]{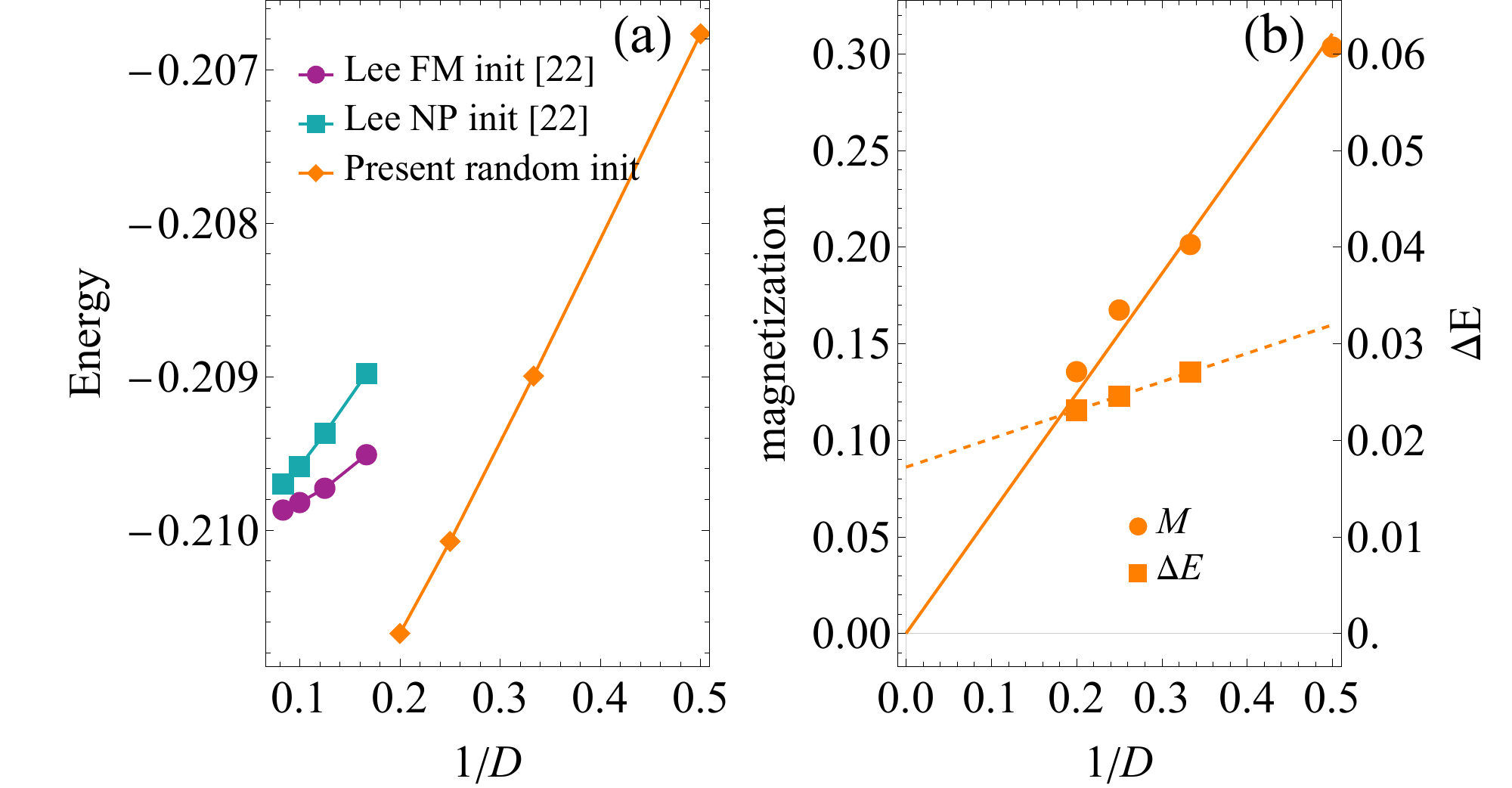}      
    \caption{Results of $K$-$\Gamma$ model with $\Gamma=0.095$,
    where the lower energies compared to previous results in (a) and zero 
    magnetization but non-zero $\Delta E$ in (b) are observed. }
    \label{fig: K-Gamma_0.095}
\end{figure}

At $\Gamma=0.095$, the previous study was not conclusive on whether the model 
shows nematic paramagnetic or ferromagnetic order~\cite{lee2020magnetic}. 
Our calculations suggest that the ground state is nematic paramagnetic with 
anisotropic bond energies, and the results are independent of initialization.
Figure~\ref{fig: K-Gamma_0.095}(a) shows that we've obtained much lower 
variational energy. Extrapolation Figure~\ref{fig: K-Gamma_0.095}(b) over the bond dimension shows that the 
obtained state has zero magnetization but finite bond energy anisotropy. 
Here we define $\Delta E = \mathrm{std}(E_x, E_y, E_ z)$ as the standard 
deviation of the bond energies. 

\begin{figure}[H] 
    \centering
    \includegraphics[width=0.45\textwidth]{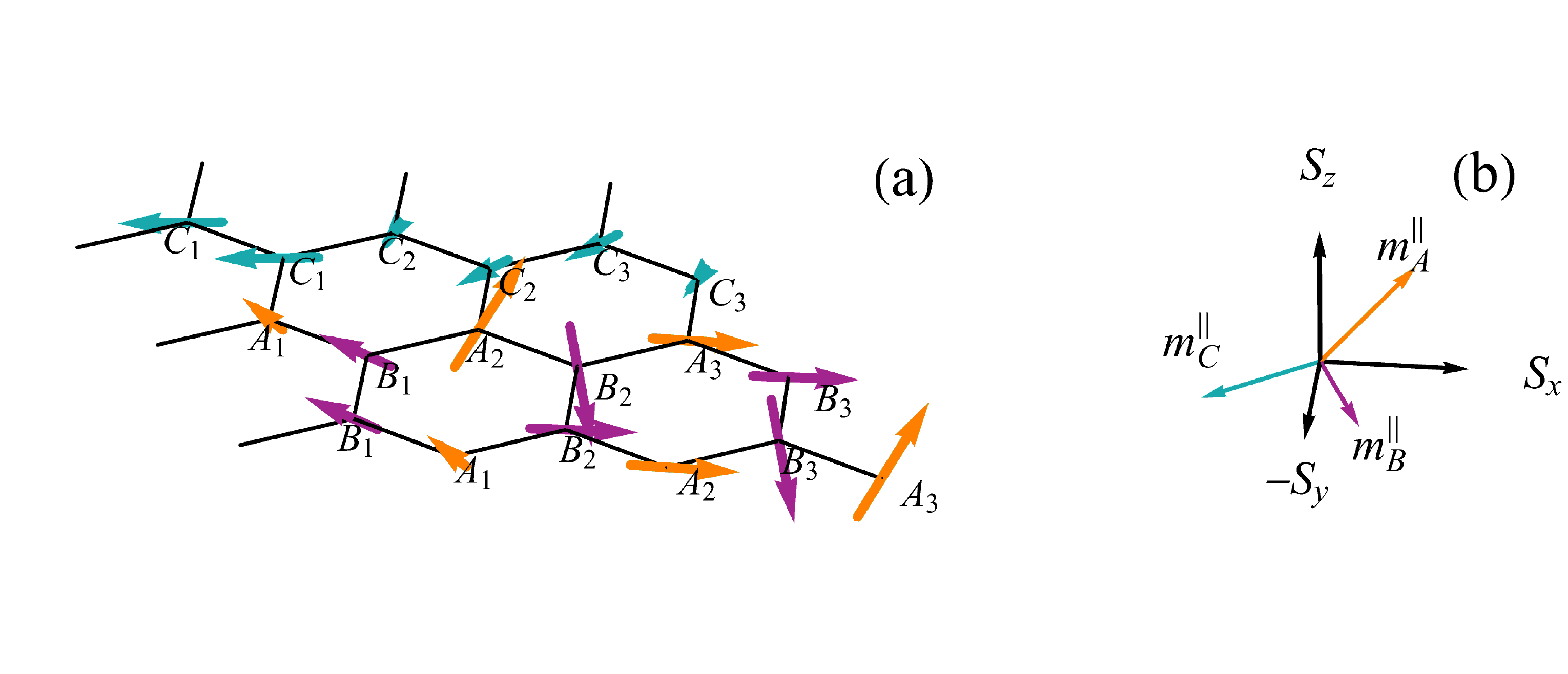}      
    \caption{$K$-$\Gamma$ configuration at $\Gamma/|K|=1.0$: the different 
    colors mean $ABC$ components in~\cref{equ: sublattice magnetization}, 
    where these components form a zigzag order in a $2\times 2$ unit cell as 
    shown Fig.~\ref{fig: 2x1_iPEPS_ansatz}(b), where they are antiparallel to 
    each other. However, here in panel (a) not all components are antiparallel,
    and some of them have an angle that deviates from perfect collinearity. 
    }
    \label{fig: 18-site_config}
\end{figure}

At $\Gamma = 1$, we obtain a complex spin-helix order with an 18-site unit cell, shown in~\cref{fig: 18-site_config}(a). Such results confirm previous findings~\cite{lee2020magnetic,li2022tangle} but with lower variational energy per site: -0.3520 at $D=4$ compared with -0.3518 at $D=8$ reported in~\cite{li2022tangle}. Moreover, we obtain the configuration starting from a single randomly initialized calculation rather than repeating hundreds of initial guesses. To minimize the impact of local minima, we also conducted experiments using different unit cell sizes and random initializations to confirm the reliability of our results. When setting the unit cell to be  $2\times 2$, we will obtain a zigzag order with higher variational energy compared to the $6\times 3$ configuration shown in \cref{fig: 18-site_config}(a). On the other hand, when enlarging the unit cell to be 36 sites we still obtain the 18-site order with $10^{-5}$ lower energy per site. 

Figure~\ref{fig: 18-site_config}(b) show the sublattice magnetization 
for the three sublattices 
$\mathbf{m}_{\mu}=\frac{1}{N_{\text {cell }}} \sum_{\text {cells }} 
\mathbf{S}_{\mu} \label{equ: sublattice magnetization}$
where $\mathbf{m}_{\mu}$ captures the longitudinal components 
(or rotation axes) of an ensemble of highly structured spin helices. 

\subsection{$K$-$J$-$\Gamma$-$\Gamma^\prime$ model}
\label{subsec: K-J-Gamma-Gamma' model} Finally, we consider 
the most general form of Hamiltonian \cref{equ: Kitaev-type Hamiltonian}. 
We adopt the model parameters $K=-25\mathrm{meV}$, 
$\Gamma=0.3|K|$,  $\Gamma^\prime=-0.02|K|$ and $J=-0.1|K|$
\cite{li2021identification}, which has been found to fit well most 
experimental observations in the compound $\alpha$-$\rm{RuCl}_3$
\cite{PhysRevB.93.214431, Winter_2017, PhysRevLett.118.107203, 
PhysRevB.99.249902, PhysRevB.98.094425,PhysRevB.98.060412, 
PhysRevB.93.075144}. 

In particular, with such a parameter, DMRG calculations that the ground state shows zigzag magnetic order, which can be suppressed by external magnetic fields.

Under in-plane fields, there is a ZZ to a paramagnetic phase transition, and for an out-of-plane magnetic field there firstly occurs a transition
from the ZZ order to the NP phase, before the systems become polarized eventually. To simulate these interesting phases and phase transitions,
we introduce magnetic fields to the Hamiltonian by including the term
$h \bm{\hat{n}}\cdot \bm{S}$. The direction perpendicular to the plane is 
represented by $\bm{\hat{n}} = (1,1,1)/\sqrt{3}$ and the direction within 
the plane is represented by $\bm{\hat{n}} = (1,1,-2)/\sqrt{6}$. $\bm{S} 
= (S_x, S_y, S_z)$ is spin-1/2 operator and $h$ represents the strength 
of the field. Reference~\cite{li2021identification} studied the phase diagram with DMRG and thermal tensor network calculations on a finite cylinder and found it to be consistent with the experimental observations. 

\begin{figure}[H]
    \centering 
    \includegraphics[width=0.45\textwidth]{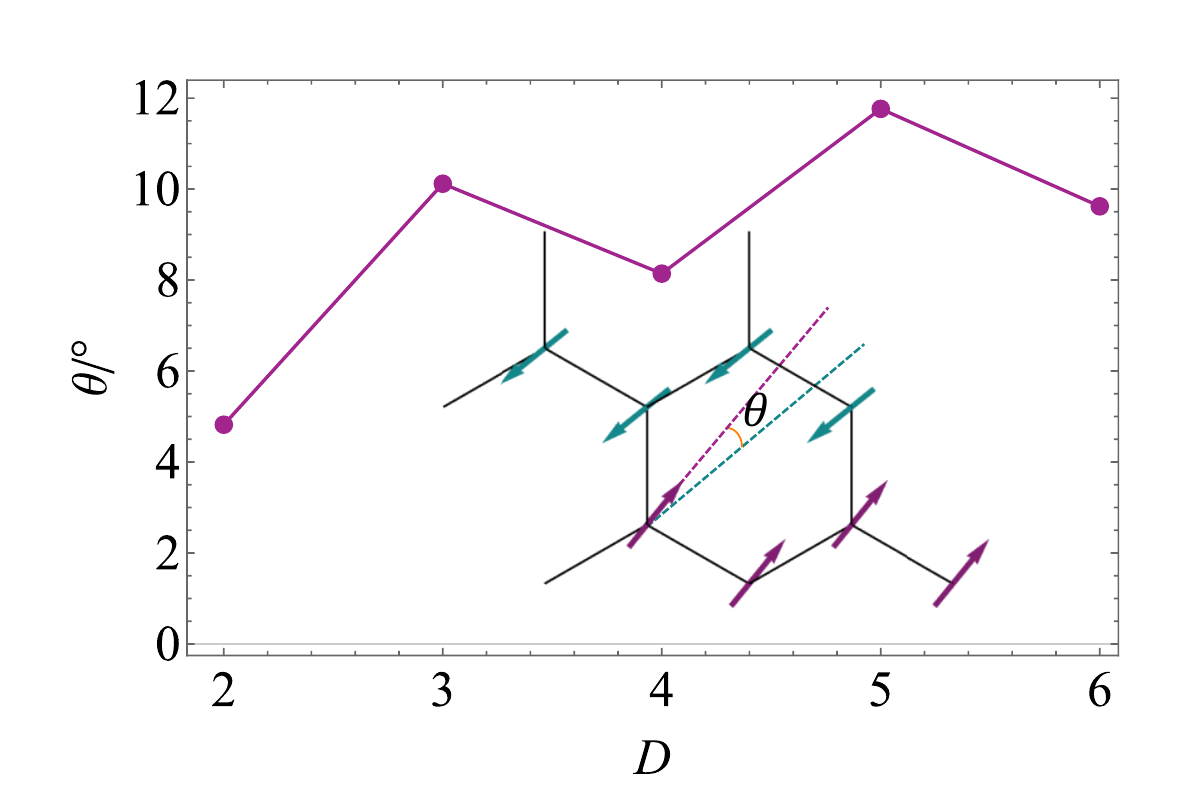}
    \caption{The non-collinear angle $\theta$ of the 
    $K$-$J$-$\Gamma$-$\Gamma^\prime$ ground state 
    versus bond dimension $D$ of iPEPS simulations. 
    Inset defines the tilting angle $\theta$.}
    \label{fig: non-collinear_angle}
\end{figure}

\begin{figure}[H]
    \centering 
    \includegraphics[width=0.45\textwidth]{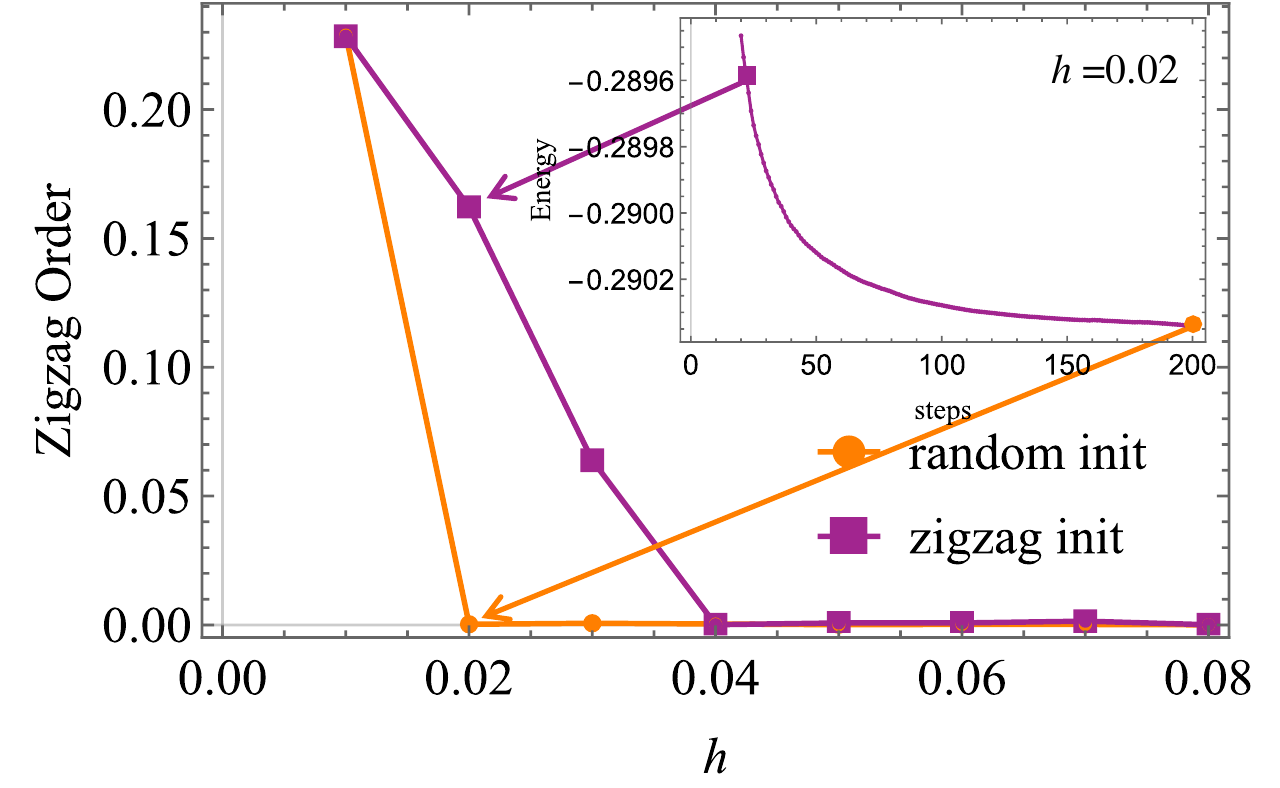}
    \caption{
    We study the effect of different initializations on the in-plane magnetization process. We compare two cases: random initialization and zigzag initialization, and find that both cases converge to the same final state, but with different optimization steps. Taking $h=0.02$ as an example, when starting from a zigzag state, the system undergoes a transition from zigzag order to a paramagnetic state at the 23rd optimization step, as 
    indicated by the purple arrow in the subfigure energy. This state has the same energy as the one reached by random initialization after 200 optimization steps, as shown by the orange arrow in the subfigure energy.
    }
    \label{fig:in-plane_mag}
\end{figure}

To start with, we compute the magnetic order under zero field $h=0$. As shown in~\cref{fig: non-collinear_angle}, indeed we find the spins form a ZZ order. Moreover, we further observe that the magnetic moments are not perfectly anti-parallel, and the conclusion holds for all different bond dimensions in the simulation, where the tilting angle increases with $D$ in~\cref{fig: non-collinear_angle}. We suppose this non-collinear spin configuration may explain the magnetodielectric effect in the $\alpha$-$\mathrm{RuCl_3}$~\cite{PhysRevB.95.245104}.

In \cref{fig:in-plane_mag}, we show the results under in-plane fields, where the zigzag order parameter vanishes when the field exceeds a certain 
critical value, revealing a ZZ-PM phase transition. This is consistent with 
previous model studies~\cite{li2021identification} and experimental observations~\cite{Banerjee2017,Zheng2017,modic2021scale}. %disappears 
Nevertheless, the critical field values obtained here show some differences from these previous studies. Starting from random iPEPS of bond dimension
$D=5$ as initial state, the critical field is found to be $h_c/K\simeq0.02$ 
that corresponds to 3.5~T (see \cref{fig:in-plane_mag} orange line), about 
half the 7-8~T transition field observed in experiments~\cite{Banerjee2017,
Zheng2017,modic2021scale} and also in previous DMRG calculations on 
the same model~\cite{li2021identification}. Instead, take the ZZ state as the initial state, we find the critical field moves to $h_c/K\simeq 0.04$ (about 7~T), 
and the obtained states are with higher energies (purple line in~\cref{fig:in-plane_mag}) in our iPEPS calculations. 

\begin{figure}[H]
    \centering 
    \includegraphics[width=0.45\textwidth]{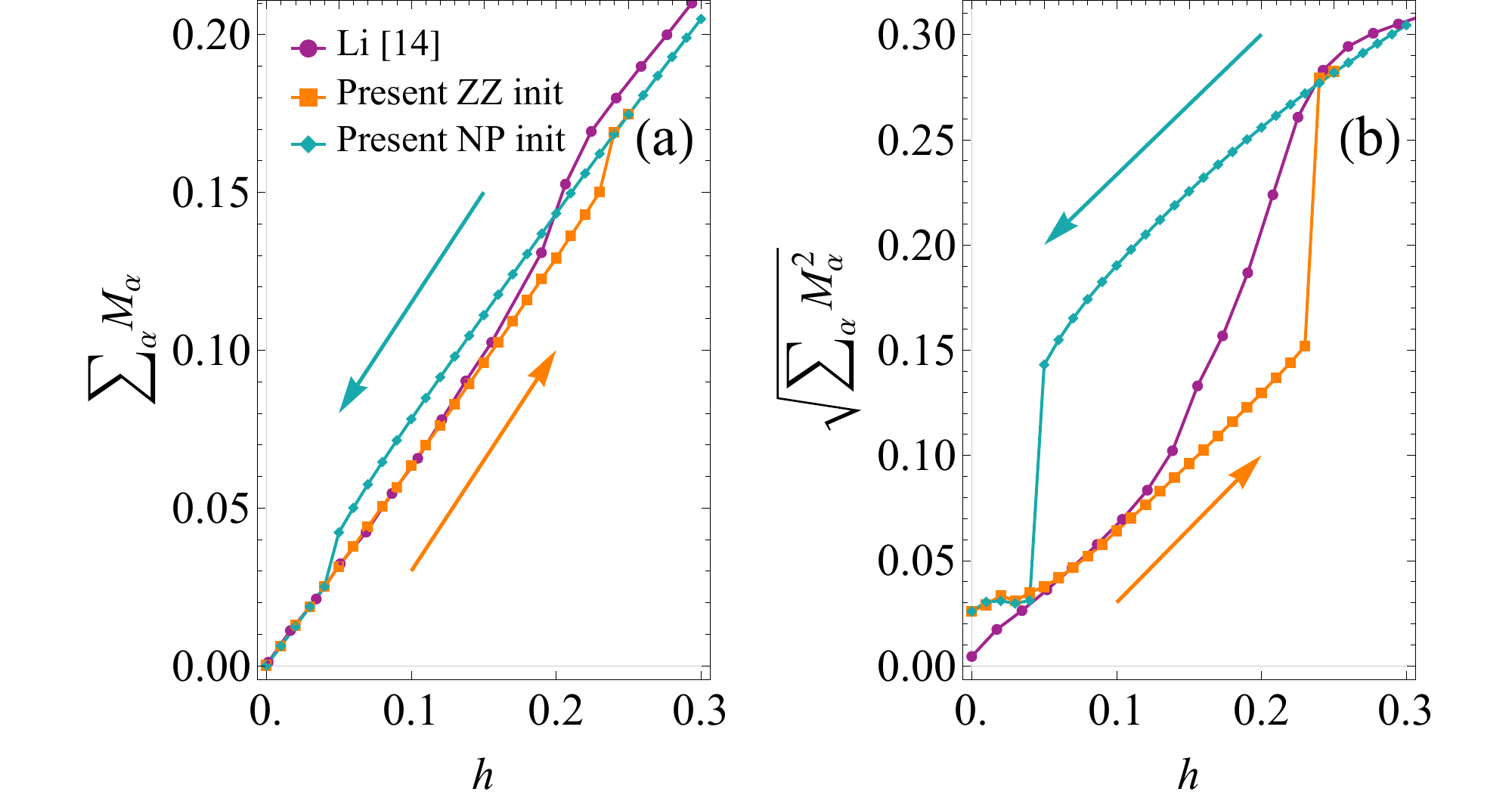}
    \caption{(a) uniform magnetization and (b) a local magnetization from 
    the zigzag and nematic initial phase with $D=5$ iPEPS calculations. The 
    first-order phase transition can be witnessed by a jump in the uniform and 
    local magnetizations. However, the two different schemes, ZZ and NB 
    initializations, produce different transition fields. Note that the local moment 
    $\sqrt{\sum_\alpha M_\alpha^2}$ is finite at the zero field, indicating the presence of ZZ order there.}
    \label{fig: K-J-Gamma-Gamma_ZZ_init_mag}
\end{figure}

An out-of-plane field causes a first-order phase transition from the ZZ to NP. We increase the magnetic field from the ZZ phase and reach a critical field of $0.23 |K|$ (43~T), consistent with DMRG result \cite{li2021identification}. The order parameter jumps sharply at the critical point [see \cref{fig: K-J-Gamma-Gamma_ZZ_init_mag}(a)]. However, in the same figure, we also find a quite different transition field when we decrease fields from the NP phase. They even form a hysteresis loop-like shape in \cref{fig: K-J-Gamma-Gamma_ZZ_init_mag}(b).

\begin{figure}[H]
    \centering 
    \includegraphics[width=0.45\textwidth]{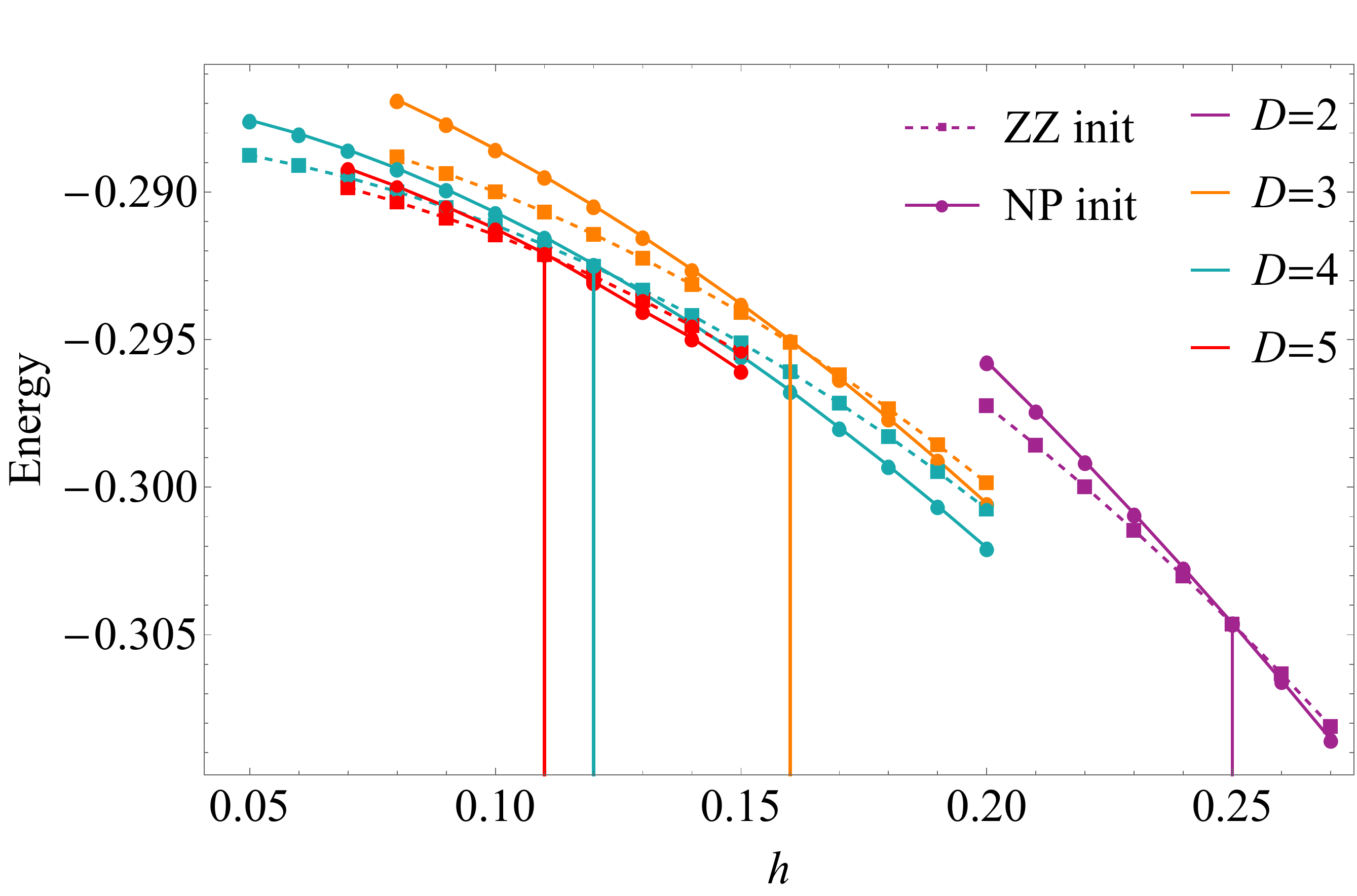}
    \caption{Initialized from ZZ or NP phase, we determine the transition point
    by the energy cross point. The energy cross between results with different initial states, ZZ and NP, which influence the energy values, and thus the cross points. The crossing points are about to converge for $D=5$ and found to be $h=0.11 |K|$.
    }
    \label{fig: K-J-Gamma-Gamma'_energy_cross}
\end{figure}

To determine the phase boundary accurately, we use a widely adopted method 
for first-order phase transition: we initialize the state as either ZZ or NP and look for the crossings of the energy curves. The crossing point converges at $D=5$ (energy difference $<10^{-4}$~\cref{fig: K-J-Gamma-Gamma'_energy_cross}), which gives a transition field strength of 0.11$|K|$. This corresponds to 21~T for RuCl$_3$, which is lower than the DMRG calculations (about 35~T). This suggests that accurate determination of the first-order transition field may be tricky and affected by finite-size effects in DMRG calculations on cylinders of limited widths.

Overall, the $K$-$J$-$\Gamma$-$\Gamma^\prime$ model captures the essential features of the magnetic order and phase transitions observed in $\alpha$-$\mathrm{RuCl_3}$. The model provides valuable insights into the behavior of this compound and its response to external magnetic fields, although the exact matching of critical field values between simulations and experiments may still pose challenges due to various factors, including the complexities of experimental fitting and the influence of finite-size/$\chi$ effects in numerical calculations. Further advances in determining the Hamiltonian parameters and experimental phase boundaries are needed to fully confirm these findings. 

\section{Discussions}
\label{sec: Discussion}
We have shown that differentiable programming tensor networks can produce 
highly accurate %excellent 
results in Kitaev-type frustrated magnet spin systems. The method provides more
reliable results compared to previous iPEPS calculations with simple or full update 
methods. For example, one can obtain a more accurate ground state even with 
smaller bond dimensions. Moreover, the findings are also less sensitive to the 
initial state used in the optimization (except for the case near first-order quantum phase transition). Differentiable programming tensor networks are 
easy to implement but perform excellently. The framework is generic and can be applied to any 2D quantum system with short-range 
interactions. 

We have also performed some calculations with fermionic tensor networks with U(1) symmetry implemented (in either spin or charge channel) and achieved consistent results from the full update calculations. However, the U(1) block structure complicates the use of our AD method in fermionic iPEPS. One limitation arises when dividing the iPEPS tensor, as different block divisions can lead to different ground state energies, necessitating manual selection of block divisions. Another limitation arises as in the VUMPS one is unable to choose block divisions for the environment. This is because VUMPS does not use SVD, which is a crucial step for selecting block divisions in other methods such as CTMRG or TEBD. We plan to address these issues in future work.

\section{Acknowledgments}
We thank Hao Xie, Qi Yang, Han Li, Shou-Shu Gong, Shun-Yao Yu, Hao-Kai Zhang, Qiang Luo, Ke Liu, and Philippe Corboz for the discussion. We thank Juan Osorio Iregui for providing the reference data shown in~\cref{fig: Kitaev_energy_mag}. This project is supported by the National Key Projects for Research and Development of China Grant. No. 2021YFA1400400 and No. 2022YFA1403900, Strategic Priority Research Program of the Chinese Academy of Sciences under Grant No. XDB30000000, National Natural Science Foundation of China (Grant Nos. T2225018, T2121001, 12222412, 11834014, 11974036), CAS Project for Young Scientists in Basic Research (Grant No. YSBR-057), China Postdoctoral Science Foundation under Grant No. 2021TQ0355, the Special Research Assistant Program of Chinese Academy of Sciences and the Youth Innovation Promotion Association CAS under Grant No. 2021004.

\appendix

\section{The convergence versus the boundary bond dimension}
\label{sec: The convergence versus the boundary bond dimension}

The figure below illustrates the energy and correlation length ($\xi$) as a function of the bond dimension $\chi$ of the boundary environment in the pure Kitaev model simulations. It clearly demonstrates the convergence behavior occurring at approximately $\chi=120$. Although the energy fluctuates by $10^{-5}$, we observe that the correlation length doubles as we increase the bond dimension from $\chi=60$ to $\chi=120$.

The maximum are $D/\chi=4/80,5/120,6/150$ in our optimization simulations and larger $\chi$ to insure the convergence of the observables.
\begin{figure}[H]
    \centering 
    \includegraphics[width=1\linewidth]{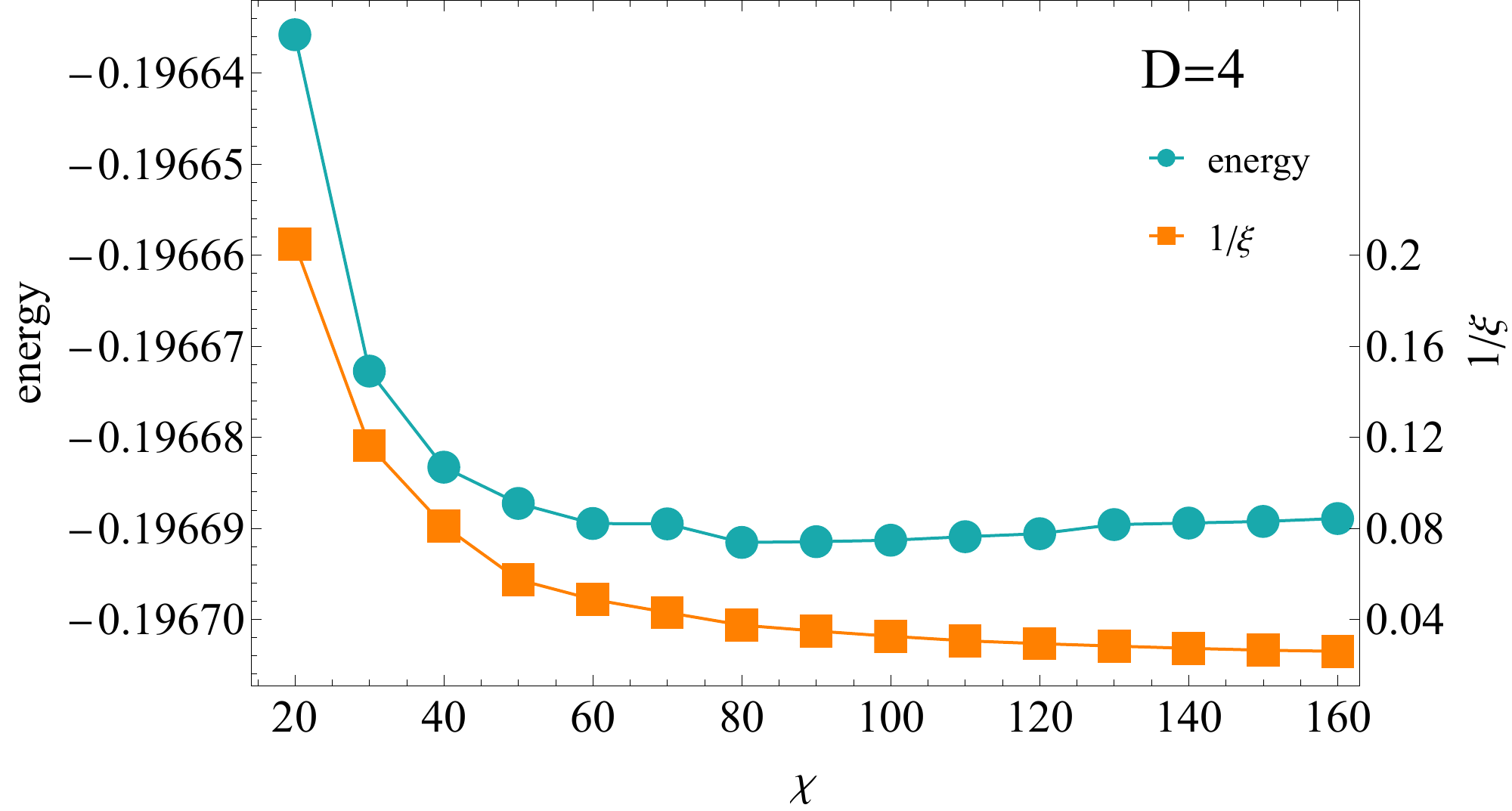}
    \label{fig: energy and corrLenD_4}
\end{figure}
\begin{figure}[H]
  \centering 
  \includegraphics[width=1\linewidth]{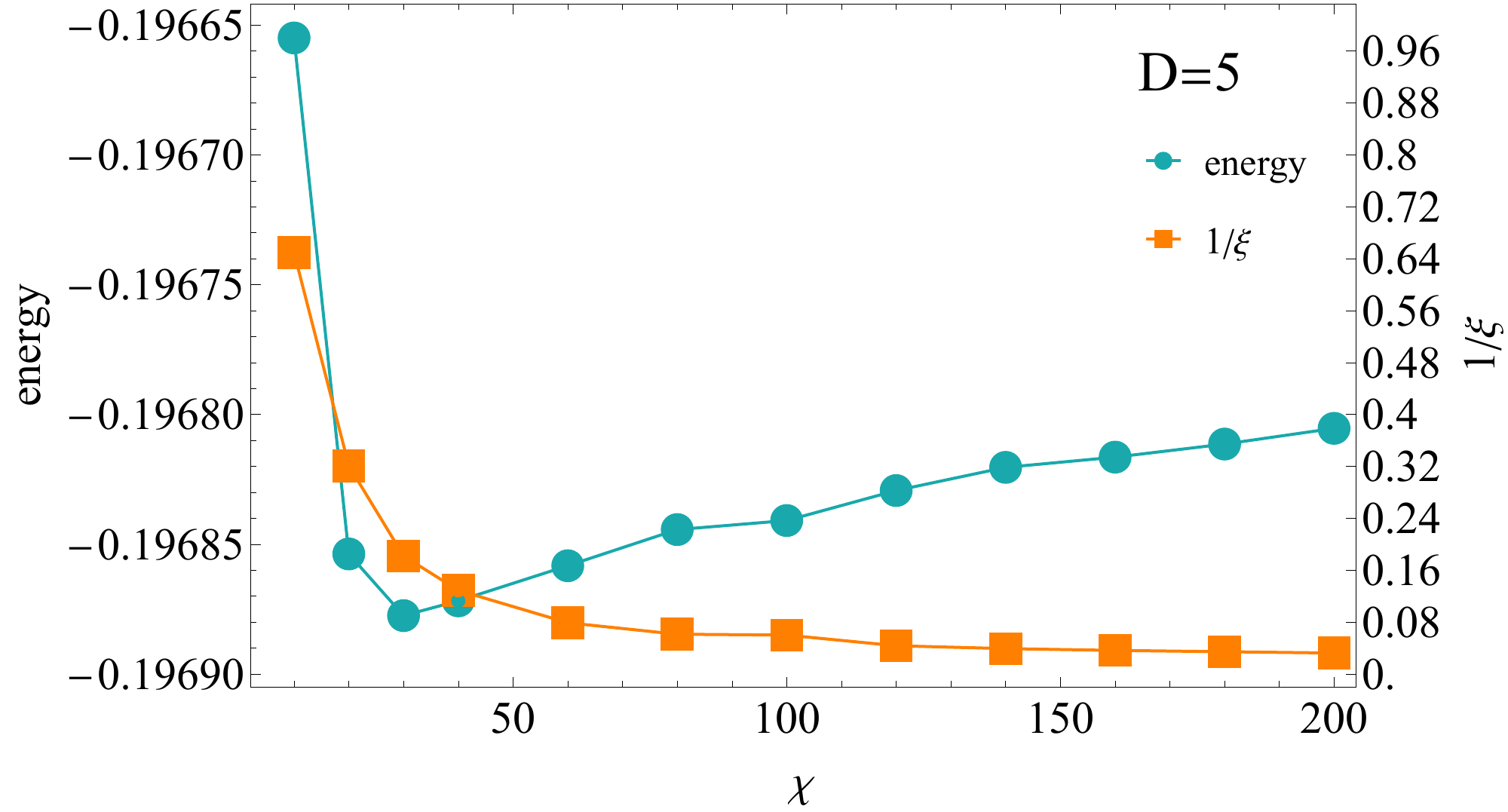}
  \label{fig: energy and corrLenD_5}
\end{figure}

\bibliography{latex}

\begin{thebibliography}{75}%
\makeatletter
\providecommand \@ifxundefined [1]{%
 \@ifx{#1\undefined}
}%
\providecommand \@ifnum [1]{%
 \ifnum #1\expandafter \@firstoftwo
 \else \expandafter \@secondoftwo
 \fi
}%
\providecommand \@ifx [1]{%
 \ifx #1\expandafter \@firstoftwo
 \else \expandafter \@secondoftwo
 \fi
}%
\providecommand \natexlab [1]{#1}%
\providecommand \enquote  [1]{``#1''}%
\providecommand \bibnamefont  [1]{#1}%
\providecommand \bibfnamefont [1]{#1}%
\providecommand \citenamefont [1]{#1}%
\providecommand \href@noop [0]{\@secondoftwo}%
\providecommand \href [0]{\begingroup \@sanitize@url \@href}%
\providecommand \@href[1]{\@@startlink{#1}\@@href}%
\providecommand \@@href[1]{\endgroup#1\@@endlink}%
\providecommand \@sanitize@url [0]{\catcode `\\12\catcode `\$12\catcode
  `\&12\catcode `\#12\catcode `\^12\catcode `\_12\catcode `\%12\relax}%
\providecommand \@@startlink[1]{}%
\providecommand \@@endlink[0]{}%
\providecommand \url  [0]{\begingroup\@sanitize@url \@url }%
\providecommand \@url [1]{\endgroup\@href {#1}{\urlprefix }}%
\providecommand \urlprefix  [0]{URL }%
\providecommand \Eprint [0]{\href }%
\providecommand \doibase [0]{https://doi.org/}%
\providecommand \selectlanguage [0]{\@gobble}%
\providecommand \bibinfo  [0]{\@secondoftwo}%
\providecommand \bibfield  [0]{\@secondoftwo}%
\providecommand \translation [1]{[#1]}%
\providecommand \BibitemOpen [0]{}%
\providecommand \bibitemStop [0]{}%
\providecommand \bibitemNoStop [0]{.\EOS\space}%
\providecommand \EOS [0]{\spacefactor3000\relax}%
\providecommand \BibitemShut  [1]{\csname bibitem#1\endcsname}%
\let\auto@bib@innerbib\@empty
%</preamble>
\bibitem [{\citenamefont {Kitaev}(2006)}]{kitaev2006anyons}%
  \BibitemOpen
  \bibfield  {author} {\bibinfo {author} {\bibfnamefont {A.}~\bibnamefont
  {Kitaev}},\ }\href
  {https://doi.org/https://doi.org/10.1016/j.aop.2005.10.005} {\bibfield
  {journal} {\bibinfo  {journal} {Annals of Physics}\ }\textbf {\bibinfo
  {volume} {321}},\ \bibinfo {pages} {2} (\bibinfo {year} {2006})},\ \bibinfo
  {note} {january Special Issue}\BibitemShut {NoStop}%
\bibitem [{\citenamefont {Jackeli}\ and\ \citenamefont
  {Khaliullin}(2009)}]{jackeli2009mott}%
  \BibitemOpen
  \bibfield  {author} {\bibinfo {author} {\bibfnamefont {G.}~\bibnamefont
  {Jackeli}}\ and\ \bibinfo {author} {\bibfnamefont {G.}~\bibnamefont
  {Khaliullin}},\ }\href {https://doi.org/10.1103/PhysRevLett.102.017205}
  {\bibfield  {journal} {\bibinfo  {journal} {Phys. Rev. Lett.}\ }\textbf
  {\bibinfo {volume} {102}},\ \bibinfo {pages} {017205} (\bibinfo {year}
  {2009})}\BibitemShut {NoStop}%
\bibitem [{\citenamefont {Winter}\ \emph {et~al.}(2016)\citenamefont {Winter},
  \citenamefont {Li}, \citenamefont {Jeschke},\ and\ \citenamefont
  {Valent\'{\i}}}]{PhysRevB.93.214431}%
  \BibitemOpen
  \bibfield  {author} {\bibinfo {author} {\bibfnamefont {S.~M.}\ \bibnamefont
  {Winter}}, \bibinfo {author} {\bibfnamefont {Y.}~\bibnamefont {Li}}, \bibinfo
  {author} {\bibfnamefont {H.~O.}\ \bibnamefont {Jeschke}},\ and\ \bibinfo
  {author} {\bibfnamefont {R.}~\bibnamefont {Valent\'{\i}}},\ }\href
  {https://doi.org/10.1103/PhysRevB.93.214431} {\bibfield  {journal} {\bibinfo
  {journal} {Phys. Rev. B}\ }\textbf {\bibinfo {volume} {93}},\ \bibinfo
  {pages} {214431} (\bibinfo {year} {2016})}\BibitemShut {NoStop}%
\bibitem [{\citenamefont {Winter}\ \emph {et~al.}(2017)\citenamefont {Winter},
  \citenamefont {Tsirlin}, \citenamefont {Daghofer}, \citenamefont {van~den
  Brink}, \citenamefont {Singh}, \citenamefont {Gegenwart},\ and\ \citenamefont
  {Valentí}}]{Winter_2017}%
  \BibitemOpen
  \bibfield  {author} {\bibinfo {author} {\bibfnamefont {S.~M.}\ \bibnamefont
  {Winter}}, \bibinfo {author} {\bibfnamefont {A.~A.}\ \bibnamefont {Tsirlin}},
  \bibinfo {author} {\bibfnamefont {M.}~\bibnamefont {Daghofer}}, \bibinfo
  {author} {\bibfnamefont {J.}~\bibnamefont {van~den Brink}}, \bibinfo {author}
  {\bibfnamefont {Y.}~\bibnamefont {Singh}}, \bibinfo {author} {\bibfnamefont
  {P.}~\bibnamefont {Gegenwart}},\ and\ \bibinfo {author} {\bibfnamefont
  {R.}~\bibnamefont {Valentí}},\ }\href
  {https://doi.org/10.1088/1361-648X/aa8cf5} {\bibfield  {journal} {\bibinfo
  {journal} {Journal of Physics: Condensed Matter}\ }\textbf {\bibinfo {volume}
  {29}},\ \bibinfo {pages} {493002} (\bibinfo {year} {2017})}\BibitemShut
  {NoStop}%
\bibitem [{\citenamefont {Hermanns}\ \emph {et~al.}(2018)\citenamefont
  {Hermanns}, \citenamefont {Kimchi},\ and\ \citenamefont
  {Knolle}}]{doi:10.1146/annurev-conmatphys-033117-053934}%
  \BibitemOpen
  \bibfield  {author} {\bibinfo {author} {\bibfnamefont {M.}~\bibnamefont
  {Hermanns}}, \bibinfo {author} {\bibfnamefont {I.}~\bibnamefont {Kimchi}},\
  and\ \bibinfo {author} {\bibfnamefont {J.}~\bibnamefont {Knolle}},\ }\href
  {https://doi.org/10.1146/annurev-conmatphys-033117-053934} {\bibfield
  {journal} {\bibinfo  {journal} {Annual Review of Condensed Matter Physics}\
  }\textbf {\bibinfo {volume} {9}},\ \bibinfo {pages} {17} (\bibinfo {year}
  {2018})},\ \Eprint
  {https://arxiv.org/abs/https://doi.org/10.1146/annurev-conmatphys-033117-053934}
  {https://doi.org/10.1146/annurev-conmatphys-033117-053934} \BibitemShut
  {NoStop}%
\bibitem [{\citenamefont {Takagi}\ \emph {et~al.}(2019)\citenamefont {Takagi},
  \citenamefont {Takayama}, \citenamefont {Jackeli}, \citenamefont
  {Khaliullin},\ and\ \citenamefont {Nagler}}]{takagi2019concept}%
  \BibitemOpen
  \bibfield  {author} {\bibinfo {author} {\bibfnamefont {H.}~\bibnamefont
  {Takagi}}, \bibinfo {author} {\bibfnamefont {T.}~\bibnamefont {Takayama}},
  \bibinfo {author} {\bibfnamefont {G.}~\bibnamefont {Jackeli}}, \bibinfo
  {author} {\bibfnamefont {G.}~\bibnamefont {Khaliullin}},\ and\ \bibinfo
  {author} {\bibfnamefont {S.~E.}\ \bibnamefont {Nagler}},\ }\href
  {https://doi.org/10.1038/s42254-019-0038-2} {\bibfield  {journal} {\bibinfo
  {journal} {Nature Reviews Physics}\ }\textbf {\bibinfo {volume} {1}},\
  \bibinfo {pages} {264} (\bibinfo {year} {2019})}\BibitemShut {NoStop}%
\bibitem [{\citenamefont {Glamazda}\ \emph {et~al.}(2017)\citenamefont
  {Glamazda}, \citenamefont {Lemmens}, \citenamefont {Do}, \citenamefont
  {Kwon},\ and\ \citenamefont {Choi}}]{PhysRevB.95.174429}%
  \BibitemOpen
  \bibfield  {author} {\bibinfo {author} {\bibfnamefont {A.}~\bibnamefont
  {Glamazda}}, \bibinfo {author} {\bibfnamefont {P.}~\bibnamefont {Lemmens}},
  \bibinfo {author} {\bibfnamefont {S.-H.}\ \bibnamefont {Do}}, \bibinfo
  {author} {\bibfnamefont {Y.~S.}\ \bibnamefont {Kwon}},\ and\ \bibinfo
  {author} {\bibfnamefont {K.-Y.}\ \bibnamefont {Choi}},\ }\href
  {https://doi.org/10.1103/PhysRevB.95.174429} {\bibfield  {journal} {\bibinfo
  {journal} {Phys. Rev. B}\ }\textbf {\bibinfo {volume} {95}},\ \bibinfo
  {pages} {174429} (\bibinfo {year} {2017})}\BibitemShut {NoStop}%
\bibitem [{\citenamefont {Wang}\ \emph {et~al.}(2017)\citenamefont {Wang},
  \citenamefont {Dong}, \citenamefont {Yu},\ and\ \citenamefont
  {Li}}]{Wei2017theoretical}%
  \BibitemOpen
  \bibfield  {author} {\bibinfo {author} {\bibfnamefont {W.}~\bibnamefont
  {Wang}}, \bibinfo {author} {\bibfnamefont {Z.-Y.}\ \bibnamefont {Dong}},
  \bibinfo {author} {\bibfnamefont {S.-L.}\ \bibnamefont {Yu}},\ and\ \bibinfo
  {author} {\bibfnamefont {J.-X.}\ \bibnamefont {Li}},\ }\href
  {https://doi.org/10.1103/PhysRevB.96.115103} {\bibfield  {journal} {\bibinfo
  {journal} {Phys. Rev. B}\ }\textbf {\bibinfo {volume} {96}},\ \bibinfo
  {pages} {115103} (\bibinfo {year} {2017})}\BibitemShut {NoStop}%
\bibitem [{\citenamefont {Ran}\ \emph {et~al.}(2017)\citenamefont {Ran},
  \citenamefont {Wang}, \citenamefont {Wang}, \citenamefont {Dong},
  \citenamefont {Ren}, \citenamefont {Bao}, \citenamefont {Li}, \citenamefont
  {Ma}, \citenamefont {Gan}, \citenamefont {Zhang}, \citenamefont {Park},
  \citenamefont {Deng}, \citenamefont {Danilkin}, \citenamefont {Yu},
  \citenamefont {Li},\ and\ \citenamefont {Wen}}]{PhysRevLett.118.107203}%
  \BibitemOpen
  \bibfield  {author} {\bibinfo {author} {\bibfnamefont {K.}~\bibnamefont
  {Ran}}, \bibinfo {author} {\bibfnamefont {J.}~\bibnamefont {Wang}}, \bibinfo
  {author} {\bibfnamefont {W.}~\bibnamefont {Wang}}, \bibinfo {author}
  {\bibfnamefont {Z.-Y.}\ \bibnamefont {Dong}}, \bibinfo {author}
  {\bibfnamefont {X.}~\bibnamefont {Ren}}, \bibinfo {author} {\bibfnamefont
  {S.}~\bibnamefont {Bao}}, \bibinfo {author} {\bibfnamefont {S.}~\bibnamefont
  {Li}}, \bibinfo {author} {\bibfnamefont {Z.}~\bibnamefont {Ma}}, \bibinfo
  {author} {\bibfnamefont {Y.}~\bibnamefont {Gan}}, \bibinfo {author}
  {\bibfnamefont {Y.}~\bibnamefont {Zhang}}, \bibinfo {author} {\bibfnamefont
  {J.~T.}\ \bibnamefont {Park}}, \bibinfo {author} {\bibfnamefont
  {G.}~\bibnamefont {Deng}}, \bibinfo {author} {\bibfnamefont {S.}~\bibnamefont
  {Danilkin}}, \bibinfo {author} {\bibfnamefont {S.-L.}\ \bibnamefont {Yu}},
  \bibinfo {author} {\bibfnamefont {J.-X.}\ \bibnamefont {Li}},\ and\ \bibinfo
  {author} {\bibfnamefont {J.}~\bibnamefont {Wen}},\ }\href
  {https://doi.org/10.1103/PhysRevLett.118.107203} {\bibfield  {journal}
  {\bibinfo  {journal} {Phys. Rev. Lett.}\ }\textbf {\bibinfo {volume} {118}},\
  \bibinfo {pages} {107203} (\bibinfo {year} {2017})}\BibitemShut {NoStop}%
\bibitem [{\citenamefont {Suzuki}\ and\ \citenamefont
  {Suga}(2019)}]{PhysRevB.99.249902}%
  \BibitemOpen
  \bibfield  {author} {\bibinfo {author} {\bibfnamefont {T.}~\bibnamefont
  {Suzuki}}\ and\ \bibinfo {author} {\bibfnamefont {S.-i.}\ \bibnamefont
  {Suga}},\ }\href {https://doi.org/10.1103/PhysRevB.99.249902} {\bibfield
  {journal} {\bibinfo  {journal} {Phys. Rev. B}\ }\textbf {\bibinfo {volume}
  {99}},\ \bibinfo {pages} {249902} (\bibinfo {year} {2019})}\BibitemShut
  {NoStop}%
\bibitem [{\citenamefont {Wu}\ \emph {et~al.}(2018)\citenamefont {Wu},
  \citenamefont {Little}, \citenamefont {Aldape}, \citenamefont {Rees},
  \citenamefont {Thewalt}, \citenamefont {Lampen-Kelley}, \citenamefont
  {Banerjee}, \citenamefont {Bridges}, \citenamefont {Yan}, \citenamefont
  {Boone}, \citenamefont {Patankar}, \citenamefont {Goldhaber-Gordon},
  \citenamefont {Mandrus}, \citenamefont {Nagler}, \citenamefont {Altman},\
  and\ \citenamefont {Orenstein}}]{PhysRevB.98.094425}%
  \BibitemOpen
  \bibfield  {author} {\bibinfo {author} {\bibfnamefont {L.}~\bibnamefont
  {Wu}}, \bibinfo {author} {\bibfnamefont {A.}~\bibnamefont {Little}}, \bibinfo
  {author} {\bibfnamefont {E.~E.}\ \bibnamefont {Aldape}}, \bibinfo {author}
  {\bibfnamefont {D.}~\bibnamefont {Rees}}, \bibinfo {author} {\bibfnamefont
  {E.}~\bibnamefont {Thewalt}}, \bibinfo {author} {\bibfnamefont
  {P.}~\bibnamefont {Lampen-Kelley}}, \bibinfo {author} {\bibfnamefont
  {A.}~\bibnamefont {Banerjee}}, \bibinfo {author} {\bibfnamefont {C.~A.}\
  \bibnamefont {Bridges}}, \bibinfo {author} {\bibfnamefont {J.-Q.}\
  \bibnamefont {Yan}}, \bibinfo {author} {\bibfnamefont {D.}~\bibnamefont
  {Boone}}, \bibinfo {author} {\bibfnamefont {S.}~\bibnamefont {Patankar}},
  \bibinfo {author} {\bibfnamefont {D.}~\bibnamefont {Goldhaber-Gordon}},
  \bibinfo {author} {\bibfnamefont {D.}~\bibnamefont {Mandrus}}, \bibinfo
  {author} {\bibfnamefont {S.~E.}\ \bibnamefont {Nagler}}, \bibinfo {author}
  {\bibfnamefont {E.}~\bibnamefont {Altman}},\ and\ \bibinfo {author}
  {\bibfnamefont {J.}~\bibnamefont {Orenstein}},\ }\href
  {https://doi.org/10.1103/PhysRevB.98.094425} {\bibfield  {journal} {\bibinfo
  {journal} {Phys. Rev. B}\ }\textbf {\bibinfo {volume} {98}},\ \bibinfo
  {pages} {094425} (\bibinfo {year} {2018})}\BibitemShut {NoStop}%
\bibitem [{\citenamefont {Cookmeyer}\ and\ \citenamefont
  {Moore}(2018)}]{PhysRevB.98.060412}%
  \BibitemOpen
  \bibfield  {author} {\bibinfo {author} {\bibfnamefont {T.}~\bibnamefont
  {Cookmeyer}}\ and\ \bibinfo {author} {\bibfnamefont {J.~E.}\ \bibnamefont
  {Moore}},\ }\href {https://doi.org/10.1103/PhysRevB.98.060412} {\bibfield
  {journal} {\bibinfo  {journal} {Phys. Rev. B}\ }\textbf {\bibinfo {volume}
  {98}},\ \bibinfo {pages} {060412} (\bibinfo {year} {2018})}\BibitemShut
  {NoStop}%
\bibitem [{\citenamefont {Sandilands}\ \emph {et~al.}(2016)\citenamefont
  {Sandilands}, \citenamefont {Tian}, \citenamefont {Reijnders}, \citenamefont
  {Kim}, \citenamefont {Plumb}, \citenamefont {Kim}, \citenamefont {Kee},\ and\
  \citenamefont {Burch}}]{PhysRevB.93.075144}%
  \BibitemOpen
  \bibfield  {author} {\bibinfo {author} {\bibfnamefont {L.~J.}\ \bibnamefont
  {Sandilands}}, \bibinfo {author} {\bibfnamefont {Y.}~\bibnamefont {Tian}},
  \bibinfo {author} {\bibfnamefont {A.~A.}\ \bibnamefont {Reijnders}}, \bibinfo
  {author} {\bibfnamefont {H.-S.}\ \bibnamefont {Kim}}, \bibinfo {author}
  {\bibfnamefont {K.~W.}\ \bibnamefont {Plumb}}, \bibinfo {author}
  {\bibfnamefont {Y.-J.}\ \bibnamefont {Kim}}, \bibinfo {author} {\bibfnamefont
  {H.-Y.}\ \bibnamefont {Kee}},\ and\ \bibinfo {author} {\bibfnamefont {K.~S.}\
  \bibnamefont {Burch}},\ }\href {https://doi.org/10.1103/PhysRevB.93.075144}
  {\bibfield  {journal} {\bibinfo  {journal} {Phys. Rev. B}\ }\textbf {\bibinfo
  {volume} {93}},\ \bibinfo {pages} {075144} (\bibinfo {year}
  {2016})}\BibitemShut {NoStop}%
\bibitem [{\citenamefont {Li}\ \emph {et~al.}(2021)\citenamefont {Li},
  \citenamefont {Zhang}, \citenamefont {Wang}, \citenamefont {Wu},
  \citenamefont {Gao}, \citenamefont {Qu}, \citenamefont {Liu}, \citenamefont
  {Gong},\ and\ \citenamefont {Li}}]{li2021identification}%
  \BibitemOpen
  \bibfield  {author} {\bibinfo {author} {\bibfnamefont {H.}~\bibnamefont
  {Li}}, \bibinfo {author} {\bibfnamefont {H.-K.}\ \bibnamefont {Zhang}},
  \bibinfo {author} {\bibfnamefont {J.}~\bibnamefont {Wang}}, \bibinfo {author}
  {\bibfnamefont {H.-Q.}\ \bibnamefont {Wu}}, \bibinfo {author} {\bibfnamefont
  {Y.}~\bibnamefont {Gao}}, \bibinfo {author} {\bibfnamefont {D.-W.}\
  \bibnamefont {Qu}}, \bibinfo {author} {\bibfnamefont {Z.-X.}\ \bibnamefont
  {Liu}}, \bibinfo {author} {\bibfnamefont {S.-S.}\ \bibnamefont {Gong}},\ and\
  \bibinfo {author} {\bibfnamefont {W.}~\bibnamefont {Li}},\ }\href
  {https://www.nature.com/articles/s41467-021-24257-8} {\bibfield  {journal}
  {\bibinfo  {journal} {Nature Communications}\ }\textbf {\bibinfo {volume}
  {12}},\ \bibinfo {pages} {1} (\bibinfo {year} {2021})}\BibitemShut {NoStop}%
\bibitem [{\citenamefont {Jiang}\ \emph {et~al.}(2011)\citenamefont {Jiang},
  \citenamefont {Gu}, \citenamefont {Qi},\ and\ \citenamefont
  {Trebst}}]{PhysRevB.83.245104}%
  \BibitemOpen
  \bibfield  {author} {\bibinfo {author} {\bibfnamefont {H.-C.}\ \bibnamefont
  {Jiang}}, \bibinfo {author} {\bibfnamefont {Z.-C.}\ \bibnamefont {Gu}},
  \bibinfo {author} {\bibfnamefont {X.-L.}\ \bibnamefont {Qi}},\ and\ \bibinfo
  {author} {\bibfnamefont {S.}~\bibnamefont {Trebst}},\ }\href
  {https://doi.org/10.1103/PhysRevB.83.245104} {\bibfield  {journal} {\bibinfo
  {journal} {Phys. Rev. B}\ }\textbf {\bibinfo {volume} {83}},\ \bibinfo
  {pages} {245104} (\bibinfo {year} {2011})}\BibitemShut {NoStop}%
\bibitem [{\citenamefont {Vodola}\ \emph {et~al.}(2015)\citenamefont {Vodola},
  \citenamefont {Lepori}, \citenamefont {Ercolessi},\ and\ \citenamefont
  {Pupillo}}]{vodola2015long}%
  \BibitemOpen
  \bibfield  {author} {\bibinfo {author} {\bibfnamefont {D.}~\bibnamefont
  {Vodola}}, \bibinfo {author} {\bibfnamefont {L.}~\bibnamefont {Lepori}},
  \bibinfo {author} {\bibfnamefont {E.}~\bibnamefont {Ercolessi}},\ and\
  \bibinfo {author} {\bibfnamefont {G.}~\bibnamefont {Pupillo}},\ }\href
  {https://doi.org/10.1088/1367-2630/18/1/015001} {\bibfield  {journal}
  {\bibinfo  {journal} {New Journal of Physics}\ }\textbf {\bibinfo {volume}
  {18}},\ \bibinfo {pages} {015001} (\bibinfo {year} {2015})}\BibitemShut
  {NoStop}%
\bibitem [{\citenamefont {Gergs}\ \emph {et~al.}(2016)\citenamefont {Gergs},
  \citenamefont {Fritz},\ and\ \citenamefont {Schuricht}}]{PhysRevB.93.075129}%
  \BibitemOpen
  \bibfield  {author} {\bibinfo {author} {\bibfnamefont {N.~M.}\ \bibnamefont
  {Gergs}}, \bibinfo {author} {\bibfnamefont {L.}~\bibnamefont {Fritz}},\ and\
  \bibinfo {author} {\bibfnamefont {D.}~\bibnamefont {Schuricht}},\ }\href
  {https://doi.org/10.1103/PhysRevB.93.075129} {\bibfield  {journal} {\bibinfo
  {journal} {Phys. Rev. B}\ }\textbf {\bibinfo {volume} {93}},\ \bibinfo
  {pages} {075129} (\bibinfo {year} {2016})}\BibitemShut {NoStop}%
\bibitem [{\citenamefont {Gohlke}\ \emph {et~al.}(2017)\citenamefont {Gohlke},
  \citenamefont {Verresen}, \citenamefont {Moessner},\ and\ \citenamefont
  {Pollmann}}]{gohlke2017dynamics}%
  \BibitemOpen
  \bibfield  {author} {\bibinfo {author} {\bibfnamefont {M.}~\bibnamefont
  {Gohlke}}, \bibinfo {author} {\bibfnamefont {R.}~\bibnamefont {Verresen}},
  \bibinfo {author} {\bibfnamefont {R.}~\bibnamefont {Moessner}},\ and\
  \bibinfo {author} {\bibfnamefont {F.}~\bibnamefont {Pollmann}},\ }\href
  {https://link.aps.org/doi/10.1103/PhysRevLett.119.157203} {\bibfield
  {journal} {\bibinfo  {journal} {Phys. Rev. Lett.}\ }\textbf {\bibinfo
  {volume} {119}},\ \bibinfo {pages} {157203} (\bibinfo {year}
  {2017})}\BibitemShut {NoStop}%
\bibitem [{\citenamefont {Gohlke}\ \emph
  {et~al.}(2018{\natexlab{a}})\citenamefont {Gohlke}, \citenamefont
  {Moessner},\ and\ \citenamefont {Pollmann}}]{PhysRevB.98.014418}%
  \BibitemOpen
  \bibfield  {author} {\bibinfo {author} {\bibfnamefont {M.}~\bibnamefont
  {Gohlke}}, \bibinfo {author} {\bibfnamefont {R.}~\bibnamefont {Moessner}},\
  and\ \bibinfo {author} {\bibfnamefont {F.}~\bibnamefont {Pollmann}},\ }\href
  {https://doi.org/10.1103/PhysRevB.98.014418} {\bibfield  {journal} {\bibinfo
  {journal} {Phys. Rev. B}\ }\textbf {\bibinfo {volume} {98}},\ \bibinfo
  {pages} {014418} (\bibinfo {year} {2018}{\natexlab{a}})}\BibitemShut
  {NoStop}%
\bibitem [{\citenamefont {Zhu}\ \emph {et~al.}(2018)\citenamefont {Zhu},
  \citenamefont {Kimchi}, \citenamefont {Sheng},\ and\ \citenamefont
  {Fu}}]{PhysRevB.97.241110}%
  \BibitemOpen
  \bibfield  {author} {\bibinfo {author} {\bibfnamefont {Z.}~\bibnamefont
  {Zhu}}, \bibinfo {author} {\bibfnamefont {I.}~\bibnamefont {Kimchi}},
  \bibinfo {author} {\bibfnamefont {D.~N.}\ \bibnamefont {Sheng}},\ and\
  \bibinfo {author} {\bibfnamefont {L.}~\bibnamefont {Fu}},\ }\href
  {https://doi.org/10.1103/PhysRevB.97.241110} {\bibfield  {journal} {\bibinfo
  {journal} {Phys. Rev. B}\ }\textbf {\bibinfo {volume} {97}},\ \bibinfo
  {pages} {241110} (\bibinfo {year} {2018})}\BibitemShut {NoStop}%
\bibitem [{\citenamefont {Gohlke}\ \emph
  {et~al.}(2018{\natexlab{b}})\citenamefont {Gohlke}, \citenamefont {Wachtel},
  \citenamefont {Yamaji}, \citenamefont {Pollmann},\ and\ \citenamefont
  {Kim}}]{gohlke2018quantum}%
  \BibitemOpen
  \bibfield  {author} {\bibinfo {author} {\bibfnamefont {M.}~\bibnamefont
  {Gohlke}}, \bibinfo {author} {\bibfnamefont {G.}~\bibnamefont {Wachtel}},
  \bibinfo {author} {\bibfnamefont {Y.}~\bibnamefont {Yamaji}}, \bibinfo
  {author} {\bibfnamefont {F.}~\bibnamefont {Pollmann}},\ and\ \bibinfo
  {author} {\bibfnamefont {Y.~B.}\ \bibnamefont {Kim}},\ }\href
  {https://link.aps.org/doi/10.1103/PhysRevB.97.075126} {\bibfield  {journal}
  {\bibinfo  {journal} {Phys. Rev. B}\ }\textbf {\bibinfo {volume} {97}},\
  \bibinfo {pages} {075126} (\bibinfo {year} {2018}{\natexlab{b}})}\BibitemShut
  {NoStop}%
\bibitem [{\citenamefont {Lee}\ \emph {et~al.}(2020)\citenamefont {Lee},
  \citenamefont {Kaneko}, \citenamefont {Chern}, \citenamefont {Okubo},
  \citenamefont {Yamaji}, \citenamefont {Kawashima},\ and\ \citenamefont
  {Kim}}]{lee2020magnetic}%
  \BibitemOpen
  \bibfield  {author} {\bibinfo {author} {\bibfnamefont {H.-Y.}\ \bibnamefont
  {Lee}}, \bibinfo {author} {\bibfnamefont {R.}~\bibnamefont {Kaneko}},
  \bibinfo {author} {\bibfnamefont {L.~E.}\ \bibnamefont {Chern}}, \bibinfo
  {author} {\bibfnamefont {T.}~\bibnamefont {Okubo}}, \bibinfo {author}
  {\bibfnamefont {Y.}~\bibnamefont {Yamaji}}, \bibinfo {author} {\bibfnamefont
  {N.}~\bibnamefont {Kawashima}},\ and\ \bibinfo {author} {\bibfnamefont
  {Y.~B.}\ \bibnamefont {Kim}},\ }\href
  {https://doi.org/10.1038/s41467-020-15320-x} {\bibfield  {journal} {\bibinfo
  {journal} {Nature Communications}\ }\textbf {\bibinfo {volume} {11}},\
  \bibinfo {pages} {1639} (\bibinfo {year} {2020})}\BibitemShut {NoStop}%
\bibitem [{\citenamefont {Verstraete}\ and\ \citenamefont
  {Cirac}(2004)}]{verstraete2004renormalization}%
  \BibitemOpen
  \bibfield  {author} {\bibinfo {author} {\bibfnamefont {F.}~\bibnamefont
  {Verstraete}}\ and\ \bibinfo {author} {\bibfnamefont {J.~I.}\ \bibnamefont
  {Cirac}},\ }\href@noop {} {\bibinfo {title} {Renormalization algorithms for
  quantum-many body systems in two and higher dimensions}} (\bibinfo {year}
  {2004}),\ \Eprint {https://arxiv.org/abs/cond-mat/0407066}
  {arXiv:cond-mat/0407066 [cond-mat.str-el]} \BibitemShut {NoStop}%
\bibitem [{\citenamefont {Osorio~Iregui}\ \emph {et~al.}(2014)\citenamefont
  {Osorio~Iregui}, \citenamefont {Corboz},\ and\ \citenamefont
  {Troyer}}]{iregui2014probing}%
  \BibitemOpen
  \bibfield  {author} {\bibinfo {author} {\bibfnamefont {J.}~\bibnamefont
  {Osorio~Iregui}}, \bibinfo {author} {\bibfnamefont {P.}~\bibnamefont
  {Corboz}},\ and\ \bibinfo {author} {\bibfnamefont {M.}~\bibnamefont
  {Troyer}},\ }\href {https://doi.org/10.1103/PhysRevB.90.195102} {\bibfield
  {journal} {\bibinfo  {journal} {Phys. Rev. B}\ }\textbf {\bibinfo {volume}
  {90}},\ \bibinfo {pages} {195102} (\bibinfo {year} {2014})}\BibitemShut
  {NoStop}%
\bibitem [{\citenamefont {Gotfryd}\ \emph {et~al.}(2017)\citenamefont
  {Gotfryd}, \citenamefont {Rusna\ifmmode~\check{c}\else \v{c}\fi{}ko},
  \citenamefont {Wohlfeld}, \citenamefont {Jackeli}, \citenamefont
  {Chaloupka},\ and\ \citenamefont {Ole\ifmmode~\acute{s}\else
  \'{s}\fi{}}}]{PhysRevB.95.024426}%
  \BibitemOpen
  \bibfield  {author} {\bibinfo {author} {\bibfnamefont {D.}~\bibnamefont
  {Gotfryd}}, \bibinfo {author} {\bibfnamefont {J.}~\bibnamefont
  {Rusna\ifmmode~\check{c}\else \v{c}\fi{}ko}}, \bibinfo {author}
  {\bibfnamefont {K.}~\bibnamefont {Wohlfeld}}, \bibinfo {author}
  {\bibfnamefont {G.}~\bibnamefont {Jackeli}}, \bibinfo {author} {\bibfnamefont
  {J.~c.~v.}\ \bibnamefont {Chaloupka}},\ and\ \bibinfo {author} {\bibfnamefont
  {A.~M.}\ \bibnamefont {Ole\ifmmode~\acute{s}\else \'{s}\fi{}}},\ }\href
  {https://doi.org/10.1103/PhysRevB.95.024426} {\bibfield  {journal} {\bibinfo
  {journal} {Phys. Rev. B}\ }\textbf {\bibinfo {volume} {95}},\ \bibinfo
  {pages} {024426} (\bibinfo {year} {2017})}\BibitemShut {NoStop}%
\bibitem [{\citenamefont {Chen}\ \emph {et~al.}(2018)\citenamefont {Chen},
  \citenamefont {Vanderstraeten}, \citenamefont {Capponi},\ and\ \citenamefont
  {Poilblanc}}]{PhysRevB.98.184409}%
  \BibitemOpen
  \bibfield  {author} {\bibinfo {author} {\bibfnamefont {J.-Y.}\ \bibnamefont
  {Chen}}, \bibinfo {author} {\bibfnamefont {L.}~\bibnamefont
  {Vanderstraeten}}, \bibinfo {author} {\bibfnamefont {S.}~\bibnamefont
  {Capponi}},\ and\ \bibinfo {author} {\bibfnamefont {D.}~\bibnamefont
  {Poilblanc}},\ }\href {https://doi.org/10.1103/PhysRevB.98.184409} {\bibfield
   {journal} {\bibinfo  {journal} {Phys. Rev. B}\ }\textbf {\bibinfo {volume}
  {98}},\ \bibinfo {pages} {184409} (\bibinfo {year} {2018})}\BibitemShut
  {NoStop}%
\bibitem [{\citenamefont {Czarnik}\ \emph {et~al.}(2019)\citenamefont
  {Czarnik}, \citenamefont {Francuz},\ and\ \citenamefont
  {Dziarmaga}}]{PhysRevB.100.165147}%
  \BibitemOpen
  \bibfield  {author} {\bibinfo {author} {\bibfnamefont {P.}~\bibnamefont
  {Czarnik}}, \bibinfo {author} {\bibfnamefont {A.}~\bibnamefont {Francuz}},\
  and\ \bibinfo {author} {\bibfnamefont {J.}~\bibnamefont {Dziarmaga}},\ }\href
  {https://doi.org/10.1103/PhysRevB.100.165147} {\bibfield  {journal} {\bibinfo
   {journal} {Phys. Rev. B}\ }\textbf {\bibinfo {volume} {100}},\ \bibinfo
  {pages} {165147} (\bibinfo {year} {2019})}\BibitemShut {NoStop}%
\bibitem [{\citenamefont {Nishino}\ and\ \citenamefont
  {Okunishi}(1997)}]{nishino1997corner}%
  \BibitemOpen
  \bibfield  {author} {\bibinfo {author} {\bibfnamefont {T.}~\bibnamefont
  {Nishino}}\ and\ \bibinfo {author} {\bibfnamefont {K.}~\bibnamefont
  {Okunishi}},\ }\href {https://doi.org/10.1143/JPSJ.66.3040} {\bibfield
  {journal} {\bibinfo  {journal} {Journal of the Physical Society of Japan}\
  }\textbf {\bibinfo {volume} {66}},\ \bibinfo {pages} {3040} (\bibinfo {year}
  {1997})},\ \Eprint
  {https://arxiv.org/abs/https://doi.org/10.1143/JPSJ.66.3040}
  {https://doi.org/10.1143/JPSJ.66.3040} \BibitemShut {NoStop}%
\bibitem [{\citenamefont {Or\'us}\ and\ \citenamefont
  {Vidal}(2009)}]{PhysRevB.80.094403}%
  \BibitemOpen
  \bibfield  {author} {\bibinfo {author} {\bibfnamefont {R.}~\bibnamefont
  {Or\'us}}\ and\ \bibinfo {author} {\bibfnamefont {G.}~\bibnamefont {Vidal}},\
  }\href {https://doi.org/10.1103/PhysRevB.80.094403} {\bibfield  {journal}
  {\bibinfo  {journal} {Phys. Rev. B}\ }\textbf {\bibinfo {volume} {80}},\
  \bibinfo {pages} {094403} (\bibinfo {year} {2009})}\BibitemShut {NoStop}%
\bibitem [{\citenamefont {Zauner-Stauber}\ \emph {et~al.}(2018)\citenamefont
  {Zauner-Stauber}, \citenamefont {Vanderstraeten}, \citenamefont {Fishman},
  \citenamefont {Verstraete},\ and\ \citenamefont
  {Haegeman}}]{zauner2018variational}%
  \BibitemOpen
  \bibfield  {author} {\bibinfo {author} {\bibfnamefont {V.}~\bibnamefont
  {Zauner-Stauber}}, \bibinfo {author} {\bibfnamefont {L.}~\bibnamefont
  {Vanderstraeten}}, \bibinfo {author} {\bibfnamefont {M.~T.}\ \bibnamefont
  {Fishman}}, \bibinfo {author} {\bibfnamefont {F.}~\bibnamefont
  {Verstraete}},\ and\ \bibinfo {author} {\bibfnamefont {J.}~\bibnamefont
  {Haegeman}},\ }\href {https://doi.org/10.1103/PhysRevB.97.045145} {\bibfield
  {journal} {\bibinfo  {journal} {Phys. Rev. B}\ }\textbf {\bibinfo {volume}
  {97}},\ \bibinfo {pages} {045145} (\bibinfo {year} {2018})}\BibitemShut
  {NoStop}%
\bibitem [{\citenamefont {Fishman}\ \emph {et~al.}(2018)\citenamefont
  {Fishman}, \citenamefont {Vanderstraeten}, \citenamefont {Zauner-Stauber},
  \citenamefont {Haegeman},\ and\ \citenamefont
  {Verstraete}}]{fishman2018faster}%
  \BibitemOpen
  \bibfield  {author} {\bibinfo {author} {\bibfnamefont {M.~T.}\ \bibnamefont
  {Fishman}}, \bibinfo {author} {\bibfnamefont {L.}~\bibnamefont
  {Vanderstraeten}}, \bibinfo {author} {\bibfnamefont {V.}~\bibnamefont
  {Zauner-Stauber}}, \bibinfo {author} {\bibfnamefont {J.}~\bibnamefont
  {Haegeman}},\ and\ \bibinfo {author} {\bibfnamefont {F.}~\bibnamefont
  {Verstraete}},\ }\href {https://doi.org/10.1103/PhysRevB.98.235148}
  {\bibfield  {journal} {\bibinfo  {journal} {Phys. Rev. B}\ }\textbf {\bibinfo
  {volume} {98}},\ \bibinfo {pages} {235148} (\bibinfo {year}
  {2018})}\BibitemShut {NoStop}%
\bibitem [{\citenamefont {Vanderstraeten}\ \emph {et~al.}(2019)\citenamefont
  {Vanderstraeten}, \citenamefont {Haegeman},\ and\ \citenamefont
  {Verstraete}}]{10.21468/SciPostPhysLectNotes.7}%
  \BibitemOpen
  \bibfield  {author} {\bibinfo {author} {\bibfnamefont {L.}~\bibnamefont
  {Vanderstraeten}}, \bibinfo {author} {\bibfnamefont {J.}~\bibnamefont
  {Haegeman}},\ and\ \bibinfo {author} {\bibfnamefont {F.}~\bibnamefont
  {Verstraete}},\ }\href {https://doi.org/10.21468/SciPostPhysLectNotes.7}
  {\bibfield  {journal} {\bibinfo  {journal} {SciPost Phys. Lect. Notes}\ ,\
  \bibinfo {pages} {7}} (\bibinfo {year} {2019})}\BibitemShut {NoStop}%
\bibitem [{\citenamefont {Corboz}\ \emph {et~al.}(2014)\citenamefont {Corboz},
  \citenamefont {Rice},\ and\ \citenamefont {Troyer}}]{corboz2014competing}%
  \BibitemOpen
  \bibfield  {author} {\bibinfo {author} {\bibfnamefont {P.}~\bibnamefont
  {Corboz}}, \bibinfo {author} {\bibfnamefont {T.~M.}\ \bibnamefont {Rice}},\
  and\ \bibinfo {author} {\bibfnamefont {M.}~\bibnamefont {Troyer}},\ }\href
  {https://doi.org/10.1103/PhysRevLett.113.046402} {\bibfield  {journal}
  {\bibinfo  {journal} {Phys. Rev. Lett.}\ }\textbf {\bibinfo {volume} {113}},\
  \bibinfo {pages} {046402} (\bibinfo {year} {2014})}\BibitemShut {NoStop}%
\bibitem [{\citenamefont {Nietner}\ \emph {et~al.}(2020)\citenamefont
  {Nietner}, \citenamefont {Vanhecke}, \citenamefont {Verstraete},
  \citenamefont {Eisert},\ and\ \citenamefont
  {Vanderstraeten}}]{nietner2020efficient}%
  \BibitemOpen
  \bibfield  {author} {\bibinfo {author} {\bibfnamefont {A.}~\bibnamefont
  {Nietner}}, \bibinfo {author} {\bibfnamefont {B.}~\bibnamefont {Vanhecke}},
  \bibinfo {author} {\bibfnamefont {F.}~\bibnamefont {Verstraete}}, \bibinfo
  {author} {\bibfnamefont {J.}~\bibnamefont {Eisert}},\ and\ \bibinfo {author}
  {\bibfnamefont {L.}~\bibnamefont {Vanderstraeten}},\ }\href
  {https://doi.org/10.22331/q-2020-09-21-328} {\bibfield  {journal} {\bibinfo
  {journal} {{Quantum}}\ }\textbf {\bibinfo {volume} {4}},\ \bibinfo {pages}
  {328} (\bibinfo {year} {2020})}\BibitemShut {NoStop}%
\bibitem [{\citenamefont {Vanderstraeten}\ \emph {et~al.}(2022)\citenamefont
  {Vanderstraeten}, \citenamefont {Burgelman}, \citenamefont {Ponsioen},
  \citenamefont {Van~Damme}, \citenamefont {Vanhecke}, \citenamefont {Corboz},
  \citenamefont {Haegeman},\ and\ \citenamefont
  {Verstraete}}]{vanderstraeten2022variational}%
  \BibitemOpen
  \bibfield  {author} {\bibinfo {author} {\bibfnamefont {L.}~\bibnamefont
  {Vanderstraeten}}, \bibinfo {author} {\bibfnamefont {L.}~\bibnamefont
  {Burgelman}}, \bibinfo {author} {\bibfnamefont {B.}~\bibnamefont {Ponsioen}},
  \bibinfo {author} {\bibfnamefont {M.}~\bibnamefont {Van~Damme}}, \bibinfo
  {author} {\bibfnamefont {B.}~\bibnamefont {Vanhecke}}, \bibinfo {author}
  {\bibfnamefont {P.}~\bibnamefont {Corboz}}, \bibinfo {author} {\bibfnamefont
  {J.}~\bibnamefont {Haegeman}},\ and\ \bibinfo {author} {\bibfnamefont
  {F.}~\bibnamefont {Verstraete}},\ }\href
  {https://doi.org/10.1103/PhysRevB.105.195140} {\bibfield  {journal} {\bibinfo
   {journal} {Phys. Rev. B}\ }\textbf {\bibinfo {volume} {105}},\ \bibinfo
  {pages} {195140} (\bibinfo {year} {2022})}\BibitemShut {NoStop}%
\bibitem [{\citenamefont {Jiang}\ \emph {et~al.}(2008)\citenamefont {Jiang},
  \citenamefont {Weng},\ and\ \citenamefont {Xiang}}]{PhysRevLett.101.090603}%
  \BibitemOpen
  \bibfield  {author} {\bibinfo {author} {\bibfnamefont {H.~C.}\ \bibnamefont
  {Jiang}}, \bibinfo {author} {\bibfnamefont {Z.~Y.}\ \bibnamefont {Weng}},\
  and\ \bibinfo {author} {\bibfnamefont {T.}~\bibnamefont {Xiang}},\ }\href
  {https://doi.org/10.1103/PhysRevLett.101.090603} {\bibfield  {journal}
  {\bibinfo  {journal} {Phys. Rev. Lett.}\ }\textbf {\bibinfo {volume} {101}},\
  \bibinfo {pages} {090603} (\bibinfo {year} {2008})}\BibitemShut {NoStop}%
\bibitem [{\citenamefont {Corboz}\ \emph {et~al.}(2010)\citenamefont {Corboz},
  \citenamefont {Or\'us}, \citenamefont {Bauer},\ and\ \citenamefont
  {Vidal}}]{PhysRevB.81.165104}%
  \BibitemOpen
  \bibfield  {author} {\bibinfo {author} {\bibfnamefont {P.}~\bibnamefont
  {Corboz}}, \bibinfo {author} {\bibfnamefont {R.}~\bibnamefont {Or\'us}},
  \bibinfo {author} {\bibfnamefont {B.}~\bibnamefont {Bauer}},\ and\ \bibinfo
  {author} {\bibfnamefont {G.}~\bibnamefont {Vidal}},\ }\href
  {https://doi.org/10.1103/PhysRevB.81.165104} {\bibfield  {journal} {\bibinfo
  {journal} {Phys. Rev. B}\ }\textbf {\bibinfo {volume} {81}},\ \bibinfo
  {pages} {165104} (\bibinfo {year} {2010})}\BibitemShut {NoStop}%
\bibitem [{\citenamefont {Phien}\ \emph {et~al.}(2015)\citenamefont {Phien},
  \citenamefont {Bengua}, \citenamefont {Tuan}, \citenamefont {Corboz},\ and\
  \citenamefont {Or\'us}}]{PhysRevB.92.035142}%
  \BibitemOpen
  \bibfield  {author} {\bibinfo {author} {\bibfnamefont {H.~N.}\ \bibnamefont
  {Phien}}, \bibinfo {author} {\bibfnamefont {J.~A.}\ \bibnamefont {Bengua}},
  \bibinfo {author} {\bibfnamefont {H.~D.}\ \bibnamefont {Tuan}}, \bibinfo
  {author} {\bibfnamefont {P.}~\bibnamefont {Corboz}},\ and\ \bibinfo {author}
  {\bibfnamefont {R.}~\bibnamefont {Or\'us}},\ }\href
  {https://doi.org/10.1103/PhysRevB.92.035142} {\bibfield  {journal} {\bibinfo
  {journal} {Phys. Rev. B}\ }\textbf {\bibinfo {volume} {92}},\ \bibinfo
  {pages} {035142} (\bibinfo {year} {2015})}\BibitemShut {NoStop}%
\bibitem [{\citenamefont {Rayyan}\ \emph {et~al.}(2021)\citenamefont {Rayyan},
  \citenamefont {Luo},\ and\ \citenamefont {Kee}}]{PhysRevB.104.094431}%
  \BibitemOpen
  \bibfield  {author} {\bibinfo {author} {\bibfnamefont {A.}~\bibnamefont
  {Rayyan}}, \bibinfo {author} {\bibfnamefont {Q.}~\bibnamefont {Luo}},\ and\
  \bibinfo {author} {\bibfnamefont {H.-Y.}\ \bibnamefont {Kee}},\ }\href
  {https://doi.org/10.1103/PhysRevB.104.094431} {\bibfield  {journal} {\bibinfo
   {journal} {Phys. Rev. B}\ }\textbf {\bibinfo {volume} {104}},\ \bibinfo
  {pages} {094431} (\bibinfo {year} {2021})}\BibitemShut {NoStop}%
\bibitem [{\citenamefont {Li}\ \emph {et~al.}(2022)\citenamefont {Li},
  \citenamefont {Rao}, \citenamefont {von Delft}, \citenamefont {Pollet},\ and\
  \citenamefont {Liu}}]{li2022tangle}%
  \BibitemOpen
  \bibfield  {author} {\bibinfo {author} {\bibfnamefont {J.-W.}\ \bibnamefont
  {Li}}, \bibinfo {author} {\bibfnamefont {N.}~\bibnamefont {Rao}}, \bibinfo
  {author} {\bibfnamefont {J.}~\bibnamefont {von Delft}}, \bibinfo {author}
  {\bibfnamefont {L.}~\bibnamefont {Pollet}},\ and\ \bibinfo {author}
  {\bibfnamefont {K.}~\bibnamefont {Liu}},\ }\href@noop {} {\bibinfo {title}
  {Tangle of spin double helices in the honeycomb kitaev-$\gamma$ model}}
  (\bibinfo {year} {2022}),\ \Eprint {https://arxiv.org/abs/2206.08946}
  {arXiv:2206.08946 [cond-mat.str-el]} \BibitemShut {NoStop}%
\bibitem [{\citenamefont {Liao}\ \emph {et~al.}(2019)\citenamefont {Liao},
  \citenamefont {Liu}, \citenamefont {Wang},\ and\ \citenamefont
  {Xiang}}]{liao2019differentiable}%
  \BibitemOpen
  \bibfield  {author} {\bibinfo {author} {\bibfnamefont {H.-J.}\ \bibnamefont
  {Liao}}, \bibinfo {author} {\bibfnamefont {J.-G.}\ \bibnamefont {Liu}},
  \bibinfo {author} {\bibfnamefont {L.}~\bibnamefont {Wang}},\ and\ \bibinfo
  {author} {\bibfnamefont {T.}~\bibnamefont {Xiang}},\ }\href
  {https://doi.org/10.1103/PhysRevX.9.031041} {\bibfield  {journal} {\bibinfo
  {journal} {Phys. Rev. X}\ }\textbf {\bibinfo {volume} {9}},\ \bibinfo {pages}
  {031041} (\bibinfo {year} {2019})}\BibitemShut {NoStop}%
\bibitem [{\citenamefont {Hasik}\ \emph {et~al.}(2021)\citenamefont {Hasik},
  \citenamefont {Poilblanc},\ and\ \citenamefont
  {Becca}}]{10.21468/SciPostPhys.10.1.012}%
  \BibitemOpen
  \bibfield  {author} {\bibinfo {author} {\bibfnamefont {J.}~\bibnamefont
  {Hasik}}, \bibinfo {author} {\bibfnamefont {D.}~\bibnamefont {Poilblanc}},\
  and\ \bibinfo {author} {\bibfnamefont {F.}~\bibnamefont {Becca}},\ }\href
  {https://doi.org/10.21468/SciPostPhys.10.1.012} {\bibfield  {journal}
  {\bibinfo  {journal} {SciPost Phys.}\ }\textbf {\bibinfo {volume} {10}},\
  \bibinfo {pages} {012} (\bibinfo {year} {2021})}\BibitemShut {NoStop}%
\bibitem [{\citenamefont {Xi}\ \emph {et~al.}(2021)\citenamefont {Xi},
  \citenamefont {Chen}, \citenamefont {Xie},\ and\ \citenamefont
  {Yu}}]{https://doi.org/10.48550/arxiv.2111.07368}%
  \BibitemOpen
  \bibfield  {author} {\bibinfo {author} {\bibfnamefont {N.}~\bibnamefont
  {Xi}}, \bibinfo {author} {\bibfnamefont {H.}~\bibnamefont {Chen}}, \bibinfo
  {author} {\bibfnamefont {Z.~Y.}\ \bibnamefont {Xie}},\ and\ \bibinfo {author}
  {\bibfnamefont {R.}~\bibnamefont {Yu}},\ }\href
  {https://doi.org/10.48550/ARXIV.2111.07368} {\bibinfo {title} {First-order
  transition between the plaquette valence bond solid and antiferromagnetic
  phases of the shastry-sutherland model}} (\bibinfo {year} {2021})\BibitemShut
  {NoStop}%
\bibitem [{\citenamefont {Liao}(2019)}]{Liao2019talk}%
  \BibitemOpen
  \bibfield  {author} {\bibinfo {author} {\bibfnamefont {H.-J.}\ \bibnamefont
  {Liao}},\ }\href@noop {} {\bibfield  {journal} {\bibinfo  {journal} {talk at
  Tokyo ISSP}\ } (\bibinfo {year} {2019})},\ \bibinfo {note}
  {\url{https://www.youtube.com/watch?v=Z4kTT6SKtVk} Accessed March 28,
  2023}\BibitemShut {NoStop}%
\bibitem [{\citenamefont {Hasik}\ \emph {et~al.}(2022)\citenamefont {Hasik},
  \citenamefont {Van~Damme}, \citenamefont {Poilblanc},\ and\ \citenamefont
  {Vanderstraeten}}]{PhysRevLett.129.177201}%
  \BibitemOpen
  \bibfield  {author} {\bibinfo {author} {\bibfnamefont {J.}~\bibnamefont
  {Hasik}}, \bibinfo {author} {\bibfnamefont {M.}~\bibnamefont {Van~Damme}},
  \bibinfo {author} {\bibfnamefont {D.}~\bibnamefont {Poilblanc}},\ and\
  \bibinfo {author} {\bibfnamefont {L.}~\bibnamefont {Vanderstraeten}},\ }\href
  {https://doi.org/10.1103/PhysRevLett.129.177201} {\bibfield  {journal}
  {\bibinfo  {journal} {Phys. Rev. Lett.}\ }\textbf {\bibinfo {volume} {129}},\
  \bibinfo {pages} {177201} (\bibinfo {year} {2022})}\BibitemShut {NoStop}%
\bibitem [{\citenamefont {Corboz}(2016)}]{corboz2016variational}%
  \BibitemOpen
  \bibfield  {author} {\bibinfo {author} {\bibfnamefont {P.}~\bibnamefont
  {Corboz}},\ }\href {https://doi.org/10.1103/PhysRevB.94.035133} {\bibfield
  {journal} {\bibinfo  {journal} {Phys. Rev. B}\ }\textbf {\bibinfo {volume}
  {94}},\ \bibinfo {pages} {035133} (\bibinfo {year} {2016})}\BibitemShut
  {NoStop}%
\bibitem [{\citenamefont {Vanderstraeten}\ \emph {et~al.}(2016)\citenamefont
  {Vanderstraeten}, \citenamefont {Haegeman}, \citenamefont {Corboz},\ and\
  \citenamefont {Verstraete}}]{vanderstraeten2016gradient}%
  \BibitemOpen
  \bibfield  {author} {\bibinfo {author} {\bibfnamefont {L.}~\bibnamefont
  {Vanderstraeten}}, \bibinfo {author} {\bibfnamefont {J.}~\bibnamefont
  {Haegeman}}, \bibinfo {author} {\bibfnamefont {P.}~\bibnamefont {Corboz}},\
  and\ \bibinfo {author} {\bibfnamefont {F.}~\bibnamefont {Verstraete}},\
  }\href {https://doi.org/10.1103/PhysRevB.94.155123} {\bibfield  {journal}
  {\bibinfo  {journal} {Phys. Rev. B}\ }\textbf {\bibinfo {volume} {94}},\
  \bibinfo {pages} {155123} (\bibinfo {year} {2016})}\BibitemShut {NoStop}%
\bibitem [{\citenamefont {Zhang}(2022{\natexlab{a}})}]{TeneT.jl}%
  \BibitemOpen
  \bibfield  {author} {\bibinfo {author} {\bibfnamefont {X.-Y.}\ \bibnamefont
  {Zhang}},\ }\href@noop {} {\bibinfo {title} {Vumps julia package}},\ \bibinfo
  {howpublished} {[EB/OL]} (\bibinfo {year} {2022}{\natexlab{a}}),\ \bibinfo
  {note} {\url{https://github.com/XingyuZhang2018/TeneT.jl} Accessed March 28,
  2022}\BibitemShut {NoStop}%
\bibitem [{\citenamefont {Zhang}(2022{\natexlab{b}})}]{AD-Kitaev}%
  \BibitemOpen
  \bibfield  {author} {\bibinfo {author} {\bibfnamefont {X.-Y.}\ \bibnamefont
  {Zhang}},\ }\href@noop {} {\bibinfo {title} {applications in section iii}},\
  \bibinfo {howpublished} {[EB/OL]} (\bibinfo {year} {2022}{\natexlab{b}}),\
  \bibinfo {note} {\url{https://github.com/XingyuZhang2018/AD-Kitaev} Accessed
  March 28, 2022}\BibitemShut {NoStop}%
\bibitem [{\citenamefont {Eisert}\ \emph {et~al.}(2010)\citenamefont {Eisert},
  \citenamefont {Cramer},\ and\ \citenamefont {Plenio}}]{eisert2010colloquium}%
  \BibitemOpen
  \bibfield  {author} {\bibinfo {author} {\bibfnamefont {J.}~\bibnamefont
  {Eisert}}, \bibinfo {author} {\bibfnamefont {M.}~\bibnamefont {Cramer}},\
  and\ \bibinfo {author} {\bibfnamefont {M.~B.}\ \bibnamefont {Plenio}},\
  }\href {https://doi.org/10.1103/RevModPhys.82.277} {\bibfield  {journal}
  {\bibinfo  {journal} {Rev. Mod. Phys.}\ }\textbf {\bibinfo {volume} {82}},\
  \bibinfo {pages} {277} (\bibinfo {year} {2010})}\BibitemShut {NoStop}%
\bibitem [{\citenamefont {Schollw\"ock}(2005)}]{schollwock2005density}%
  \BibitemOpen
  \bibfield  {author} {\bibinfo {author} {\bibfnamefont {U.}~\bibnamefont
  {Schollw\"ock}},\ }\href {https://doi.org/10.1103/RevModPhys.77.259}
  {\bibfield  {journal} {\bibinfo  {journal} {Rev. Mod. Phys.}\ }\textbf
  {\bibinfo {volume} {77}},\ \bibinfo {pages} {259} (\bibinfo {year}
  {2005})}\BibitemShut {NoStop}%
\bibitem [{\citenamefont {P{\'{e}}rez{-}Garc{\'{\i}}a}\ \emph
  {et~al.}(2007)\citenamefont {P{\'{e}}rez{-}Garc{\'{\i}}a}, \citenamefont
  {Verstraete}, \citenamefont {Wolf},\ and\ \citenamefont
  {Cirac}}]{perez2006matrix}%
  \BibitemOpen
  \bibfield  {author} {\bibinfo {author} {\bibfnamefont {D.}~\bibnamefont
  {P{\'{e}}rez{-}Garc{\'{\i}}a}}, \bibinfo {author} {\bibfnamefont
  {F.}~\bibnamefont {Verstraete}}, \bibinfo {author} {\bibfnamefont {M.~M.}\
  \bibnamefont {Wolf}},\ and\ \bibinfo {author} {\bibfnamefont {J.~I.}\
  \bibnamefont {Cirac}},\ }\href {https://doi.org/10.26421/QIC7.5-6-1}
  {\bibfield  {journal} {\bibinfo  {journal} {Quantum Inf. Comput.}\ }\textbf
  {\bibinfo {volume} {7}},\ \bibinfo {pages} {401} (\bibinfo {year}
  {2007})}\BibitemShut {NoStop}%
\bibitem [{\citenamefont {Corboz}\ \emph {et~al.}(2011)\citenamefont {Corboz},
  \citenamefont {White}, \citenamefont {Vidal},\ and\ \citenamefont
  {Troyer}}]{corboz2011stripes}%
  \BibitemOpen
  \bibfield  {author} {\bibinfo {author} {\bibfnamefont {P.}~\bibnamefont
  {Corboz}}, \bibinfo {author} {\bibfnamefont {S.~R.}\ \bibnamefont {White}},
  \bibinfo {author} {\bibfnamefont {G.}~\bibnamefont {Vidal}},\ and\ \bibinfo
  {author} {\bibfnamefont {M.}~\bibnamefont {Troyer}},\ }\href
  {https://doi.org/10.1103/PhysRevB.84.041108} {\bibfield  {journal} {\bibinfo
  {journal} {Phys. Rev. B}\ }\textbf {\bibinfo {volume} {84}},\ \bibinfo
  {pages} {041108} (\bibinfo {year} {2011})}\BibitemShut {NoStop}%
\bibitem [{\citenamefont {Vanhecke}\ \emph
  {et~al.}(2021{\natexlab{a}})\citenamefont {Vanhecke}, \citenamefont {Damme},
  \citenamefont {Haegeman}, \citenamefont {Vanderstraeten},\ and\ \citenamefont
  {Verstraete}}]{vanhecke2021tangent}%
  \BibitemOpen
  \bibfield  {author} {\bibinfo {author} {\bibfnamefont {B.}~\bibnamefont
  {Vanhecke}}, \bibinfo {author} {\bibfnamefont {M.~V.}\ \bibnamefont {Damme}},
  \bibinfo {author} {\bibfnamefont {J.}~\bibnamefont {Haegeman}}, \bibinfo
  {author} {\bibfnamefont {L.}~\bibnamefont {Vanderstraeten}},\ and\ \bibinfo
  {author} {\bibfnamefont {F.}~\bibnamefont {Verstraete}},\ }\href
  {https://doi.org/10.21468/SciPostPhysCore.4.1.004} {\bibfield  {journal}
  {\bibinfo  {journal} {SciPost Phys. Core}\ }\textbf {\bibinfo {volume} {4}},\
  \bibinfo {pages} {004} (\bibinfo {year} {2021}{\natexlab{a}})}\BibitemShut
  {NoStop}%
\bibitem [{\citenamefont {Arnoldi}(1951)}]{arnoldi1951principle}%
  \BibitemOpen
  \bibfield  {author} {\bibinfo {author} {\bibfnamefont {W.~E.}\ \bibnamefont
  {Arnoldi}},\ }\href {https://doi.org/10.1090/qam/42792} {\bibfield  {journal}
  {\bibinfo  {journal} {Quarterly of applied mathematics}\ }\textbf {\bibinfo
  {volume} {9}},\ \bibinfo {pages} {17} (\bibinfo {year} {1951})}\BibitemShut
  {NoStop}%
\bibitem [{\citenamefont {Vanhecke}\ \emph
  {et~al.}(2021{\natexlab{b}})\citenamefont {Vanhecke}, \citenamefont
  {Colbois}, \citenamefont {Vanderstraeten}, \citenamefont {Verstraete},\ and\
  \citenamefont {Mila}}]{PhysRevResearch.3.013041}%
  \BibitemOpen
  \bibfield  {author} {\bibinfo {author} {\bibfnamefont {B.}~\bibnamefont
  {Vanhecke}}, \bibinfo {author} {\bibfnamefont {J.}~\bibnamefont {Colbois}},
  \bibinfo {author} {\bibfnamefont {L.}~\bibnamefont {Vanderstraeten}},
  \bibinfo {author} {\bibfnamefont {F.}~\bibnamefont {Verstraete}},\ and\
  \bibinfo {author} {\bibfnamefont {F.}~\bibnamefont {Mila}},\ }\href
  {https://doi.org/10.1103/PhysRevResearch.3.013041} {\bibfield  {journal}
  {\bibinfo  {journal} {Phys. Rev. Res.}\ }\textbf {\bibinfo {volume} {3}},\
  \bibinfo {pages} {013041} (\bibinfo {year} {2021}{\natexlab{b}})}\BibitemShut
  {NoStop}%
\bibitem [{\citenamefont {Nocedal}\ and\ \citenamefont
  {Wright}(2006)}]{nocedal2006numerical}%
  \BibitemOpen
  \bibfield  {author} {\bibinfo {author} {\bibfnamefont {J.}~\bibnamefont
  {Nocedal}}\ and\ \bibinfo {author} {\bibfnamefont {S.}~\bibnamefont
  {Wright}},\ }\href@noop {} {\bibfield  {journal} {\bibinfo  {journal} {New
  York}\ } (\bibinfo {year} {2006})}\BibitemShut {NoStop}%
\bibitem [{\citenamefont {Baur}\ and\ \citenamefont
  {Strassen}(1983)}]{BAUR1983317}%
  \BibitemOpen
  \bibfield  {author} {\bibinfo {author} {\bibfnamefont {W.}~\bibnamefont
  {Baur}}\ and\ \bibinfo {author} {\bibfnamefont {V.}~\bibnamefont
  {Strassen}},\ }\href
  {https://doi.org/https://doi.org/10.1016/0304-3975(83)90110-X} {\bibfield
  {journal} {\bibinfo  {journal} {Theoretical Computer Science}\ }\textbf
  {\bibinfo {volume} {22}},\ \bibinfo {pages} {317} (\bibinfo {year}
  {1983})}\BibitemShut {NoStop}%
\bibitem [{\citenamefont {Paszke}\ \emph {et~al.}(2017)\citenamefont {Paszke},
  \citenamefont {Gross}, \citenamefont {Chintala}, \citenamefont {Chanan},
  \citenamefont {Yang}, \citenamefont {DeVito}, \citenamefont {Lin},
  \citenamefont {Desmaison}, \citenamefont {Antiga},\ and\ \citenamefont
  {Lerer}}]{paszke2017automatic}%
  \BibitemOpen
  \bibfield  {author} {\bibinfo {author} {\bibfnamefont {A.}~\bibnamefont
  {Paszke}}, \bibinfo {author} {\bibfnamefont {S.}~\bibnamefont {Gross}},
  \bibinfo {author} {\bibfnamefont {S.}~\bibnamefont {Chintala}}, \bibinfo
  {author} {\bibfnamefont {G.}~\bibnamefont {Chanan}}, \bibinfo {author}
  {\bibfnamefont {E.}~\bibnamefont {Yang}}, \bibinfo {author} {\bibfnamefont
  {Z.}~\bibnamefont {DeVito}}, \bibinfo {author} {\bibfnamefont
  {Z.}~\bibnamefont {Lin}}, \bibinfo {author} {\bibfnamefont {A.}~\bibnamefont
  {Desmaison}}, \bibinfo {author} {\bibfnamefont {L.}~\bibnamefont {Antiga}},\
  and\ \bibinfo {author} {\bibfnamefont {A.}~\bibnamefont {Lerer}},\ }\href
  {https://openreview.net/pdf?id=BJJsrmfCZ} {\bibinfo {title} {Automatic
  differentiation in pytorch}} (\bibinfo {year} {2017})\BibitemShut {NoStop}%
\bibitem [{\citenamefont {Frostig}\ \emph {et~al.}(2018)\citenamefont
  {Frostig}, \citenamefont {Johnson},\ and\ \citenamefont
  {Leary}}]{frostig2018compiling}%
  \BibitemOpen
  \bibfield  {author} {\bibinfo {author} {\bibfnamefont {R.}~\bibnamefont
  {Frostig}}, \bibinfo {author} {\bibfnamefont {M.~J.}\ \bibnamefont
  {Johnson}},\ and\ \bibinfo {author} {\bibfnamefont {C.}~\bibnamefont
  {Leary}},\ }\href {https://mlsys.org/Conferences/doc/2018/146.pdf} {\bibfield
   {journal} {\bibinfo  {journal} {Systems for Machine Learning}\ }\textbf
  {\bibinfo {volume} {4}} (\bibinfo {year} {2018})}\BibitemShut {NoStop}%
\bibitem [{\citenamefont {Innes}(2019)}]{innes2018don}%
  \BibitemOpen
  \bibfield  {author} {\bibinfo {author} {\bibfnamefont {M.}~\bibnamefont
  {Innes}},\ }\href@noop {} {\bibinfo {title} {Don't unroll adjoint:
  Differentiating ssa-form programs}} (\bibinfo {year} {2019}),\ \Eprint
  {https://arxiv.org/abs/1810.07951} {arXiv:1810.07951 [cs.PL]} \BibitemShut
  {NoStop}%
\bibitem [{\citenamefont {Seeger}\ \emph {et~al.}(2019)\citenamefont {Seeger},
  \citenamefont {Hetzel}, \citenamefont {Dai}, \citenamefont {Meissner},\ and\
  \citenamefont {Lawrence}}]{seeger2017auto}%
  \BibitemOpen
  \bibfield  {author} {\bibinfo {author} {\bibfnamefont {M.}~\bibnamefont
  {Seeger}}, \bibinfo {author} {\bibfnamefont {A.}~\bibnamefont {Hetzel}},
  \bibinfo {author} {\bibfnamefont {Z.}~\bibnamefont {Dai}}, \bibinfo {author}
  {\bibfnamefont {E.}~\bibnamefont {Meissner}},\ and\ \bibinfo {author}
  {\bibfnamefont {N.~D.}\ \bibnamefont {Lawrence}},\ }\href@noop {} {\bibinfo
  {title} {Auto-differentiating linear algebra}} (\bibinfo {year} {2019}),\
  \Eprint {https://arxiv.org/abs/1710.08717} {arXiv:1710.08717 [cs.MS]}
  \BibitemShut {NoStop}%
\bibitem [{\citenamefont {Xie}\ \emph {et~al.}(2020)\citenamefont {Xie},
  \citenamefont {Liu},\ and\ \citenamefont {Wang}}]{xie2020automatic}%
  \BibitemOpen
  \bibfield  {author} {\bibinfo {author} {\bibfnamefont {H.}~\bibnamefont
  {Xie}}, \bibinfo {author} {\bibfnamefont {J.-G.}\ \bibnamefont {Liu}},\ and\
  \bibinfo {author} {\bibfnamefont {L.}~\bibnamefont {Wang}},\ }\href
  {https://doi.org/10.1103/PhysRevB.101.245139} {\bibfield  {journal} {\bibinfo
   {journal} {Phys. Rev. B}\ }\textbf {\bibinfo {volume} {101}},\ \bibinfo
  {pages} {245139} (\bibinfo {year} {2020})}\BibitemShut {NoStop}%
\bibitem [{\citenamefont {Saad}\ and\ \citenamefont
  {Schultz}(1986)}]{saad1986gmres}%
  \BibitemOpen
  \bibfield  {author} {\bibinfo {author} {\bibfnamefont {Y.}~\bibnamefont
  {Saad}}\ and\ \bibinfo {author} {\bibfnamefont {M.~H.}\ \bibnamefont
  {Schultz}},\ }\href {https://doi.org/10.1137/0907058} {\bibfield  {journal}
  {\bibinfo  {journal} {SIAM Journal on Scientific and Statistical Computing}\
  }\textbf {\bibinfo {volume} {7}},\ \bibinfo {pages} {856} (\bibinfo {year}
  {1986})}\BibitemShut {NoStop}%
\bibitem [{\citenamefont {Lehoucq}\ \emph {et~al.}(1998)\citenamefont
  {Lehoucq}, \citenamefont {Sorensen},\ and\ \citenamefont
  {Yang}}]{lehoucq1998arpack}%
  \BibitemOpen
  \bibfield  {author} {\bibinfo {author} {\bibfnamefont {R.~B.}\ \bibnamefont
  {Lehoucq}}, \bibinfo {author} {\bibfnamefont {D.~C.}\ \bibnamefont
  {Sorensen}},\ and\ \bibinfo {author} {\bibfnamefont {C.}~\bibnamefont
  {Yang}},\ }\href {https://doi.org/10.1137/1.9780898719628} {\emph {\bibinfo
  {title} {ARPACK Users' Guide}}}\ (\bibinfo  {publisher} {Society for
  Industrial and Applied Mathematics},\ \bibinfo {address} {University City,
  Philadelphia},\ \bibinfo {year} {1998})\ \Eprint
  {https://arxiv.org/abs/https://epubs.siam.org/doi/pdf/10.1137/1.9780898719628}
  {https://epubs.siam.org/doi/pdf/10.1137/1.9780898719628} \BibitemShut
  {NoStop}%
\bibitem [{\citenamefont {Haegeman}(2022)}]{KrylovKit.jl}%
  \BibitemOpen
  \bibfield  {author} {\bibinfo {author} {\bibfnamefont {J.}~\bibnamefont
  {Haegeman}},\ }\href@noop {} {\bibinfo {title} {julia package for krylov
  space method}},\ \bibinfo {howpublished} {[EB/OL]} (\bibinfo {year} {2022}),\
  \bibinfo {note} {\url{https://github.com/Jutho/KrylovKit.jl} Accessed March
  28, 2022}\BibitemShut {NoStop}%
\bibitem [{\citenamefont {Rau}\ \emph {et~al.}(2014)\citenamefont {Rau},
  \citenamefont {Lee},\ and\ \citenamefont {Kee}}]{PhysRevLett.112.077204}%
  \BibitemOpen
  \bibfield  {author} {\bibinfo {author} {\bibfnamefont {J.~G.}\ \bibnamefont
  {Rau}}, \bibinfo {author} {\bibfnamefont {E.~K.-H.}\ \bibnamefont {Lee}},\
  and\ \bibinfo {author} {\bibfnamefont {H.-Y.}\ \bibnamefont {Kee}},\ }\href
  {https://doi.org/10.1103/PhysRevLett.112.077204} {\bibfield  {journal}
  {\bibinfo  {journal} {Phys. Rev. Lett.}\ }\textbf {\bibinfo {volume} {112}},\
  \bibinfo {pages} {077204} (\bibinfo {year} {2014})}\BibitemShut {NoStop}%
\bibitem [{\citenamefont {Chaloupka}\ \emph {et~al.}(2010)\citenamefont
  {Chaloupka}, \citenamefont {Jackeli},\ and\ \citenamefont
  {Khaliullin}}]{PhysRevLett.105.027204}%
  \BibitemOpen
  \bibfield  {author} {\bibinfo {author} {\bibfnamefont {J.~c.~v.}\
  \bibnamefont {Chaloupka}}, \bibinfo {author} {\bibfnamefont {G.}~\bibnamefont
  {Jackeli}},\ and\ \bibinfo {author} {\bibfnamefont {G.}~\bibnamefont
  {Khaliullin}},\ }\href {https://doi.org/10.1103/PhysRevLett.105.027204}
  {\bibfield  {journal} {\bibinfo  {journal} {Phys. Rev. Lett.}\ }\textbf
  {\bibinfo {volume} {105}},\ \bibinfo {pages} {027204} (\bibinfo {year}
  {2010})}\BibitemShut {NoStop}%
\bibitem [{Note1()}]{Note1}%
  \BibitemOpen
  \bibinfo {note} {In contrast with brick-wall lattice~\cite
  {iregui2014probing}, this transformation merging two sites contains twice the
  physical bonds but half the number of unit cells. In practice, we find such a
  scheme is more effective and thus use it in later simulations.}\BibitemShut
  {Stop}%
\bibitem [{\citenamefont {Lee}\ \emph {et~al.}(2019)\citenamefont {Lee},
  \citenamefont {Kaneko}, \citenamefont {Okubo},\ and\ \citenamefont
  {Kawashima}}]{PhysRevLett.123.087203}%
  \BibitemOpen
  \bibfield  {author} {\bibinfo {author} {\bibfnamefont {H.-Y.}\ \bibnamefont
  {Lee}}, \bibinfo {author} {\bibfnamefont {R.}~\bibnamefont {Kaneko}},
  \bibinfo {author} {\bibfnamefont {T.}~\bibnamefont {Okubo}},\ and\ \bibinfo
  {author} {\bibfnamefont {N.}~\bibnamefont {Kawashima}},\ }\href
  {https://doi.org/10.1103/PhysRevLett.123.087203} {\bibfield  {journal}
  {\bibinfo  {journal} {Phys. Rev. Lett.}\ }\textbf {\bibinfo {volume} {123}},\
  \bibinfo {pages} {087203} (\bibinfo {year} {2019})}\BibitemShut {NoStop}%
\bibitem [{\citenamefont {Lukin}\ and\ \citenamefont
  {Sotnikov}(2023)}]{PhysRevB.107.054424}%
  \BibitemOpen
  \bibfield  {author} {\bibinfo {author} {\bibfnamefont {I.~V.}\ \bibnamefont
  {Lukin}}\ and\ \bibinfo {author} {\bibfnamefont {A.~G.}\ \bibnamefont
  {Sotnikov}},\ }\href {https://doi.org/10.1103/PhysRevB.107.054424} {\bibfield
   {journal} {\bibinfo  {journal} {Phys. Rev. B}\ }\textbf {\bibinfo {volume}
  {107}},\ \bibinfo {pages} {054424} (\bibinfo {year} {2023})}\BibitemShut
  {NoStop}%
\bibitem [{\citenamefont {Aoyama}\ \emph {et~al.}(2017)\citenamefont {Aoyama},
  \citenamefont {Hasegawa}, \citenamefont {Kimura}, \citenamefont {Kimura},\
  and\ \citenamefont {Ohgushi}}]{PhysRevB.95.245104}%
  \BibitemOpen
  \bibfield  {author} {\bibinfo {author} {\bibfnamefont {T.}~\bibnamefont
  {Aoyama}}, \bibinfo {author} {\bibfnamefont {Y.}~\bibnamefont {Hasegawa}},
  \bibinfo {author} {\bibfnamefont {S.}~\bibnamefont {Kimura}}, \bibinfo
  {author} {\bibfnamefont {T.}~\bibnamefont {Kimura}},\ and\ \bibinfo {author}
  {\bibfnamefont {K.}~\bibnamefont {Ohgushi}},\ }\href
  {https://doi.org/10.1103/PhysRevB.95.245104} {\bibfield  {journal} {\bibinfo
  {journal} {Phys. Rev. B}\ }\textbf {\bibinfo {volume} {95}},\ \bibinfo
  {pages} {245104} (\bibinfo {year} {2017})}\BibitemShut {NoStop}%
\bibitem [{\citenamefont {{Banerjee}}\ \emph {et~al.}(2017)\citenamefont
  {{Banerjee}}, \citenamefont {{Yan}}, \citenamefont {{Knolle}}, \citenamefont
  {{Bridges}}, \citenamefont {{Stone}}, \citenamefont {{Lumsden}},
  \citenamefont {{Mandrus}}, \citenamefont {{Tennant}}, \citenamefont
  {{Moessner}},\ and\ \citenamefont {{Nagler}}}]{Banerjee2017}%
  \BibitemOpen
  \bibfield  {author} {\bibinfo {author} {\bibfnamefont {A.}~\bibnamefont
  {{Banerjee}}}, \bibinfo {author} {\bibfnamefont {J.}~\bibnamefont {{Yan}}},
  \bibinfo {author} {\bibfnamefont {J.}~\bibnamefont {{Knolle}}}, \bibinfo
  {author} {\bibfnamefont {C.~A.}\ \bibnamefont {{Bridges}}}, \bibinfo {author}
  {\bibfnamefont {M.~B.}\ \bibnamefont {{Stone}}}, \bibinfo {author}
  {\bibfnamefont {M.~D.}\ \bibnamefont {{Lumsden}}}, \bibinfo {author}
  {\bibfnamefont {D.~G.}\ \bibnamefont {{Mandrus}}}, \bibinfo {author}
  {\bibfnamefont {D.~A.}\ \bibnamefont {{Tennant}}}, \bibinfo {author}
  {\bibfnamefont {R.}~\bibnamefont {{Moessner}}},\ and\ \bibinfo {author}
  {\bibfnamefont {S.~E.}\ \bibnamefont {{Nagler}}},\ }\href
  {https://doi.org/10.1126/science.aah6015} {\bibfield  {journal} {\bibinfo
  {journal} {Science}\ }\textbf {\bibinfo {volume} {356}},\ \bibinfo {pages}
  {1055} (\bibinfo {year} {2017})}\BibitemShut {NoStop}%
\bibitem [{\citenamefont {Zheng}\ \emph {et~al.}(2017)\citenamefont {Zheng},
  \citenamefont {Ran}, \citenamefont {Li}, \citenamefont {Wang}, \citenamefont
  {Wang}, \citenamefont {Liu}, \citenamefont {Liu}, \citenamefont {Normand},
  \citenamefont {Wen},\ and\ \citenamefont {Yu}}]{Zheng2017}%
  \BibitemOpen
  \bibfield  {author} {\bibinfo {author} {\bibfnamefont {J.}~\bibnamefont
  {Zheng}}, \bibinfo {author} {\bibfnamefont {K.}~\bibnamefont {Ran}}, \bibinfo
  {author} {\bibfnamefont {T.}~\bibnamefont {Li}}, \bibinfo {author}
  {\bibfnamefont {J.}~\bibnamefont {Wang}}, \bibinfo {author} {\bibfnamefont
  {P.}~\bibnamefont {Wang}}, \bibinfo {author} {\bibfnamefont {B.}~\bibnamefont
  {Liu}}, \bibinfo {author} {\bibfnamefont {Z.-X.}\ \bibnamefont {Liu}},
  \bibinfo {author} {\bibfnamefont {B.}~\bibnamefont {Normand}}, \bibinfo
  {author} {\bibfnamefont {J.}~\bibnamefont {Wen}},\ and\ \bibinfo {author}
  {\bibfnamefont {W.}~\bibnamefont {Yu}},\ }\href
  {https://doi.org/10.1103/PhysRevLett.119.227208} {\bibfield  {journal}
  {\bibinfo  {journal} {Phys. Rev. Lett.}\ }\textbf {\bibinfo {volume} {119}},\
  \bibinfo {pages} {227208} (\bibinfo {year} {2017})}\BibitemShut {NoStop}%
\bibitem [{\citenamefont {Modic}\ \emph {et~al.}(2021)\citenamefont {Modic},
  \citenamefont {McDonald}, \citenamefont {Ruff}, \citenamefont {Bachmann},
  \citenamefont {Lai}, \citenamefont {Palmstrom}, \citenamefont {Graf},
  \citenamefont {Chan}, \citenamefont {Balakirev}, \citenamefont {Betts},
  \citenamefont {Boebinger}, \citenamefont {Schmidt}, \citenamefont {Lawler},
  \citenamefont {Sokolov}, \citenamefont {Moll}, \citenamefont {Ramshaw},\ and\
  \citenamefont {Shekhter}}]{modic2021scale}%
  \BibitemOpen
  \bibfield  {author} {\bibinfo {author} {\bibfnamefont {K.~A.}\ \bibnamefont
  {Modic}}, \bibinfo {author} {\bibfnamefont {R.~D.}\ \bibnamefont {McDonald}},
  \bibinfo {author} {\bibfnamefont {J.~P.~C.}\ \bibnamefont {Ruff}}, \bibinfo
  {author} {\bibfnamefont {M.~D.}\ \bibnamefont {Bachmann}}, \bibinfo {author}
  {\bibfnamefont {Y.}~\bibnamefont {Lai}}, \bibinfo {author} {\bibfnamefont
  {J.~C.}\ \bibnamefont {Palmstrom}}, \bibinfo {author} {\bibfnamefont
  {D.}~\bibnamefont {Graf}}, \bibinfo {author} {\bibfnamefont {M.~K.}\
  \bibnamefont {Chan}}, \bibinfo {author} {\bibfnamefont {F.~F.}\ \bibnamefont
  {Balakirev}}, \bibinfo {author} {\bibfnamefont {J.~B.}\ \bibnamefont
  {Betts}}, \bibinfo {author} {\bibfnamefont {G.~S.}\ \bibnamefont
  {Boebinger}}, \bibinfo {author} {\bibfnamefont {M.}~\bibnamefont {Schmidt}},
  \bibinfo {author} {\bibfnamefont {M.~J.}\ \bibnamefont {Lawler}}, \bibinfo
  {author} {\bibfnamefont {D.~A.}\ \bibnamefont {Sokolov}}, \bibinfo {author}
  {\bibfnamefont {P.~J.~W.}\ \bibnamefont {Moll}}, \bibinfo {author}
  {\bibfnamefont {B.~J.}\ \bibnamefont {Ramshaw}},\ and\ \bibinfo {author}
  {\bibfnamefont {A.}~\bibnamefont {Shekhter}},\ }\href
  {https://doi.org/10.1038/s41567-020-1028-0} {\bibfield  {journal} {\bibinfo
  {journal} {Nature Physics}\ }\textbf {\bibinfo {volume} {17}},\ \bibinfo
  {pages} {240} (\bibinfo {year} {2021})}\BibitemShut {NoStop}%
\end{thebibliography}%

\end{document}